\documentclass[11pt,a4paper]{article}

\renewcommand{\baselinestretch}{1}

\usepackage{fullpage}
\usepackage{titlesec}
\titleformat*{\section}{\large\bfseries}

\usepackage{latexsym,amssymb,amsthm,amsmath,amsfonts,graphicx,booktabs} 
\usepackage[round]{natbib}
\usepackage[linesnumbered,lined,ruled,longend]{algorithm2e}

\newcommand{\setV}{\mathcal{V}}
\newcommand{\diag}{\text{diag}}
\newcommand{\Var}{\text{var}}
\newcommand{\Cov}{\text{cov}}
\newcommand{\X}{\mathcal{X}}
\newcommand{\Tc}{\mathcal{T}}
\newcommand{\T}{\mathrm{\scriptscriptstyle T}}
\newcommand{\tr}{\text{tr}}
\newcommand{\supp}{\text{supp}}

\newtheorem{theorem}{Theorem}
\newtheorem{proposition}[theorem]{Proposition}

\usepackage{collab}
\collabAuthor{wm}{red!85}{Werner}
\collabAuthor{mh}{magenta!85}{Markus}
\collabAuthor{ap}{blue!85}{Andrej}

\usepackage{xcolor}
\definecolor{rev1}{rgb}{0.82,0.1,0.26}

\begin{document}

\begin{center}
\renewcommand{\baselinestretch}{1}
{ \Large \bf Practical aspects of the virtual noise convex optimum design approach for correlated responses
}

\vspace{0.5cm}

{\large
Markus~Hainy$^a$\footnote{Corresponding author: \texttt{markus.hainy@jku.at}},
Werner~G.~M\"uller$^a$ and
Andrej~P\'azman$^b$ \\[3ex]
$^a$ Institute of Applied Statistics, Johannes Kepler University, Altenberger Stra{\ss}e 69, 4040 Linz, Austria  \\[1ex]
$^b$ Faculty of Mathematics, Physics and Informatics, Comenius University, \v{S}af\'{a}rikovo~n\'{a}mestie~6, 814 99 Bratislava, Slovakia
}\end{center}

\renewcommand{\baselinestretch}{1}

\noindent Abstract: 
In this paper we present several practically-oriented extensions and considerations for the virtual noise method in optimal design under correlation. First we introduce a slightly modified virtual noise representation which further illuminates the parallels to the classical design approach for uncorrelated observations. We suggest more efficient algorithms to obtain the design measures. Furthermore, we show that various convex relaxation methods used for sensor selection are special cases of our approach and can be solved within our framework. Finally, we provide practical guidelines on how to generally approach a design problem with correlated observations and demonstrate how to utilize the virtual noise method in this context in a meaningful way.

\noindent Keywords: Correlated response; design algorithm; Gaussian processes; sensor selection.

\section{Introduction}

Optimal design of experiments for Gaussian process regression is an increasingly relevant and active research topic. For recent reviews with attention to this issue see e.g. \cite{lopez-fidalgo_optimal_2023} and \cite{huan_optimal_2024}. While there has been remarkable progress on the topic stemming from the work of Dette, Zhigljavsky and coauthors culminating in \cite{dette_blue_2019}, another successful approach is the virtual noise method. In \cite{pazman_correlated_2022} it has recently been complemented by a convex formulation leading to an equivalence theorem comparable to the uncorrelated case. Its main advantage is that the corresponding algorithm gives an upper performance bound against which alternative design methods can be judged. The design measure obtained through this approach can also be used to generate exact designs.

Section~\ref{sec2} reviews the general setup and presents a slight modification which allows a direct connection to classical design measures in the uncorrelated case. Section~\ref{sec3} explores relations to methods of convex relaxation that have recently been proposed and shows that those can be considered as special cases of the virtual noise method. Section~\ref{sec4} describes and compares several computational algorithms that are available for the optimization task and Section~\ref{sec5} exemplifies the methods in some typical situations. Section~\ref{sec6} eventually provides guidelines on how to proceed in an experiment where correlated observations have to be expected.

\section{Model}

\label{sec2}

\subsection{Regression model with correlated observations}

As in \cite{pazman_correlated_2022}, we consider optimum experimental design for estimating the parameter $\theta$ in the regression model
\begin{equation}
y(x) = f^{\T} (x) \, \theta + \varepsilon(x); \quad x \in \X = \{x_1,\ldots,x_N\}, \: \theta \in \mathbb{R}^p, \label{eq:model_orig}
\end{equation}
where the discrete space $\X$ is assumed to be finite with $N \in \mathbb{N}$ elements, $\theta$ is the $p$-dimensional real-valued parameter, and $f(x)$ is a known $p$-dimensional function defined on $x \in \X$. The error terms $\varepsilon(x)$ are assumed to have zero mean, so $E\{\varepsilon(x)\} = 0$ for all $x \in \X$, but in contrast to the standard regression model there are no restrictions such as homogeneous variances or zero correlations placed on the covariance structure of the points in the set $\X$. The covariance structure is defined by the symmetric and positive 
definite $N \times N$ covariance matrix $C$ with entries
\begin{equation*}
C_{ij} = \Cov\{\varepsilon(x_i), \varepsilon(x_j)\}; \qquad i,j = 1,\ldots,N.
\end{equation*}
With the normality assumption this setup has become a popular choice for approximating all kinds of phenomena under the name of Gaussian process regression, for a good overview see \cite{rasmussen_gaussian_2005} or \cite{gramacy_surrogates_2020}.

Throughout this paper, we assume that the covariance matrix is known and does not need to be estimated. When $C$ is unknown, which is the common case in practice, one often substitutes $C$ with an estimated variance-covariance matrix obtained, for example, by variogram estimation from preliminary data \citep[see][]{cressie_spatial_1993}.

Given a fixed budget of $n \ll N$ observations that might be taken at the $n$ distinct points $x_{i_1},\ldots,x_{i_n} \in \X$, the goal of experimental design is to select those $n$ design points where some increasing and concave criterion $\Phi$ of the information matrix is maximized. Let $\tau = \{x_{i_1},\ldots,x_{i_n}\}$ denote the exact design, let $F(\tau) = (f(x_{i_1}) \cdots f(x_{i_n}))^{\T}$ be the $n \times p$ model matrix and let $C(\tau)$ be the $n \times n$ covariance matrix with entries $C_{jk} = \Cov\{\varepsilon(x_{i_j}), \varepsilon(x_{i_k})\}$ for $j,k = 1,\ldots,n$. The information matrix of the best linear unbiased estimator (BLUE) of $\theta$ for the design $\tau$ is then given by
\begin{equation}
M(\tau) = F(\tau)^{\T} C(\tau)^{-1} F(\tau), \label{eq:information_matrix_exact}
\end{equation}
see \cite{nather_effective_1985}.
If $\tau = \X$, we will omit the explicit dependency on $\tau$ and write $F(\X) = F$ and $C(\X) = C$.

Therefore, to find the optimal exact $n$-point design, one has to maximize the criterion
\begin{equation*}
\Phi\{M(\tau)\}
\end{equation*}
with respect to all $\tau \in \Tc_n$, where $\Tc_n$ is the set of all $\tau \subseteq \X$ of size $n$. This is a challenging combinatorial problem \citep[cf.][]{atkinson_optimum_2007-1}. Common criteria are $\Phi(M) = \log \det(M)$ (D-optimality) or $\Phi(M) = -\tr(M^{-1})$ (A-optimality).

\subsection{The original virtual noise model}

In \cite{pazman_new_1998} and \cite{muller_measures_2003}, the authors introduced design measures for the correlated setting.
Unfortunately, their approach did not yield criteria that were concave in the design measures, so in general their approach could only be used to find designs which are locally optimum. 
However, this problem was resolved in \cite{pazman_correlated_2022}, where a new formulation was introduced which leads to a concave criterion.

In these approaches a so-called ``virtual noise'' component $u_{\xi}(x)$ is added to the original model \eqref{eq:model_orig} for purely computational purposes. The modified model is
\begin{equation}
y(x) = f^{\T} (x) \, \theta + \varepsilon(x) + u_{\xi}(x); \quad x \in \X, \: \theta \in \mathbb{R}^p. \label{eq:model_virtual_noise}
\end{equation}
In \cite{pazman_correlated_2022}, the following explicit form is used for the variance of $u_{\xi}(x)$:
\begin{equation}
\Var\{u_{\xi}(x)\} = \kappa \, \frac{1/n - \xi(x)}{\xi(x)} = \kappa \, \left[\frac{1}{n \, \xi(x)} - 1 \right], \label{eq:variance_virtual_noise_original}
\end{equation}
where the measure $\xi = (\xi(x_1),\ldots,\xi(x_N))$ is taken from the restricted set of probability measures
\begin{equation*}
\Xi_+ = \left\{ \xi: \: \sum_{x \in \X} \xi(x) = 1, \; \forall_{x \in \X} \: 0 < \xi(x) \leq 1/n \right\}. 
\end{equation*}
Furthermore, $\Cov\{u_{\xi}(x_i), u_{\xi}(x_j)\} = 0$ for $x_i \neq x_j$ and $u_{\xi}(x)$ and $\varepsilon(x)$ are independent. 

The variance-covariance matrix of $\varepsilon(x) + u_{\xi}(x)$ over the set $\X = \{x_1,\ldots,x_N\}$ is given by $C + W_{\kappa}(\xi)$, where $W_{\kappa}(\xi)$ is a diagonal matrix whose diagonal elements are the virtual noise variances \eqref{eq:variance_virtual_noise_original}, i.e.,
\begin{equation}
\left[W_{\kappa}(\xi)\right]_{i,i} = \Var\{u_{\xi}(x_i)\} = \kappa \, \frac{1/n - \xi(x_i)}{\xi(x_i)}, \quad i = 1,\ldots,N. \label{eq:virtual_noise_variance_matrix}
\end{equation}

The information matrix of this model can then be expressed as
\begin{equation}
M(\xi) = F^{\T} \left[C + W_{\kappa}(\xi) \right]^{-1} F. \label{eq:infmatrix_vn_fullsupport}
\end{equation}

The definition can be extended to also include design measures with a value of exactly $0$ at some points $x \in \X$, so that
\begin{equation}
\xi \in \Xi = \left\{ \xi: \: \sum_{x \in \X} \xi(x) = 1, \; \forall_{x \in \X} \: 0 \leq \xi(x) \leq 1/n \right\}.  \label{eq:Xi_set}
\end{equation}

Let $C'$ denote the submatrix of $C$ and $W_{\kappa}'(\xi)$ the submatrix of $W_{\kappa}(\xi)$ where all rows and columns corresponding to design points $x$ with $\xi(x) = 0$ are removed, and let $F'$ denote the submatrix of $F$ where all rows corresponding to points $x$ with $\xi(x) = 0$ are removed. Then the information matrix of the virtual noise model is given by
\begin{equation}
M(\xi) = (F')^{\T} \left[C' + W_{\kappa}'(\xi) \right]^{-1} F'. \label{eq:infmatrix_vn_orig}
\end{equation}

It can be shown that if $\kappa$ is not larger than the smallest eigenvalue of $C$ and $\Phi(M)$ is a concave, increasing, and continuous function of the matrix $M$, then the mapping $\xi \in \Xi \rightarrow \Phi\{M(\xi)\}$ is continuous and concave, see Theorem 1 in \cite{pazman_correlated_2022} with a detailed proof.

For design points $x$ with $\xi(x) = 1/n$, the observation is not perturbed by the virtual noise at all. If $\xi(x) \rightarrow 0$, then $\Var\{u_{\xi}(x)\} \rightarrow \infty$, and no observation will be made at that design point. The higher the design measure $\xi(x)$, the lower the amount of signal suppression at the point $x$. Therefore, the measure $\xi(x)$ can be interpreted as signifying the ``importance'' of the point $x$. Unlike in the uncorrelated setting, however, this notion of importance does not translate into the interpretation of the design measures as being proportional to the number of replications at $x$, since no replications are possible at any design point $x$.

If $\xi(x) = 1/n$ at the $n$ points $x \in \tau = \{x_{i_1},\ldots,x_{i_n}\}$ and $\xi(x) = 0$ for all $x \in \X \backslash \tau$, then $\Var\{u_{\xi}(x)\} = 0$ for all $x \in \tau$, and all the rows of $F$ and the rows and columns of $C$ and $W_{\kappa}(\xi)$ corresponding to all $x \in \X \backslash \tau$ are removed. Therefore, the information matrix $M(\xi)$ given by Eq.~\eqref{eq:infmatrix_vn_orig} turns into the information matrix $M(\tau)$ of the exact design $\tau$ given by Eq.~\eqref{eq:information_matrix_exact}. 

Note that when using the virtual noise variance \eqref{eq:variance_virtual_noise_original} and the corresponding design measure set~\eqref{eq:Xi_set} with upper limit $1/n$, it is not possible to obtain the information matrix for any exact design $\tau_{\dagger} \in \mathcal{T}_m$ for $m \neq n$ as a special case of the information matrix \eqref{eq:infmatrix_vn_orig}.
Therefore, optimizing $\Phi\{M(\xi)\}$ over $\xi \in \Xi$ yields an upper bound to the criterion for the optimal exact $n$-point design given by  $\Phi\{M(\tau^*)\}$ with $\tau^* = \underset{\tau \in \Tc_n}{\arg \max} \: \Phi\{M(\tau)\}$. Furthermore, the optimal design $\xi^* =  \underset{\xi \in \Xi}{\arg \max} \: \Phi\{M(\xi)\}$ can be used to help find good exact designs, for example by randomly selecting $n$ design points without replacement according to the design measure $\xi^*$.

\subsection{The modified virtual noise variance}

In this paper, we suggest a slightly modified variance formulation of the virtual noise component because of its property that the classical information matrix for the uncorrelated setting is a special case. Therefore, the design measures we obtain can be understood as a generalization of Kiefer's design measures in the classical uncorrelated setting \citep{kiefer_optimum_1959}. In addition, the new formulation can increase the computational stability and efficiency in some examples.

The model we consider is again the virtual noise model \eqref{eq:model_virtual_noise}. Let the diagonal elements of the variance-covariance matrix $C$ be denoted by $\sigma^2(x_1),\ldots,\sigma^2(x_N)$ and let the diagonal matrix $\Sigma$ have diagonal elements $\Sigma_{i,i} = \sigma^2(x_i)$ for $i = 1,\ldots,N$. The correlation matrix of the error term $\varepsilon(x)$, $x \in \X$, can then be calculated as $K = \Sigma^{-1/2} C \Sigma^{-1/2}$. Again, we denote the submatrices of $C$, $K$, $\Sigma$, and $F$ corresponding to points $x \in \supp(\xi)$ by $C'$, $K'$, $\Sigma'$, and $F'$, respectively.

We replace the variance formulation \eqref{eq:variance_virtual_noise_original} with the following variance:
\begin{equation}
\Var\{u_{\xi}(x)\} = \tilde{\kappa} \, \sigma^2(x) \, \frac{1/n - \xi(x)}{\xi(x)} = \tilde{\kappa} \, \sigma^2(x) \left[\frac{1}{n \, \xi(x)} - 1 \right], \quad x \in \supp(\xi), \label{eq:variance_virtual_noise_modified}
\end{equation}
where the design measure $\xi(x)$ is taken from the set $\Xi$ given by \eqref{eq:Xi_set}. Define $\widetilde{W}_{\tilde{\kappa}}'(\xi)$ to be the diagonal matrix whose diagonal entries consist of the modified virtual noise variances \eqref{eq:variance_virtual_noise_modified} for all $x \in \supp(\xi)$.
The information matrix for this virtual noise variance formulation is
\begin{equation}
\widetilde{M}(\xi) = (F')^{\T} \left(C' + \widetilde{W}_{\tilde{\kappa}}'(\xi) \right)^{-1} F'. \label{eq:information_matrix_modified}
\end{equation}

\begin{theorem}
	\label{th:modified_concave}
	If  $
	\tilde{\kappa} \le \lambda_{\min }\left( K\right), 
	$
	the minimal eigenvalue of the correlation matrix $K$, and if $\Phi(M) $ is any optimality criterion expressed as a
	concave, increasing, and continuous function on the set of all positive semi-definite matrices $M$, then the mapping  
	\[
	\xi \in \Xi \rightarrow \Phi \left\{ \widetilde{M}( \xi ) \right\}
	\]
	is concave as well, with $\widetilde{M}( \xi )$ defined in \eqref{eq:information_matrix_modified}.
\end{theorem}

\textit{Proof}: We can write
\begin{eqnarray}
\widetilde{M}(\xi) & = & (F')^{\T} \left\{C' + \widetilde{W}_{\tilde{\kappa}}'(\xi) \right\}^{-1} F' \notag \\
& = & (F')^{\T} \left\{(\Sigma')^{1/2} \left[ K' + W_{\tilde{\kappa}}'(\xi) \right] (\Sigma')^{1/2} \right\}^{-1} F' \notag \\
& = & \left((\Sigma')^{-1/2} F'\right)^{\T} \left[ K' + W_{\tilde{\kappa}}'(\xi) \right]^{-1} \left((\Sigma')^{-1/2} F'\right) \notag \\
& = & (\widetilde{F}')^{\T}  \left[ K' + W_{\tilde{\kappa}}'(\xi) \right]^{-1} \widetilde{F}', \label{eq:information_matrix_modified2}
\end{eqnarray}
where $W_{\tilde{\kappa}}'(\xi)$ is given by \eqref{eq:virtual_noise_variance_matrix} (with $\kappa$ replaced by $\tilde{\kappa}$ and $x \in \supp(\xi)$) and the rows of $\widetilde{F}'$ are composed of $f(x)^{\T}/\sigma(x)$ for $x \in \supp(\xi)$. Eq.~\eqref{eq:information_matrix_modified2} can be interpreted as the information matrix of model~\eqref{eq:model_virtual_noise} with the original virtual noise variance formulation \eqref{eq:variance_virtual_noise_original}, where $F'$ is set to the scaled model matrix $\widetilde{F}'$, the covariance matrix $C'$ is equal to the correlation matrix $K'$, and $\tilde{\kappa}$ is used instead of $\kappa$. According to Theorem~1 in \cite{pazman_correlated_2022}, if $\tilde{\kappa}$ is not larger than the minimal eigenvalue of $K$, then the mapping $\xi \in \Xi \rightarrow \Phi \left\{ \widetilde{M}( \xi ) \right\}$ is concave for the virtual noise model~\eqref{eq:model_virtual_noise} with the virtual noise variance \eqref{eq:variance_virtual_noise_original}, model matrix $\widetilde{F}$, and variance-covariance matrix $K$. 

\qed

From the proof of Theorem~\ref{th:modified_concave}, one can see that using the modified virtual noise variance \eqref{eq:variance_virtual_noise_modified} instead of \eqref{eq:variance_virtual_noise_original} for model~\eqref{eq:model_virtual_noise} is equivalent to using our original virtual noise variance~\eqref{eq:variance_virtual_noise_original} for a model with scaled model matrix $\widetilde{F}$ and the correlation matrix $K$ being the covariance matrix. Therefore, all the results derived for the virtual noise model with virtual noise variance function \eqref{eq:variance_virtual_noise_original} in \cite{pazman_correlated_2022} also apply to the modified virtual noise model introduced in this section.

The next theorem establishes that when the errors are uncorrelated, the information matrix of the modified virtual noise model coincides with the classical information matrix.

\begin{theorem}
	\label{th:special_case}
	In the particular case that the variance-covariance matrix $C$ is diagonal with diagonal entries $C_{i,i} = \sigma^2(x_i)$ for $i = 1,\ldots,N$, that means the observations are uncorrelated, we have that
	\begin{equation*}
	\widetilde{M}(\xi) = n \sum_{x \in \X} \frac{1}{\sigma^2(x)} \, f(x) \, f(x)^{\T} \, \xi(x),
	\end{equation*}
	which is the classical information matrix used when the observations are uncorrelated.
	{\normalfont The proof of this theorem is given in Appendix~\ref{sec:proof_special_case}.}
\end{theorem}

To keep notation simple, in the rest of the paper we will only use the notation for the virtual noise model based on the original virtual noise variance formulation \eqref{eq:variance_virtual_noise_original}. However, in the examples we prefer to use the modified virtual noise variance formulation \eqref{eq:variance_virtual_noise_modified}, because this model contains the uncorrelated case as a special case. All the algorithms and methods developed for the original virtual noise model can also be applied to the modified virtual noise model by setting $F$ to $\widetilde{F} = \Sigma^{-1/2} F$, $C$ to $K = \Sigma^{-1/2} C \Sigma^{-1/2}$, and selecting $\kappa$ such that $\kappa \leq \lambda_{\min}(K)$.

\section{Links to convex relaxation for sensor selection}

\label{sec3}

\cite{liu_sensorselection_2016} develop a convex relaxation scheme for selecting $n$ sensor locations out of a set of $N$ sensor locations for the linear regression model with correlated errors stated in Eq.~\eqref{eq:model_orig}, see also \cite{ucinski_convexrelax_2024}. In this section we will show that this can be fully embedded into the virtual noise method described in the sections above.

The vector $w$ encodes the sensor selection information. Each location $x \in \X$ might either be selected, in which case $w(x) = 1$, or it might not be selected, in which case $w(x) = 0$. The vector $w = (w(x_1),\ldots,w(x_N))$ contains these indicators for all $N$ potential locations. Since $n$ locations are selected, we have $\sum_{i=1}^N w(x_i) = n$. Let $\diag\{w\}$ denote a diagonal matrix whose $i$-th diagonal entry is equal to the $i$-th entry of the vector $w$, so $\left[\diag\{w\}\right]_{i,i} = w(x_i)$ for $i = 1,\ldots,N$ and $\left[\diag\{w\}\right]_{i,j} = 0$ for $i \neq j$.

In addition, \cite{liu_sensorselection_2016} introduce the $n \times N$ matrix $\Omega_w$, which is equal to $\diag\{w\}$ with all the rows containing only $0$ entries removed. The matrix $\Omega_w$ has the property that $\Omega_w^{\T} \Omega_w = \diag\{w\}$ and that $\Omega_w \Omega_w^{\T} = I_n$, where $I_n$ is the $n \times n$ identity matrix.

Using this notation, the information matrix for model~\eqref{eq:model_orig} for the $n$ selected locations can be written as
\begin{equation*}
M(w) = F^{\T} \Omega_w^{\T} \left(\Omega_w C \Omega_w^{\T}\right)^{-1} \Omega_w F.
\end{equation*}

\cite{liu_sensorselection_2016} actually follow a Bayesian approach and add the prior information matrix $\Gamma^{-1}_{\mathrm{prior}}$ to the information matrix $M(w)$ to ensure that the posterior information matrix is positive definite even in cases where $n < p = \dim(\theta)$. We only consider cases where $n \geq p$ and the information matrix is invertible, so we do not add $\Gamma^{-1}_{\mathrm{prior}}$.

\cite{liu_sensorselection_2016} introduce the matrix $S = C - \kappa I_N$, where $I_N$ denotes the $N \times N$ identity matrix and $\kappa$ is chosen such that $S$ is still positive definite, i.e., $\kappa < \lambda_{\min}(C)$. They obtain
\begin{eqnarray}
M(w) & = & F^{\T} \Omega_w^{\T} \left[\Omega_w \left(S + \kappa I_N\right) \Omega_w^{\T}\right]^{-1} \Omega_w F \notag \\
& = & F^{\T} \Omega_w^{\T} \left[\Omega_w S \Omega_w^{\T} + \kappa I_n\right]^{-1} \Omega_w F \notag \\
& = & F^{\T} \left[S^{-1} - S^{-1} \left(S^{-1} +\kappa^{-1} \Omega_w^{\T} \Omega_w \right)^{-1} S^{-1}\right] F \notag \\
& = & F^{\T} \left[S^{-1} - S^{-1} \left(S^{-1} +\kappa^{-1} \diag\{w\} \right)^{-1} S^{-1}\right] F. \label{eq:infomat_liu}
\end{eqnarray} 

The step from the second to the third line above makes use of the matrix inversion lemma $(A + BCD)^{-1} \equiv A^{-1} - A^{-1} B (C^{-1} + D A^{-1}B)^{-1} D A^{-1}$, from which it follows that $B(C^{-1} + D A^{-1}B)^{-1} D \equiv A - A(A + BCD)^{-1} A$. If setting $A = S^{-1}$, $B = \Omega_w^{\T}$, $C = (1/\kappa) \, I_n$, and $D = \Omega_w$, the result is obtained.

In order to select the $n$ sensor locations, \cite{liu_sensorselection_2016} consider the following optimization problem with the weights relaxed to lie in $w \in [0,1]^N$:
\begin{align*}
\min_w &\; \tr\left(M(w)^{-1}\right) \\
\text{subject to} &\;\sum_{i=1}^N w(x_i) \leq n, \\
&\; w \in [0,1]^N.
\end{align*}
They introduce some auxiliary matrices and transform this problem into a semidefinite program (SDP) containing linear matrix inequalities, which they solve using an interior-point algorithm. Some of the additional constraints involve an auxiliary weight matrix, where the purpose of these constraints is to force the solution of the weights to be close to $0$ or $1$.

The following proposition establishes that the information matrix \eqref{eq:infomat_liu} with relaxed selection indicators $w(x_i)$ is equivalent to the information matrix for our original virtual noise model with design measure $\xi(x_i) = w(x_i) / n$. This means that the relaxed sensor selection problem is a convex optimization problem for any concave and increasing criterion function. Furthermore, one may employ the optimization algorithms suggested in this paper as an efficient alternative to the semidefinite programs put forward by \cite{liu_sensorselection_2016}.

\begin{proposition}
	\label{prop:liu_equality}
	Let the selection indicators $w(x_i)$ be relaxed such that $w(x_i) \in [0,1]$ for $i = 1,\ldots,N$ and $\sum_{i=1}^N w(x_i) = n$. Then the information matrix \eqref{eq:infomat_liu} is equal to the information matrix \eqref{eq:infmatrix_vn_orig} for the virtual noise model \eqref{eq:model_virtual_noise} with virtual noise variance \eqref{eq:variance_virtual_noise_original} and the measure set to $\xi(x) = w(x)/n$ for all $x \in \X$.
\end{proposition}

\textit{Proof}: Let us first consider the case where $\xi \in \Xi_+ = \left\{ \xi: \: \sum_{x \in \X} \xi(x) = 1, \; \forall_{x \in \X} \: 0 < \xi(x) \leq 1/n \right\}$, i.e., $\xi(x) > 0$ for all $x \in \X$:

Set $w = n \xi$, so Eq.~\eqref{eq:infomat_liu} can be written as
\begin{eqnarray*}
	M(\xi) & = & F^{\T} \left\{S^{-1} - S^{-1} \left(S^{-1} +\frac{n}{\kappa} \, \diag\{\xi\} \right)^{-1} S^{-1}\right\} F \\
	& = & F^{\T} \left\{ S + \frac{\kappa}{n} \, \diag\{\xi^{-1}\} \right\}^{-1} F = F^{\T} \left\{C + \frac{\kappa}{n} \, \diag\{\xi^{-1}\} -\kappa \, I_N \right\}^{-1} F \\
	& = & F^{\T} \left\{C + W_{\kappa}(\xi) \right\}^{-1} F,
\end{eqnarray*}
which is the information matrix formula \eqref{eq:infmatrix_vn_fullsupport} for the virtual noise model \eqref{eq:model_virtual_noise} with virtual noise variance \eqref{eq:variance_virtual_noise_original} if $\xi \in \Xi_+$.
The second line above is obtained by using the matrix identity $(A + BCD)^{-1} \equiv A^{-1} - A^{-1} B \left(C^{-1} + D A^{-1} B\right)^{-1} D A^{-1}$ and setting $A = S$, $C = (\kappa/n) \, \diag\{\xi^{-1}\}$, and $B = D = I_N$.

See Appendix~\ref{sec:proof_convex_relaxation} for the proof of the technically more involved case where $\xi \in \Xi\backslash\Xi_+$, i.e.\ $\xi(x_i) = 0$ for some $i \in \{1,\ldots,N\}$. \qed


\section{Algorithms}

\label{sec4}

\label{sec:algorithms}
In this section we will review and compare several algorithms that can be used to calculate the virtual noise design measure. 

\subsection{Preliminaries: gradients of criteria}
\label{sec:algorithms_preliminaries}

For the algorithms to find a solution for
\begin{equation*}
\xi^* = \underset{\xi \in \Xi}{\arg \max} \: \Phi\{M(\xi)\},
\end{equation*}
we need the gradient of $\Phi\{M(\xi)\}$ with respect to $\xi$, denoted by
\begin{equation*}
\nabla_{\xi} \Phi\{M(\xi)\} = \left(\frac{\partial \, \Phi\{M(\xi)\}}{\partial \, \xi(x_1)}, \ldots, \frac{\partial \, \Phi\{M(\xi)\}}{\partial \, \xi(x_N)} \right)^{\T}.
\end{equation*}
For some $\bar{x} \in \X$, the derivative with respect to the measure at $\bar{x}$ is given by
\begin{equation*}
\frac{\partial \, \Phi\{M(\xi)\}}{\partial \, \xi(\bar{x})} = \tr \left( \nabla_M \Phi\{M(\xi)\} \, \frac{\partial \, M(\xi)}{\partial \, \xi(\bar{x})} \right).
\end{equation*}

In our paper, we consider two criteria, D-optimality and A-optimality. For D-optimality, we use two different formulations depending on the optimization method. They are $\Phi_{D1}(M) = \log \det(M)$ and $\Phi_{D2}(M) = \{\det(M)\}^{1/p}$ for $p > 1$. The second formulation is used when the optimization method requires that $\Phi(M) > 0$. Otherwise, the first formulation is preferred because the gradient and especially the Hessian are easier to compute. The A-optimality criterion is given by $\Phi_{A}(M) = -\tr(M^{-1})$. For optimization methods that require positive criterion values, we minimize the convex criterion $-\Phi_A(M)$ instead.
For the D-optimality criterion
\begin{equation*}
\nabla_M \Phi_D(M) = \psi_D \: M^{-1},
\end{equation*}
where $\psi_D = 1$ if $\Phi_D = \Phi_{D1}$ and $\psi_D = \{\det(M)\}^{1/p}/p$ if $\Phi_D = \Phi_{D2}$.
For the A-optimality criterion,
\begin{equation*}
\nabla_M \Phi_A(M) = \: M^{-2}.
\end{equation*}

To derive $\partial \, M(\xi) \bigl/ \partial \, \xi(\bar{x})$, we use an alternative representation of the information matrix \eqref{eq:infmatrix_vn_orig}. In Lemma~F.3 of \cite{pazman_correlated_2022}, it is shown that the matrix
\begin{equation}
L(\xi) = F^{\T} \left[\diag\{\xi\} (C - \kappa I_N) + \frac{\kappa}{n} \, I_N\right]^{-1} \, \diag\{\xi\} F
\end{equation}
is well defined and continuous and equal to $M(\xi)$ given by Eq.~\eqref{eq:infmatrix_vn_orig} on $\xi \in \Xi$. In practical applications, it is preferable to use the representation $L(\xi)$ instead of $M(\xi)$ from Eq.~\eqref{eq:infmatrix_vn_orig}, because the matrix
\begin{equation*}
Z(\xi) = \diag\{\xi\} (C - \kappa I_N) + \frac{\kappa}{n} \, I_N
\end{equation*}
is always invertible even if some of the elements of $\xi$ are $0$. This means there is no need to take special care of the case where $\xi(x_i) = 0$ for some points $x_i \in \X$. Furthermore, there are no problems with the numerical stability of the computations if $\xi(x_i) \approx 0$. Using the original representation \eqref{eq:infmatrix_vn_orig}, one needs to define some threshold $\epsilon > 0$ such that points $x_i$ with $\xi(x_i) \leq \epsilon$ are treated as if $\xi(x_i) = 0$ to ensure numerically stable computations. The same applies to the derivatives $\partial \, L(\xi) \bigl/ \partial \, \xi(\bar{x})$ for $\bar{x} \in \X$.

The derivations of $\partial \, L(\xi) \bigl/ \partial \, \xi(\bar{x})$ and subsequently of the gradients $\nabla_{\xi} \Phi\{L(\xi)\}$ for D- and A-optimality are provided in Appendix~\ref{sec:deriv_gradients_Hessians}. For D-optimality, the gradient is given by the diagonal elements of the matrix
\begin{equation*}
\psi_D \, \frac{\kappa}{n} \, \left\{ Z(\xi)^{-1} \right\}^{\T} F L(\xi)^{-1} F^{\T} Z(\xi)^{-1}.
\end{equation*}
In the case of A-optimality, the gradient is equal to the diagonal elements of the matrix
\begin{equation*}
\frac{\kappa}{n} \, \left\{ Z(\xi)^{-1} \right\}^{\T} F L(\xi)^{-2} F^{\T} Z(\xi)^{-1}.
\end{equation*}

The Hessians for the various criteria, which are needed for some optimization algorithms, are also derived in Appendix~\ref{sec:deriv_gradients_Hessians}.

\subsection{Cutting-plane method}

\label{sec:cutting_plane}

In \cite{pazman_correlated_2022}, the cutting-plane method \citep{kelley_cutting-plane_1960} was used to maximize $\Phi\{L(\xi)\}$ with respect to $\xi$. Since $\Phi\{L(\xi)\}$ is concave, each tangent plane lies above the criterion surface, so for each $\xi \in \Xi$,
\begin{equation*}
\Phi\{L(\xi)\} \leq \Phi\{L(\mu)\} + \nabla^{\T} \Phi\{L(\mu)\} \left(\xi - \mu \right) \quad \forall \: \mu \in \Xi.
\end{equation*}
Furthermore, we have that
\begin{equation}
\Phi\{L(\xi)\} = \min_{\mu \in \Xi} \left[ \Phi\{L(\mu)\} + \nabla^{\T} \Phi\{L(\mu)\} \left(\xi - \mu \right) \right]. \label{eq:crit_min_infinite}
\end{equation}
Therefore, finding the optimal $\xi^* = \underset{\xi \in \Xi}{\arg \max} \, \Phi\{L(\xi)\}$ is turned into a maximin problem. The cutting-plane method is an iterative procedure, in which minimization over the infinite-dimensional set $\Xi$ is replaced by minimization over a discrete set of measures. This discrete set is augmented in each step. After step $k$, the discrete set is composed of $k$ design measures: $\Xi^{(k)} = \{\xi^{(1)}, \ldots, \xi^{(k)}\}$. The optimization problem $\xi^* = \underset{\xi \in \Xi}{\arg \max} \, \Phi\{L(\xi)\}$ with $\Phi\{L(\xi)\}$ given by \eqref{eq:crit_min_infinite} is replaced by
\begin{equation*}
\xi^{(k+1)} = \arg \max_{\xi \in \Xi} \min_{\mu \in \Xi^{(k)}} \left[ \Phi\{L(\mu)\} + \nabla^{\T} \Phi\{L(\mu)\} \left(\xi - \mu \right) \right].
\end{equation*}
This can be formulated as a linear program in the following way:
\begin{align}
\max_{t, \, \xi} &\;\: t \label{eq:CP_LP} \\
\text{subject to} &\;\: t \leq \Phi\{L(\mu)\} + \nabla^{\T} \Phi\{L(\mu)\} \left(\xi - \mu \right), \; \mu \in \Xi^{(k)},  \notag \\
&\;\:t \geq 0, \notag \\
&\;\:\xi \in \Xi. \notag
\end{align}
Let $t^{(k+1)}$ and $\xi^{(k+1)}$ denote the solution to the linear program \eqref{eq:CP_LP} and let $\bar{\Phi}^{(k)} = \max_{\xi \in \Xi^{(k)}} \, \Phi\{L(\xi)\}$. If $t^{(k+1)} - \bar{\Phi}^{(k)} < \delta$, then $\Phi\{L(\xi^*)\} - \bar{\Phi}^{(k)} < \delta$, so $t^{(k+1)} - \bar{\Phi}^{(k)} < \delta$ for some $\delta > 0$ can serve as a stopping rule. The relations $t^{(k+1)} \geq \Phi\{L(\xi^*)\} \geq \bar{\Phi}^{(k)}$ follow from
\begin{eqnarray*}
	t^{(k+1)} & = & \max_{\xi \in \Xi} \min_{\mu \in \Xi^{(k)}} \left[ \Phi\{L(\mu)\} + \nabla^{\T} \Phi\{L(\mu)\} \left(\xi - \mu \right) \right] \\
	& \geq & \max_{\xi \in \Xi} \min_{\mu \in \Xi} \left[ \Phi\{L(\mu)\} + \nabla^{\T} \Phi\{L(\mu)\} \left(\xi - \mu \right) \right] \\
	& = & \max_{\xi \in \Xi} \Phi\{L(\xi)\} \geq \max_{\xi \in \Xi^{(k)}} \, \Phi\{L(\xi)\} = \bar{\Phi}^{(k)}.
\end{eqnarray*}
If the algorithm does not stop after iteration $k$, the solution $\xi^{(k+1)}$ is added to the set $\Xi^{(k)}$, so $\Xi^{(k+1)} = \Xi^{(k)} \cup \xi^{(k+1)}$, $k \leftarrow k + 1$, and the linear program \eqref{eq:CP_LP} is solved again.

If the criterion $\Phi\{L(\xi)\}$ is convex and needs to be minimized, the linear program analogous to \eqref{eq:CP_LP} that has to be solved when applying the cutting-plane method is
\begin{align*}
\min_{t, \, \xi} &\;\: t \\
\text{subject to} &\;\: t \geq \Phi\{L(\mu)\} + \nabla^{\T} \Phi\{L(\mu)\} \left(\xi - \mu \right), \; \mu \in \Xi^{(k)},  \\
&\;\:t \geq 0, \\
&\;\:\xi \in \Xi,
\end{align*}	
and the algorithm is stopped as soon as $\min_{\xi \in \Xi^{(k)}} \, \Phi\{L(\xi)\} - t^{(k+1)} < \delta$.

\subsection{Level method}

\label{sec:level}

The cutting-plane method often exhibits slow convergence because the next solution is determined by solving the linear program \eqref{eq:CP_LP} over the entire polyhedron defined by the restrictions. For the so-called level method \citep[see][]{nesterov_lectures_2004,pronzato_design_2014}, the design $\xi^{(k+1)}$ which is added to the set $\Xi^{(k)}$ for the next iteration is not the solution to the linear program \eqref{eq:CP_LP} but the projection of the current solution $\xi^{(k)}$ to the polyhedron 
\begin{equation}
\{\xi \in \Xi: \: L_k(\alpha) \leq \Phi\{L(\mu)\} + \nabla^{\T} \Phi\{L(\mu)\} \left(\xi - \mu \right) \; \forall \, \mu \in \Xi^{(k)}\}, \label{eq:level_polyhedron}
\end{equation}
where $\displaystyle L_k(\alpha) = (1-\alpha) t^{(k+1)} + \alpha \left[\max_{\xi \in \Xi^{(k)}} \, \Phi\{L(\xi)\} \right]$. This means that the next solution $\xi^{(k+1)}$ is found by selecting the design measure which is closest to the current solution $\xi^{(k)}$ among those measures which achieve a certain minimum improvement compared to the maximum criterion value found so far, where $\Phi\{L(\xi)\}$ is approximated by taking the minimum over a finite set of tangent planes. \cite{nesterov_lectures_2004} suggests an optimal value of $\alpha^* = 1/(2 + \sqrt{2})$.

Formally, the projection of $\xi^{(k)}$ to the polyhedron \eqref{eq:level_polyhedron} is computed by solving the following quadratic program:
\begin{align}
\min_{\xi} &\;\: \left\|\xi - \xi^{(k)}\right\|_2^2 \label{eq:level_QP} \\
\text{subject to} &\;\:  L_k(\alpha) \leq \Phi\{L(\mu)\} + \nabla^{\T} \Phi\{L(\mu)\} \left(\xi - \mu \right), \; \mu \in \Xi^{(k)},  \notag \\
&\;\:\xi \in \Xi. \notag
\end{align}

Before solving the quadratic program \eqref{eq:level_QP}, the linear program \eqref{eq:CP_LP} is run to determine $t^{(k+1)}$ and to decide whether to stop the algorithm. The stopping rule is the same as for the cutting-plane method.

In the case of minimizing a convex criterion $\Phi\{L(\xi)\}$, the constraints in the quadratic program \eqref{eq:level_QP} change to
\begin{equation*}
L_k(\alpha) \geq \Phi\{L(\mu)\} + \nabla^{\T} \Phi\{L(\mu)\} \left(\xi - \mu \right), \; \mu \in \Xi^{(k)},
\end{equation*}
and $L_k(\alpha)$ is computed as  $\displaystyle L_k(\alpha) = (1-\alpha) t^{(k+1)} + \alpha \left[\min_{\xi \in \Xi^{(k)}} \, \Phi\{L(\xi)\} \right]$.

Compared to the cutting-plane method, the level method includes an additional quadratic programming step. However, due to its typically significantly accelerated convergence, the overall computational complexity of the level method is usually much lower than for the cutting-plane method for a given numerical accuracy.

\subsection{Simplicial decomposition}

\label{sec:simplicial_general}

For optimizing concave design criteria over the polyhedral set 
\begin{equation*}
\Xi = \left\{ \xi: \: \sum_{x \in \X} \xi(x) = 1, \; \forall_{x \in \X} \: 0 \leq \xi(x) \leq 1/n \right\},
\end{equation*}
another suitable optimization method is simplicial decomposition, see \cite{patriksson_simplicial_2008} and \cite{ucinski_convexrelax_2024}. The idea of the method is to reduce the computational complexity by optimizing the criterion function over a subset of $\Xi$ spanned by a set of extreme points, which turns the problem into one with standard constraints and with usually decreased dimensionality. The set of extreme points is extended and updated efficiently in another step by solving the problem using a linear approximation to the criterion function. These steps are iterated until convergence of the criterion.

Let $\Xi^{(k)} = \left\{\tilde{\xi}^{(1)}, \ldots, \tilde{\xi}^{(k)}\right\}$ denote the set of extreme points after iteration $k$ and let $\xi^{(k)}$ denote the current solution.

A new extreme point is found by maximizing the linear Taylor approximation to $\Phi\{L(\xi)\}$ evaluated at the current solution $\xi^{(k)}$ over the set $\Xi$:
\begin{equation}
\tilde{\xi}^{(k+1)} = \underset{\xi \in \Xi}{\arg \max} \left[ \Phi\left\{L\left(\xi^{(k)}\right)\right\} + \nabla^{\T} \Phi\left\{L\left(\xi^{(k)}\right)\right\} \left(\xi - \xi^{(k)}\right) \right]. \label{eq:SD_CGP}
\end{equation}
The solution to problem \eqref{eq:SD_CGP} is very simple. One just has to set the components of $\tilde{\xi}^{(k+1)}$ corresponding to the $n$ largest elements of $\nabla \,  \Phi\left\{L\left(\xi^{(k)}\right)\right\}$ to $1/n$ and all the other components to $0$. Hence, the solution is always at a vertex of the polyhedron defined by $\Xi$. The set $\Xi^{(k)}$ is then extended to $\Xi^{(k+1)} = \Xi^{(k)} \cup \tilde{\xi}^{(k)}$. Problem \eqref{eq:SD_CGP} is therefore called the \emph{column generation problem (CGP)}.

The convex hull of the points in $\Xi^{(k+1)}$ is a subset of the design space $\Xi$. The \emph{restricted master problem (RMP)} optimizes the original criterion $\Phi\{L(\xi)\}$ over this subspace:
\begin{equation}
\xi^{(k+1)} = \underset{\xi \, \in \, \mathrm{conv}\left(\Xi^{(k+1)}\right)}{\arg \max} \: \Phi\{L(\xi)\}, \label{eq:SD_RMP1}
\end{equation}
where $\xi^{(k+1)}$ is the new current solution after step $k+1$.

The condition $\xi \, \in \, \mathrm{conv}\left(\Xi^{(k+1)}\right)$ can be reformulated as $\xi = \sum_{i=1}^{k+1} w_i \, \tilde{\xi}^{(i)}$, where 
\begin{equation*}
w = (w_1,\ldots,w_{k+1})^{\T} \in \setV_{k+1} = \left\{w: \: \sum_{i=1}^{k+1} w_i = 1, \; w_i \geq 0 \: \forall \, i = 1,\ldots,k+1 \right\}. \label{eq:setV}
\end{equation*}
Furthermore, let the columns of the $N \times (k+1)$ matrix $X_{k+1}$ be comprised of the $k+1$ elements of $\Xi^{(k+1)}$. We can then rewrite optimization problem \eqref{eq:SD_RMP1} as
\begin{equation}
\xi^{(k+1)} = \underset{w \, \in \, \setV_{k+1}}{\arg \max} \: \Phi\{L(X_{k+1} w)\}. \label{eq:SD_RMP2}
\end{equation}
The dimension $k+1$ of this problem is usually much lower than $N$. Moreover, there are many efficient algorithms available that allow for optimizing $w$ over the set $\setV_{k+1}$. In our paper, we propose to use a multiplicative algorithm or a projected Newton method to solve the RMP.

After obtaining a new solution $\xi^{(k+1)}$ in the restricted master problem, the set $\Xi^{(k+1)}$ can be purged of those extreme points $\tilde{\xi}^{(j)}$ for which the weight $w_j = 0$.
The stopping rule for the algorithm is based on the Kuhn-Karush-Tucker conditions. The details are provided in Appendix~\ref{sec:KKT_conditions}.

We implemented two different algorithms to solve the restricted master problem~\eqref{eq:SD_RMP2}. The first method is a gradient-projection method called multiplicative algorithm \citep[see, e.g.,][]{pronzato_design_2014, ucinski_convexrelax_2024}, where the parameters are chosen such that the weights are updated according to a multiplicative rule. Appendix~\ref{sec:multiplicative} contains a detailed description of this algorithm. The multiplicative algorithm is relatively easy to implement, but the implied choices for the search direction and the step size are not optimal. Furthermore, the multiplicative algorithm requires all weights $w_i > 0$, so the set $\Xi^{(k+1)}$ is never purged of irrelevant extreme points.

To improve the efficiency of the restricted master problem, we therefore also consider a more general projected Newton algorithm. In particular, we have implemented the method described in \cite{bertsekas_projected_1983}, which is suitable when the constraint set is a Cartesian product of simplexes. This algorithm was also employed in \cite{ucinski_sensor_2019} and \cite{ucinski_d-optimal_2020}. More details on this algorithm can be found in Appendix~\ref{sec:projected_Newton}.


\section{Examples} \label{sec:examples}

\label{sec5}

\subsection{Example setup}

We applied our methodology to compute the optimal design measures for the original as well as for the modified virtual noise model to a range of examples with either a one- or a two-dimensional design space. To that end, we tried three of the algorithms presented in Section~\ref{sec:algorithms}: the level method and the simplicial decomposition method using either the multiplicative algorithm or the projected Newton method for the restricted master problem, where the latter is used as a benchmark.

For each example, the optimal virtual noise measures were computed for a range of $n$ values, usually from $4$ to $20$. We then checked the quality of the criterion bounds provided by the virtual noise model by comparing the criterion values to those for the optimal exact $n$-point design. If the original and the modified virtual noise formulations did not coincide, we tried both formulations to see whether the tightness of the bounds for the exact $n$-point design's criterion value differs significantly between the formulations.

All the algorithms were implemented in R \citep{RSoftware}. The packages \texttt{lpSolve} \citep{lpSolve} and \texttt{quadprog} \citep{quadprog} were used for the linear and quadratic programming parts of the level method. The programs were run on an SGI UV 3000 cluster with 20 TB shared memory and processors of type Intel Xeon E5-4650V3 with 2.1 -- 2.8 GHz.

Since the conceptual differences between the algorithms are very large, it is difficult to compare the efficiencies of the algorithms. Therefore, in the absence of more suitable metrics we opted to compare the running times of the algorithms directly. However, we acknowledge that this metric is highly dependent on the programming language, the implementation details and the hardware on which the programs are run. The results we present should therefore only be interpreted with caution, but they can indicate situations in which some algorithm might be relatively more efficient than the others.

To make the comparison as fair as possible, we aimed to ensure that the precision of the criterion values is similar across the different algorithms. This is difficult to achieve due to the different and incompatible stopping rules employed by the different algorithms. The simplicial decomposition method using the projected Newton method, henceforth abbreviated to SD-PN, proved to be fairly efficient in general. However, depending on the setting of the tuning parameters, numerical errors frequently occurred that caused the program to abort. For example, choosing values of the tolerances that are too low may lead to numerical errors when computing the Hessians at some point. We therefore had to select tuning parameter values where the program could be completed while maintaining the best-possible precision. We varied two parameters which are crucial for the precision of the results. The first parameter was the tolerance $\epsilon$ when checking the Kuhn-Karush-Tucker conditions about whether to stop the simplicial decomposition algorithm, see Appendix~\ref{sec:KKT_conditions}. The tolerance is provided relative to $\lambda_{\mathrm{int},\mathrm{avg}}$ if interior design points are available or relative to $\lambda_{\mathrm{up},\min}$ if no interior design points are available. Let this relative tolerance be denoted by $\rho_{\mathrm{SD}}$. The second parameter is the relative tolerance for checking the Kuhn-Karush-Tucker conditions for convergence of the projected Newton step inside the simplicial decomposition method, see Appendix~\ref{sec:projected_Newton}. We denote this relative tolerance by $\rho_{\mathrm{PN}}$.

To goal was to find those feasible settings of $\rho_{\mathrm{SD}}$ and $\rho_{\mathrm{PN}}$ that yielded the highest-achievable precision while being as efficient as possible. To that end, we tried different values for $\rho_{\mathrm{SD}}$ and $\rho_{\mathrm{PN}}$. The relative tolerance $\rho_{\mathrm{SD}}$ was reduced step by step according to the geometric sequence $(1/4)^2, (1/4)^3, (1/4)^4,\ldots$, up to at most $(1/4)^9$. For each value of $\rho_{\mathrm{SD}}$, the values of  $\rho_{\mathrm{PN}}$ were set to the geometric sequence $\rho_{\mathrm{PN}} = \rho_{\mathrm{SD}}, \: \rho_{\mathrm{SD}} \cdot (1/4), \: \rho_{\mathrm{SD}} \cdot (1/4)^2$. The algorithm was stopped early if for some $\rho_{\mathrm{SD}}$ all values of $\rho_{\mathrm{PN}}$ resulted in the SD-PN algorithm to abort.

Among the successfully completed runs of the SD-PN algorithm, we identified the run achieving the highest criterion value. This run is likely to be associated with low tolerance values. However, it might be possible to increase the tolerance values to achieve substantial efficiency gains while hardly compromising the precision. We considered all the runs where the relative deviation of the criterion value from the highest criterion value was less than $10^{-6}$. Among the runs with negligible deviation from the highest criterion value, we first selected the runs with the highest relative tolerance $\rho_{\mathrm{SD}}$, and among those runs we finally selected the run with the highest relative tolerance $\rho_{\mathrm{PN}}$.

Using these target values for $\rho_{\mathrm{SD}}$ and $\rho_{\mathrm{PN}}$ found for each example, we re-ran the SD-PN method for each example with exactly these target tolerances. We then used the target tolerances $\rho_{\mathrm{SD}}$ to run the simplicial decomposition method using the multiplicative algorithm, abbreviated to SD-M, for each example.

The level method has a different stopping criterion. To get results for the level method with a precision comparable to SD-PN and SD-M, we modified the algorithm outlined in Sections~\ref{sec:cutting_plane} and \ref{sec:level} to stop as soon as the criterion value exceeds the criterion value obtained by SD-PN when using the selected tolerances.

To find exact $n$-point designs, we employ a simple greedy exchange-type algorithm given by Algorithm~2 of \cite{pazman_correlated_2022}, where we use the criterion update functions described in \cite{brimkulov_numerical_1980} and \cite{fedorov_design_1996} for D-optimality and in \cite{liu_sensorselection_2016} for A-optimality. 
For details about the criterion update functions see Appendix~\ref{sec:criterion_updates}.

For each example defined by the function $f(x)$ and the kernel function, we run our algorithms using the D- and the A-optimality criterion. Except for the third example, 
where the original and the modified virtual noise formulation are the same, we compute the optimal virtual noise design measures under both formulations. In summary, we consider all the settings given by the combinations with respect to the example, the criterion, and the virtual noise formulation. For each setting and each $n = 4,\ldots,20$, we plot the running times of the algorithms relative to SD-PN. 

For all our examples, the D- and A-optimality criteria led to very similar design measures. Furthermore, the differences between the design measures using the original and the modified virtual noise formulation were only minor as well. In all our examples, we therefore illustrate the design measures for $n = 5$ and $n = 20$ only for the D-optimality criterion using the modified virtual noise formulation.

The efficiency of an exact $n$-point design $\tau$ with positive criterion value $\Phi_e\{M(\tau)\}$ relative to the optimal virtual noise design measure $\xi^*$ with positive criterion value $\Phi_e\{M(\xi^*)\}$ is defined as
\begin{equation}
\mathrm{eff}(\tau,\xi^*) = \frac{\Phi_e\{M(\tau)\}}{\Phi_e\{M(\xi^*)\}} \leq 1, \label{eq:design_efficiency}
\end{equation}
cf.~\cite{lopez-fidalgo_optimal_2023}. For D-optimality, we use $\Phi_e(M) = \det(M)^{1/p}$, for A-optimality, we use $\Phi_e(M) = 1 \bigr/ \tr\left(M^{-1}\right)$. Evaluating this design efficiency at $\tau^*$, the optimal exact $n$-point design, provides information about the tightness of the criterion bound implied by the optimal virtual noise measure $\xi^*$ on the optimal criterion values for exact $n$-point designs.

The trajectories of the design efficiencies of the exact $n$-point designs across the design sizes appeared very similar between the D- and A-optimality criteria, so we only show the design efficiencies for D-optimality.

In all examples, the values of $\kappa$ and $\tilde{\kappa}$ were set by rounding down the minimum eigenvalue of the covariance or correlation matrix over the design grid $\mathcal{X}$ to four significant digits.


\subsection{Example 1: two-parameter model in one design dimension with triangular kernel}\label{sec:example1}

For our first example, we use the following model:
\begin{eqnarray*}
	f(x)^{\T} & = & \left(1, \: 1 + 0.5 \cos(2 \pi x)\right), \\
	\Cov\{\varepsilon(x),\varepsilon( x')\} & = & \begin{cases}
		x^2 x' \qquad x \leq x' \\
		x (x')^2 \quad x > x', 
	\end{cases} \\
	\mathcal{X} & = & \{1, 1.01, 1.02, \ldots, 1.99, 2\}.
\end{eqnarray*}

This model is inspired by the model used by \cite{sacks_designs_1966}, \cite{dette_optimal_2016-1}, and \cite{pazman_correlated_2022}. Their model does not have an intercept and uses the sine function instead of the cosine function.

\begin{figure}[hbtp!]
	\centering
	\begin{tabular}{cc}
		\includegraphics[width=0.45\textwidth]{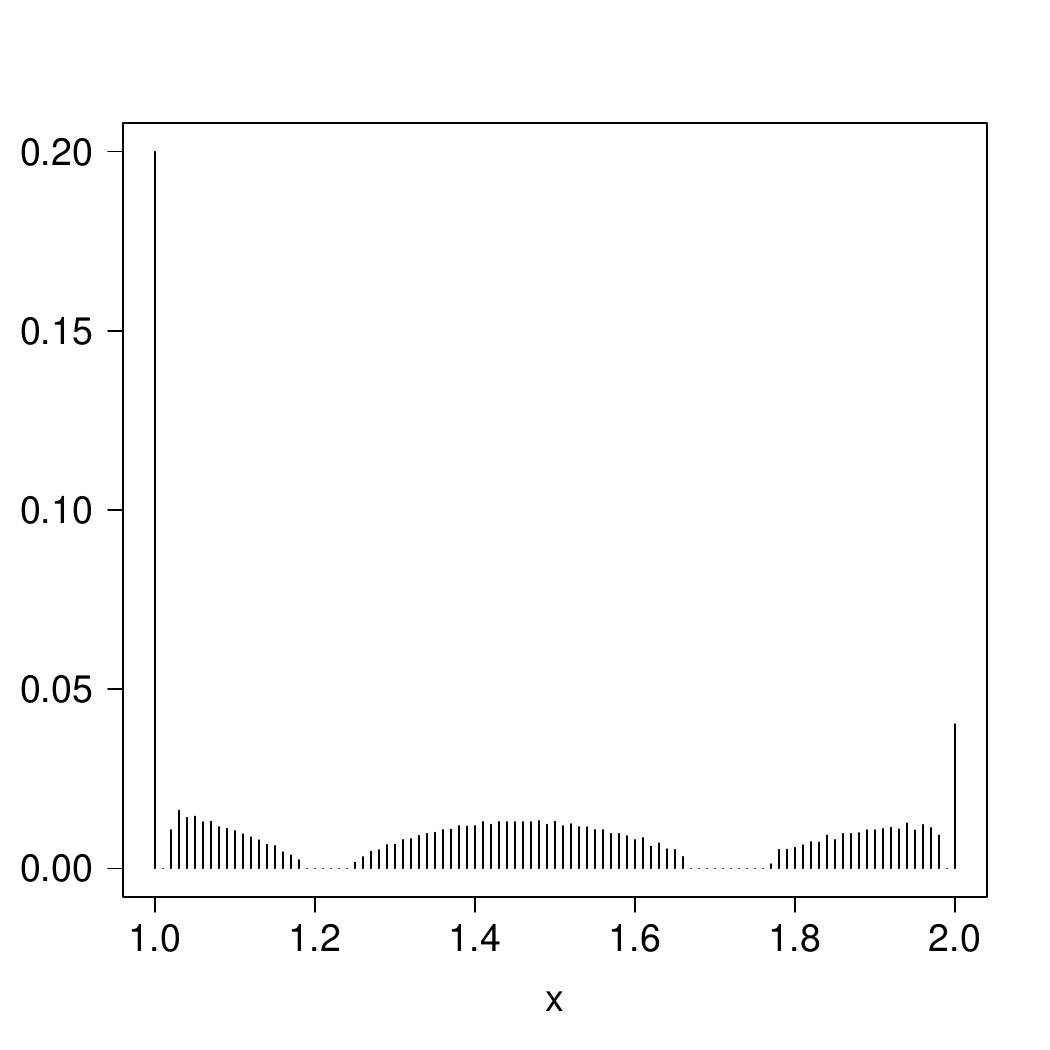} & 
		\includegraphics[width=0.45\textwidth]{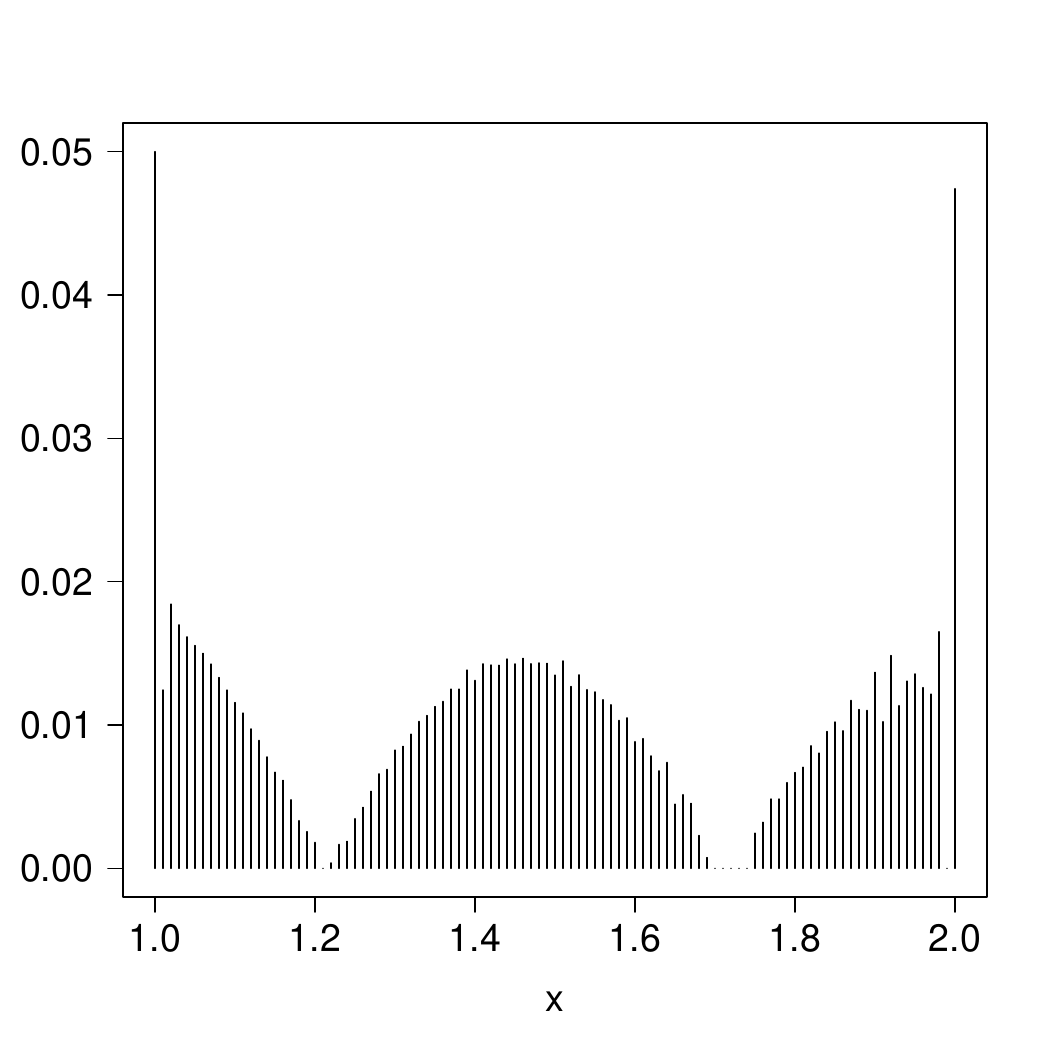} 		
	\end{tabular}
	\caption{Example~1: virtual noise design measures for $n = 5$ (left) and $n = 20$ (right) for D-optimality using the modified formulation. \label{fig:measures_example1}}
\end{figure}

The D-optimal design measure for the modified virtual noise formulation found using the SD-PN method is displayed in Figure~\ref{fig:measures_example1} for $n = 5$ and $n = 20$. At $n = 5$, the lower bound $x = 1$ has the largest possible permitted mass, which is $1/n = 0.2$. One can see that the mass from this point is evenly distributed across the design space as $n$ rises and therefore the upper limit for the permitted mass at each single point decreases.

\begin{figure}[hbtp!]
	\centering
	\begin{tabular}{cc}
		\includegraphics[width=0.45\textwidth]{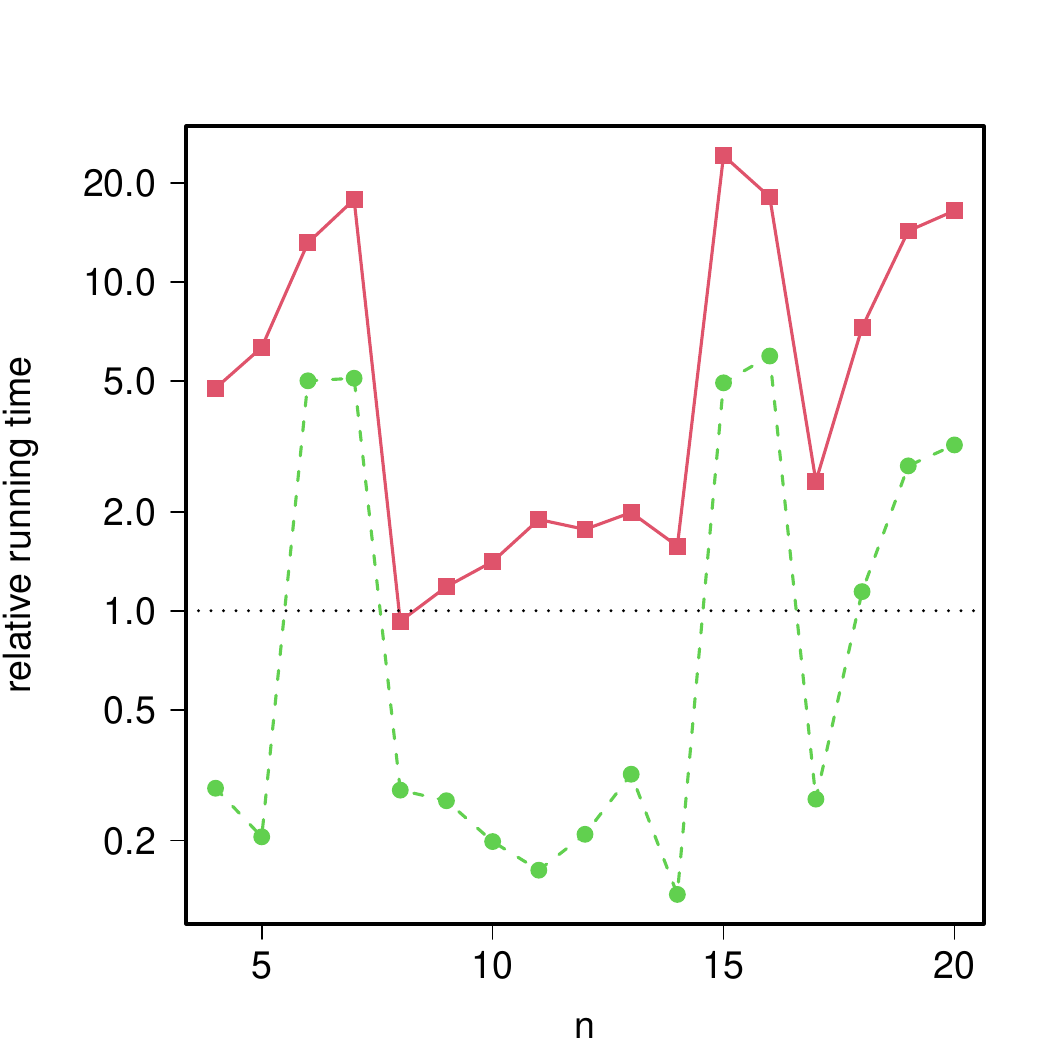} & 
		\includegraphics[width=0.45\textwidth]{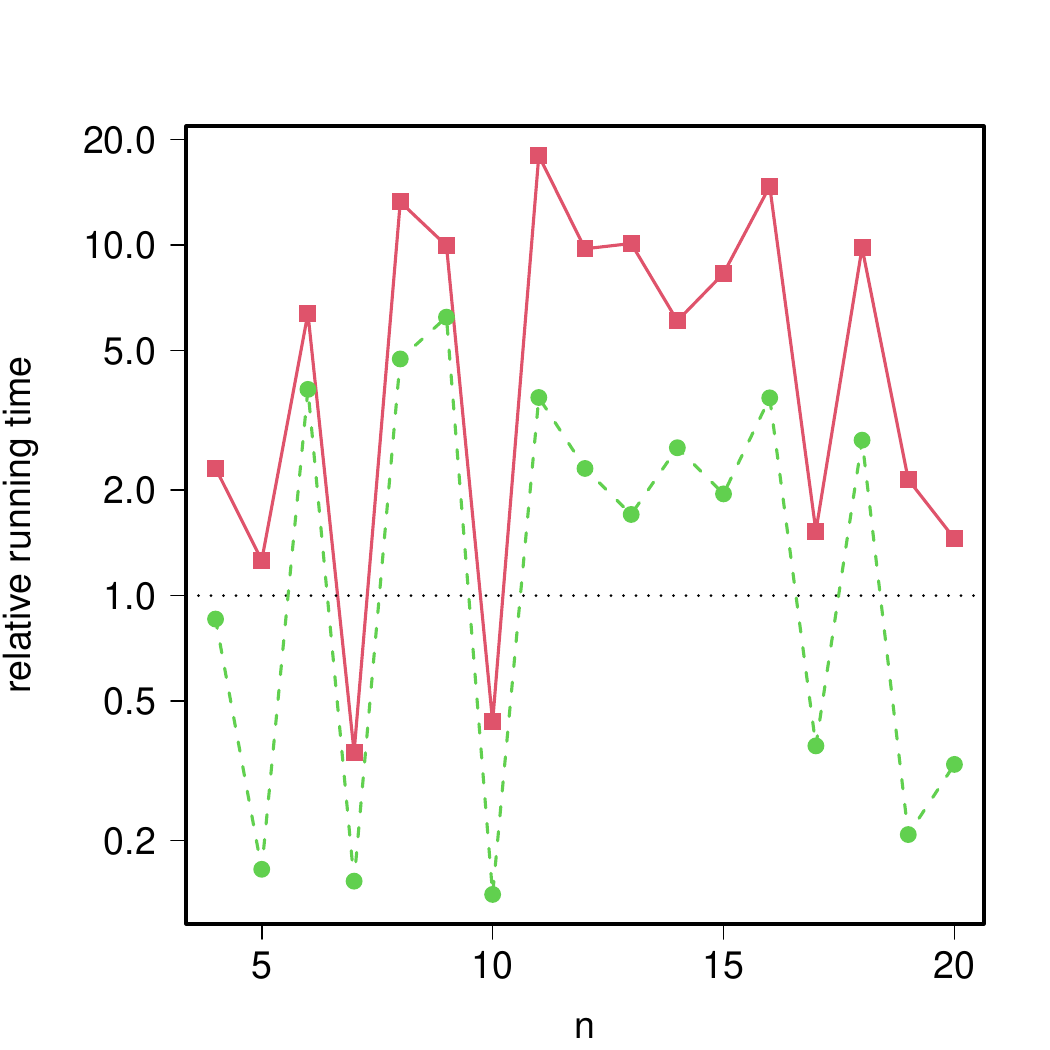} \\
		\includegraphics[width=0.45\textwidth]{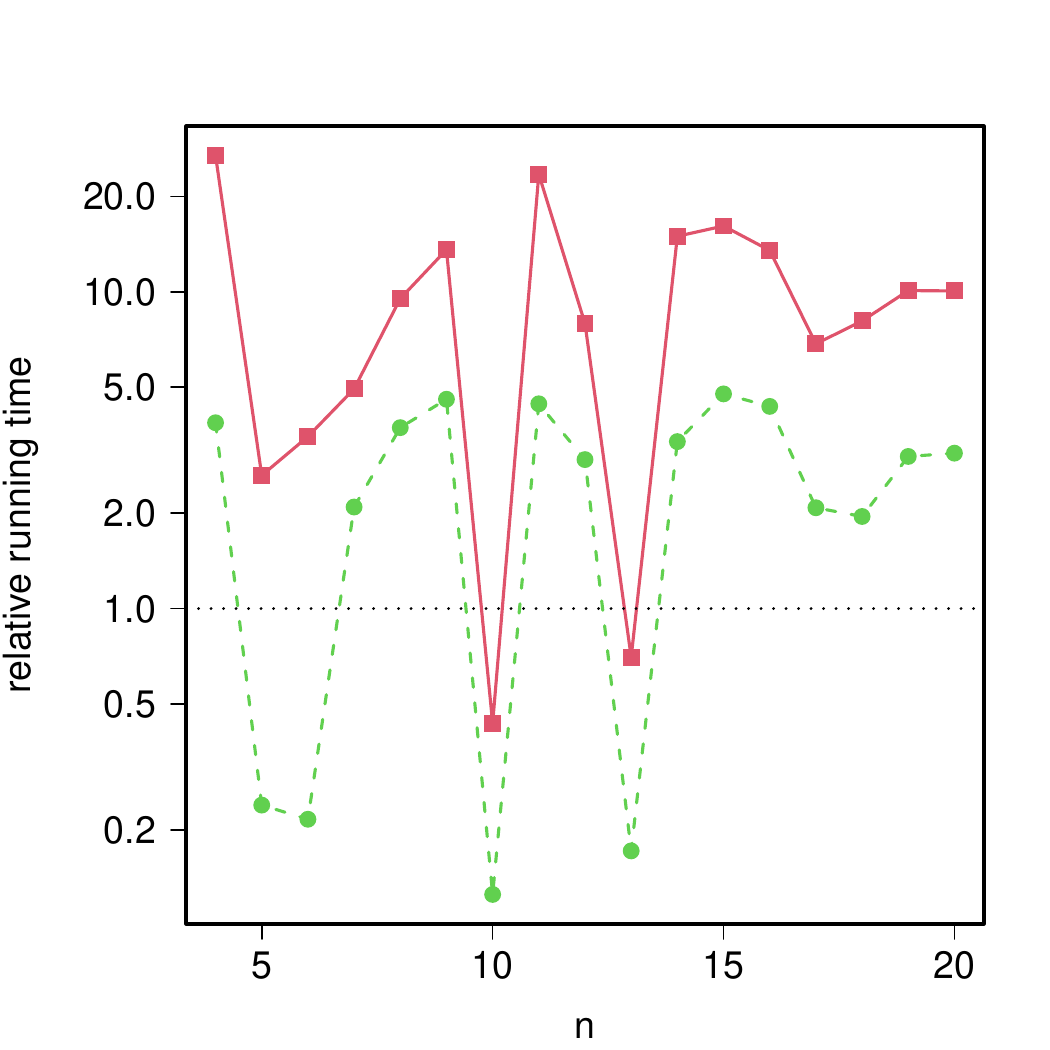} & 
		\includegraphics[width=0.45\textwidth]{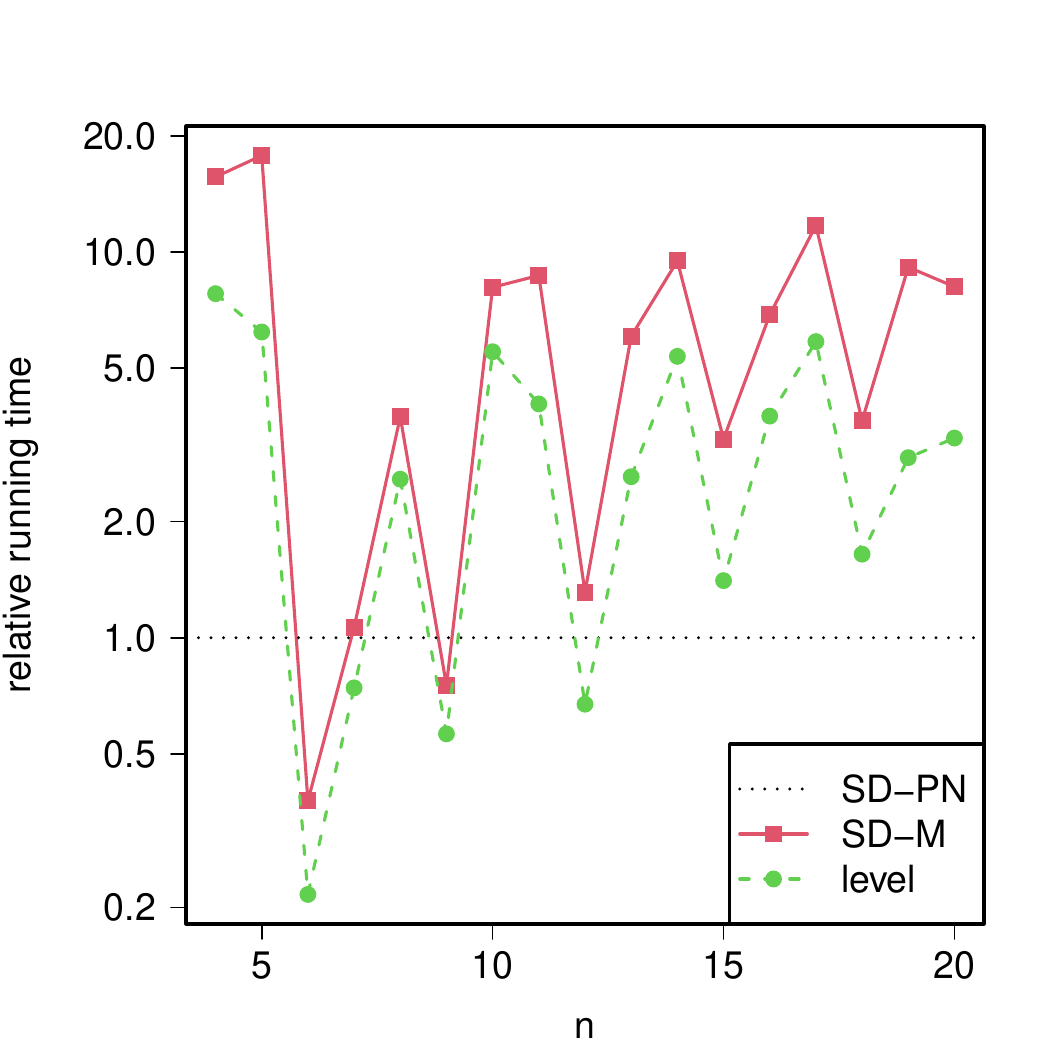} \\		
	\end{tabular}
	\caption{Example~1: running times relative to SD-PN. Left: D-optimality, right: A-optimality, top: original formulation, bottom: modified formulation. Note that the y-axis uses a logarithmic scale. SD-PN is represented by the horizontal line at $1$ since it serves as the benchmark. \label{fig:runtimes_example1}}
\end{figure}

The relative running times for the various criteria and virtual noise formulations are shown in Figure~\ref{fig:runtimes_example1}. The relative running times exhibit strong fluctuations as the design size $n$ rises, but in general for this example among the two simplicial decomposition variants the projected Newton method outperforms the multiplicative algorithm. There is no clear advantage for either SD-PN or the level method for this example.

\begin{figure}[hbtp!]
	\centering
	\includegraphics[width=0.55\textwidth]{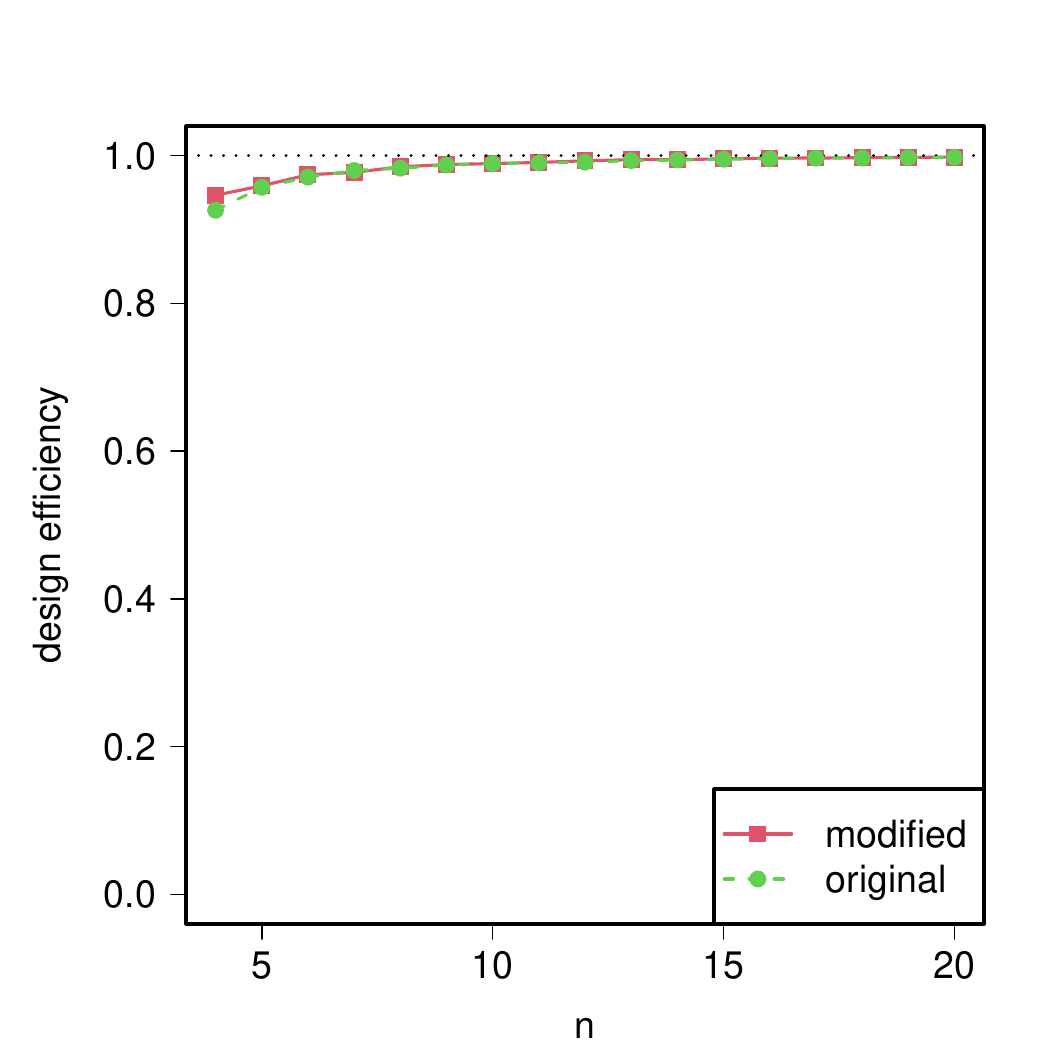} 
	
	\caption{Example~1: efficiencies of optimal exact $n$-point design relative to optimal virtual noise measure for original and modified formulation for D-optimality. \label{fig:efficiencies_example1}}
\end{figure}

Figure~\ref{fig:efficiencies_example1} demonstrates that the virtual noise formulation makes hardly any difference with respect to the efficiencies of the optimal exact $n$-point designs relative to the optimal virtual noise measure.

\subsection{Example 2: Example 1 with once-differentiable kernel}\label{sec:example2}

Similar to \cite{pazman_correlated_2022}, for the second example we use the same model as for Example~1 but replace the triangular kernel with a more challenging covariance structure,
\begin{equation*}
\Cov\{\varepsilon(x),\varepsilon( x')\}  = \left\{\min(x,x')\right\}^2 \left\{ 3 \max(x,x') - \min(x,x') \right\} / 6.
\end{equation*}
This is the kernel of the integrated Brownian motion, which is once continuously differentiable and therefore much smoother than the kernel of Example~1. Due to this smoothness, the minimum eigenvalues of the covariance and correlation matrices over the whole design grid are very low. Especially for low values of $n$, the programs aborted sometimes due to numerical problems involving operations across the whole grid, but for most settings reasonable design measures could be found using our suggested algorithms.

Notwithstanding these numerical issues, which might be overcome by modifying the coarseness of the design grid, the virtual noise method offers a generalized approach to obtain a design measure from which to construct criterion bounds for exact designs. 
If available, an alternative criterion bound could be derived from computing the covariance matrix of the BLUE of the continuous version of the regression model~\eqref{eq:model_orig}. However, as shown by \cite{dette_blue_2019}, for the integrated Brownian motion kernel the computations for the BLUE of the continuous model need to include the first derivatives of the observations, so it can be expected that this bound performs poorly for the discrete BLUE with no access to observation derivatives. Furthermore, it is not clear how to generalize this approach to higher design dimensions.

\begin{figure}[hbtp!]
	\centering
	\begin{tabular}{cc}
		\includegraphics[width=0.45\textwidth]{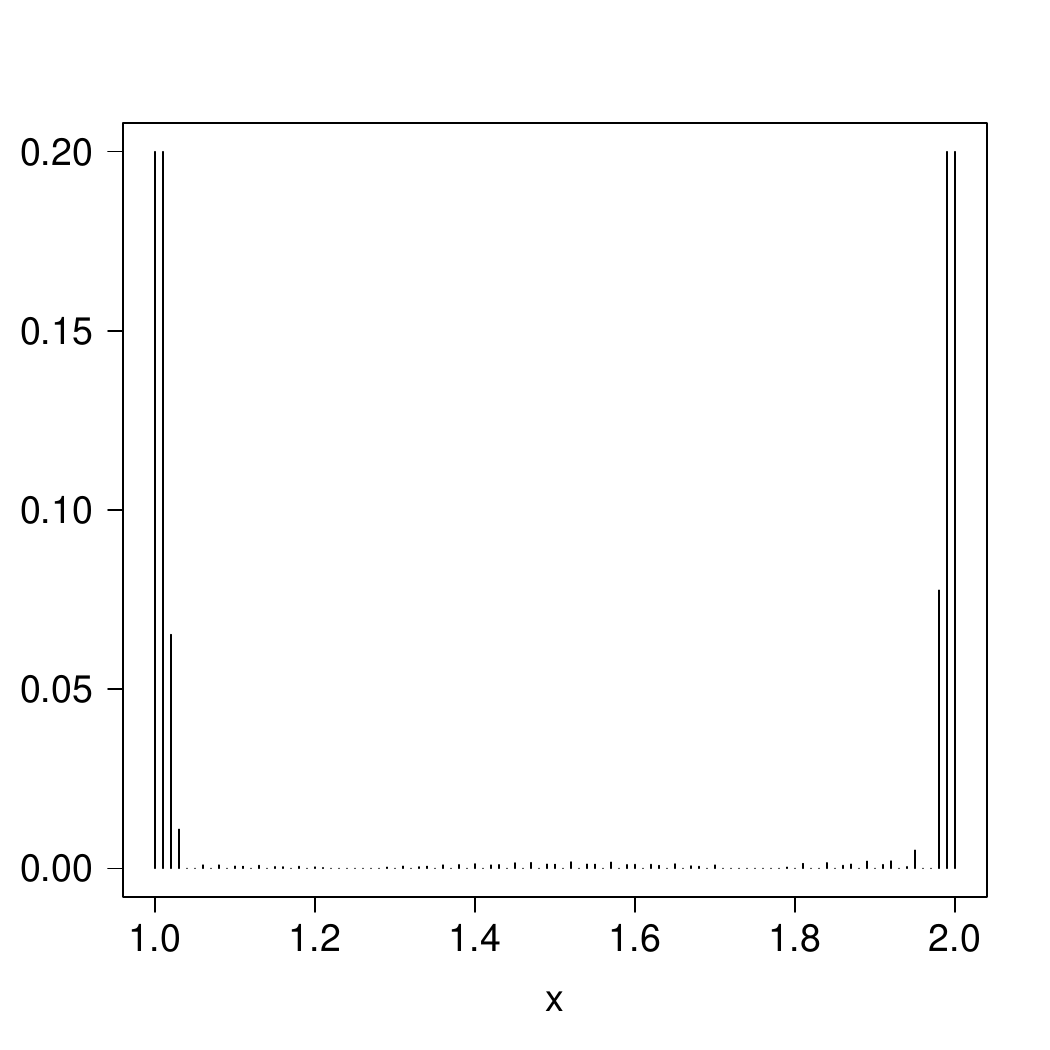} & 
		\includegraphics[width=0.45\textwidth]{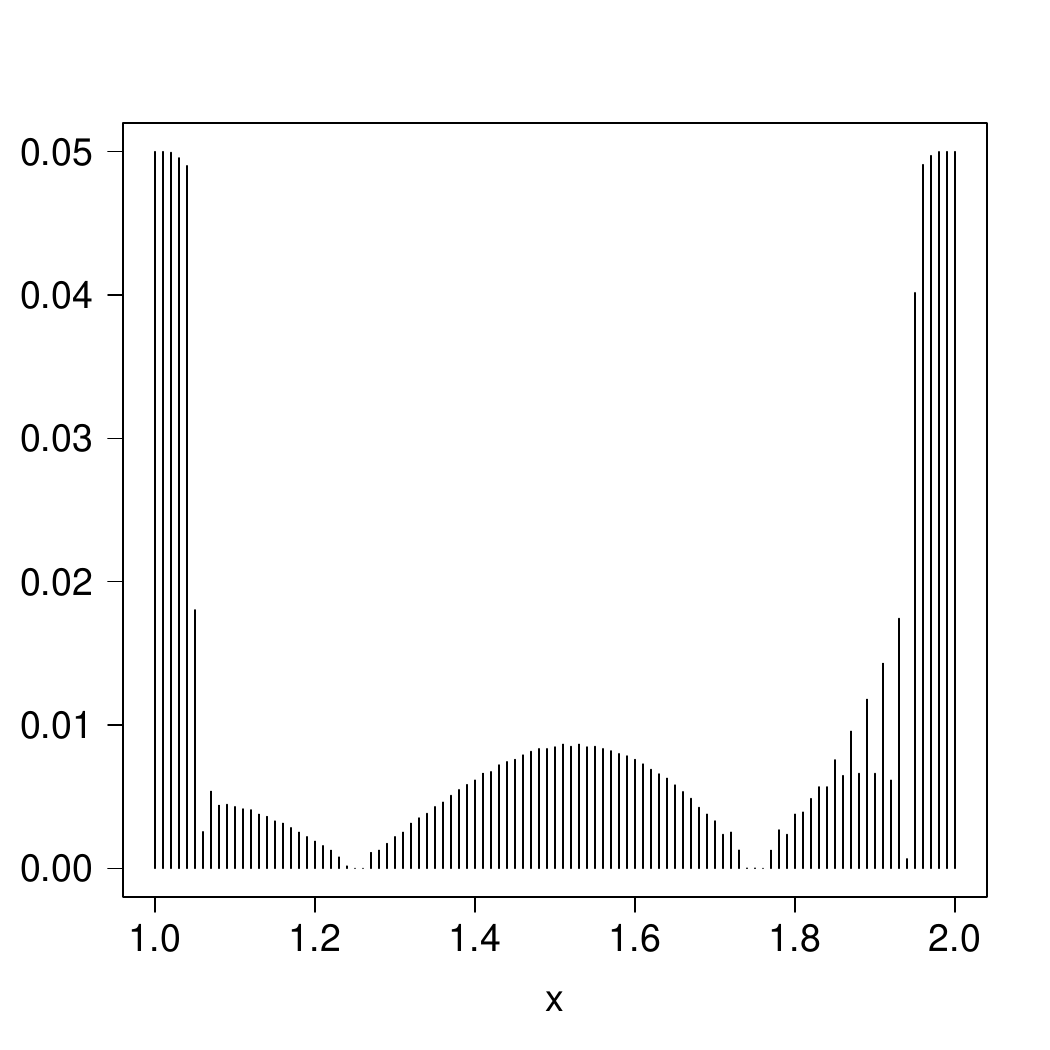}
	\end{tabular}
	\caption{Example~2: virtual noise design measures for $n = 5$ (left) and $n = 20$ (right) for D-optimality using the modified formulation. \label{fig:measures_example2}}
\end{figure}

Figure~\ref{fig:measures_example2} illustrates the D-optimal design measures for this example, again using the modified virtual noise formulation for $n = 5$ and $n = 20$. Compared to Figure~\ref{fig:measures_example1} for the first example, the largest permissible mass is not just allocated to the border locations but also to the available locations next to the border locations. This can be interpreted as allowing for implicitly estimating the derivative information that the continuous BLUE would require.

\begin{figure}[hbtp!]
	\centering
	\begin{tabular}{cc}
		\includegraphics[width=0.45\textwidth]{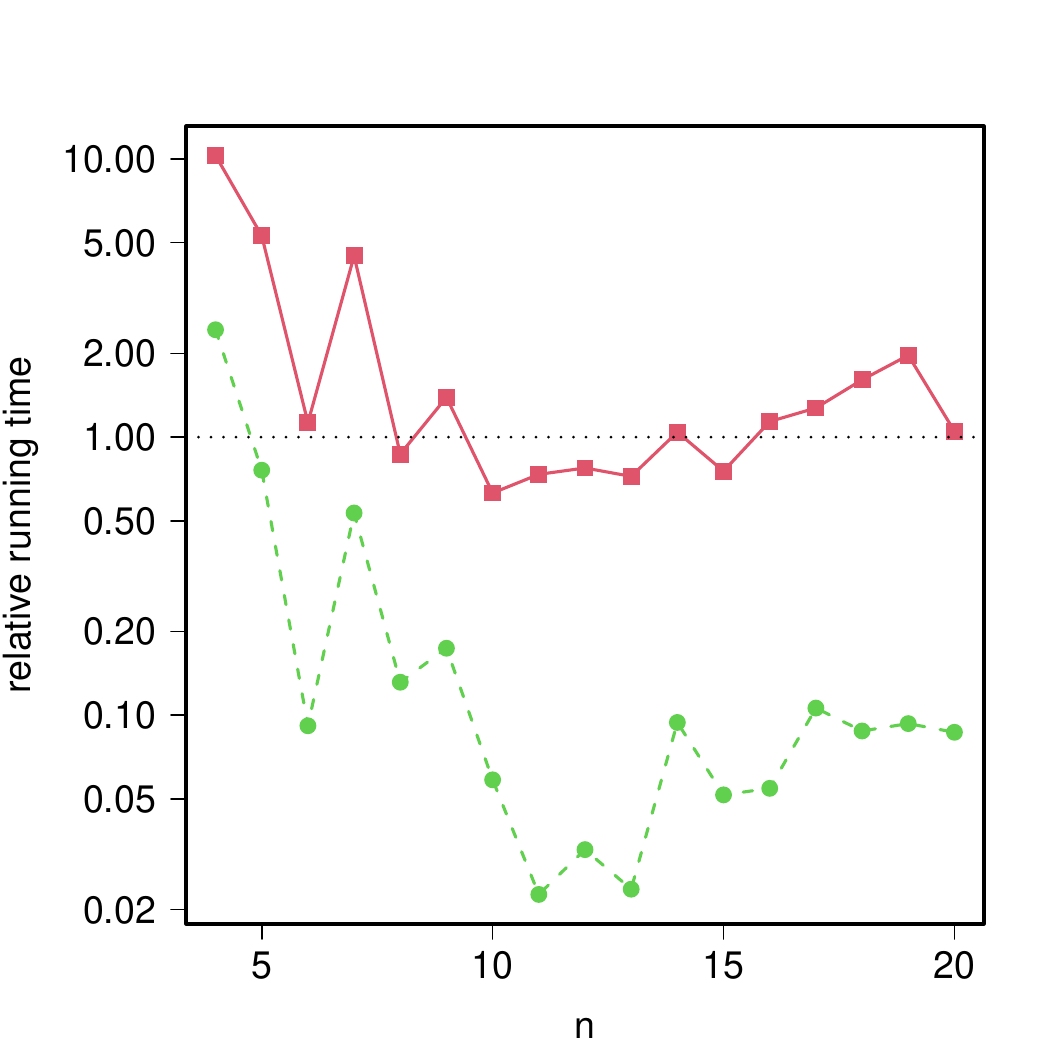} & 
		\includegraphics[width=0.45\textwidth]{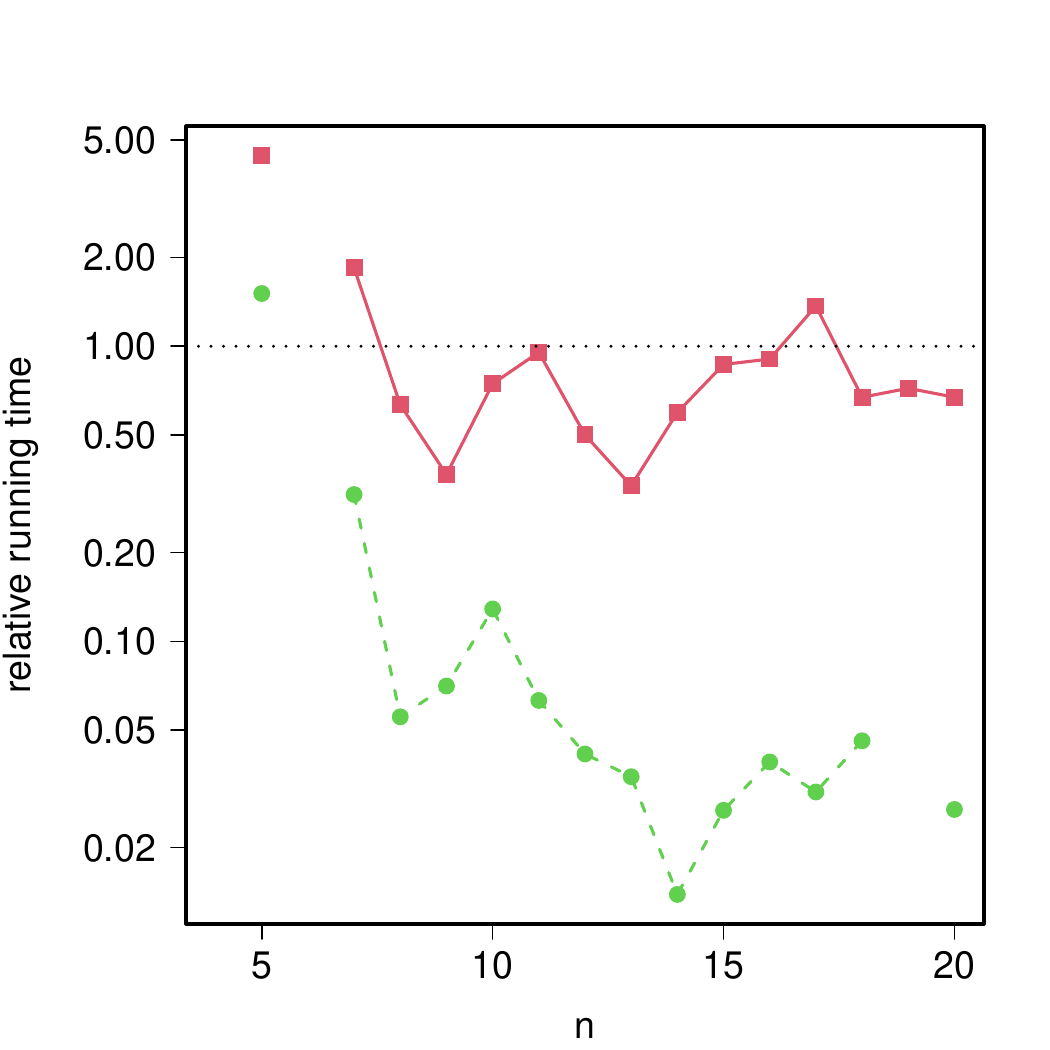} \\
		\includegraphics[width=0.45\textwidth]{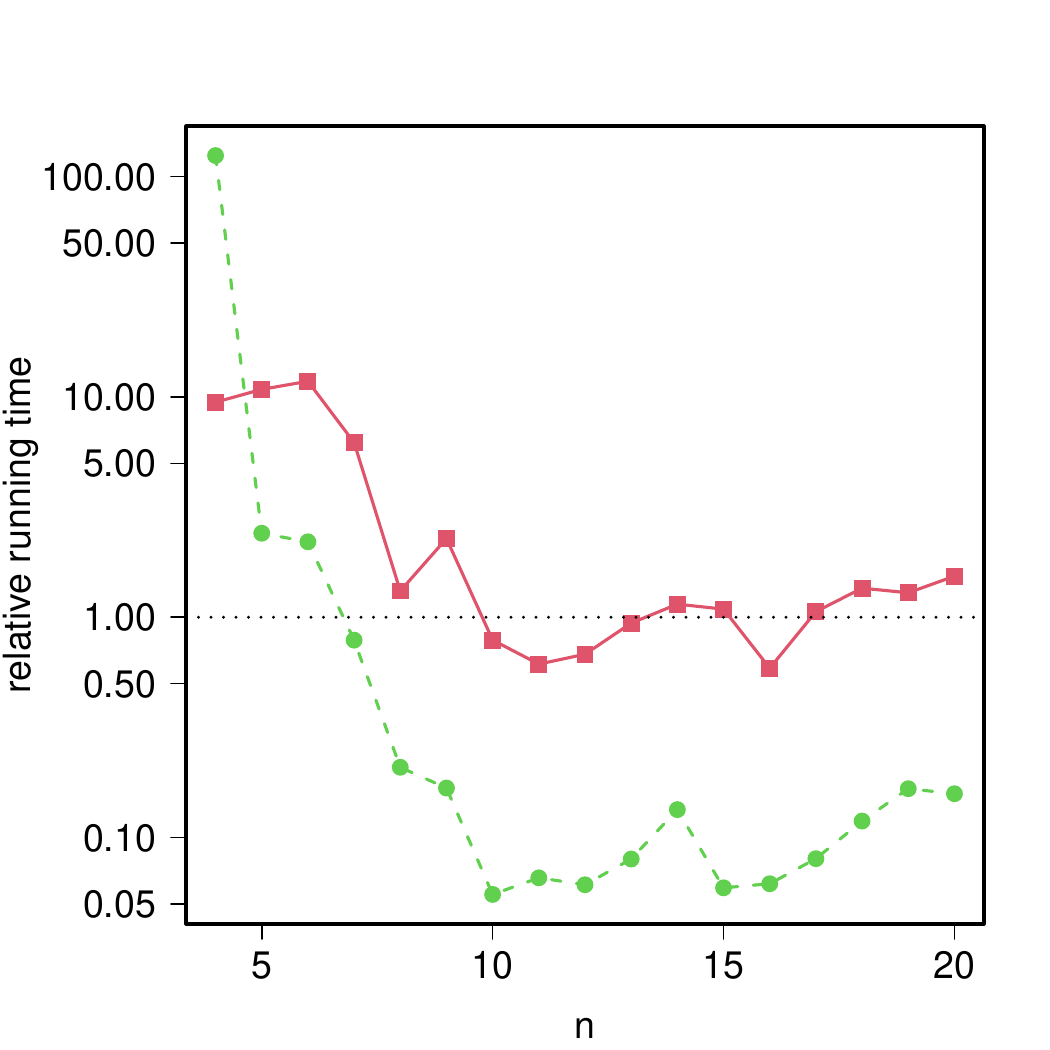} & 
		\includegraphics[width=0.45\textwidth]{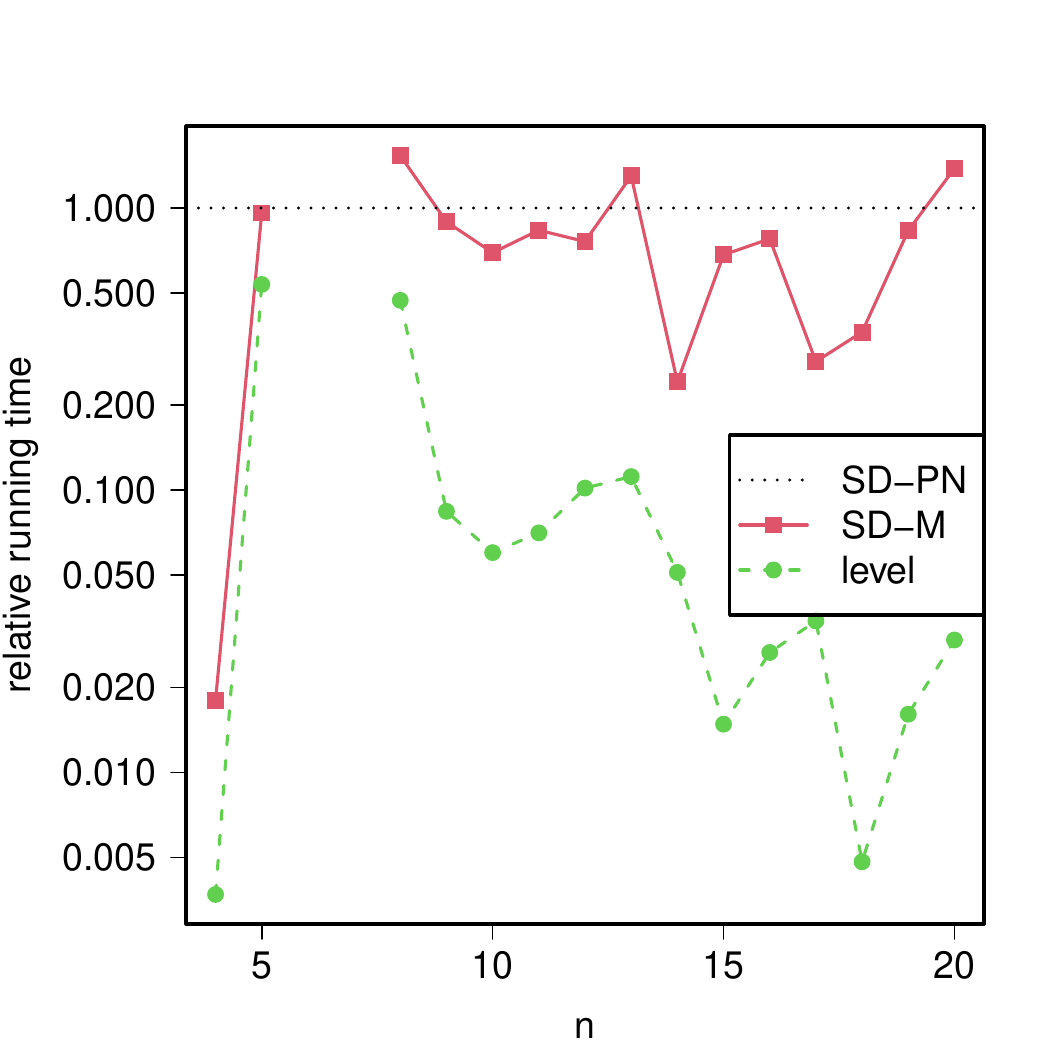} \\		
	\end{tabular}
	\caption{Example~2: running times relative to SD-PN. Left: D-optimality, right: A-optimality, top: original formulation, bottom: modified formulation. Note that the y-axis uses a logarithmic scale. SD-PN is represented by the horizontal line at $1$ since it serves as the benchmark. \label{fig:runtimes_example2}}
\end{figure}

From Figure~\ref{fig:runtimes_example2}, one can see that the level method generally performs best for this example. The two variants of the simplicial decomposition algorithm perform quite similarly, with SD-PM being mostly preferred for D-optimality and SD-M being mostly preferred for A-optimality.

\begin{figure}[hbtp!]
	\centering
	\includegraphics[width=0.55\textwidth]{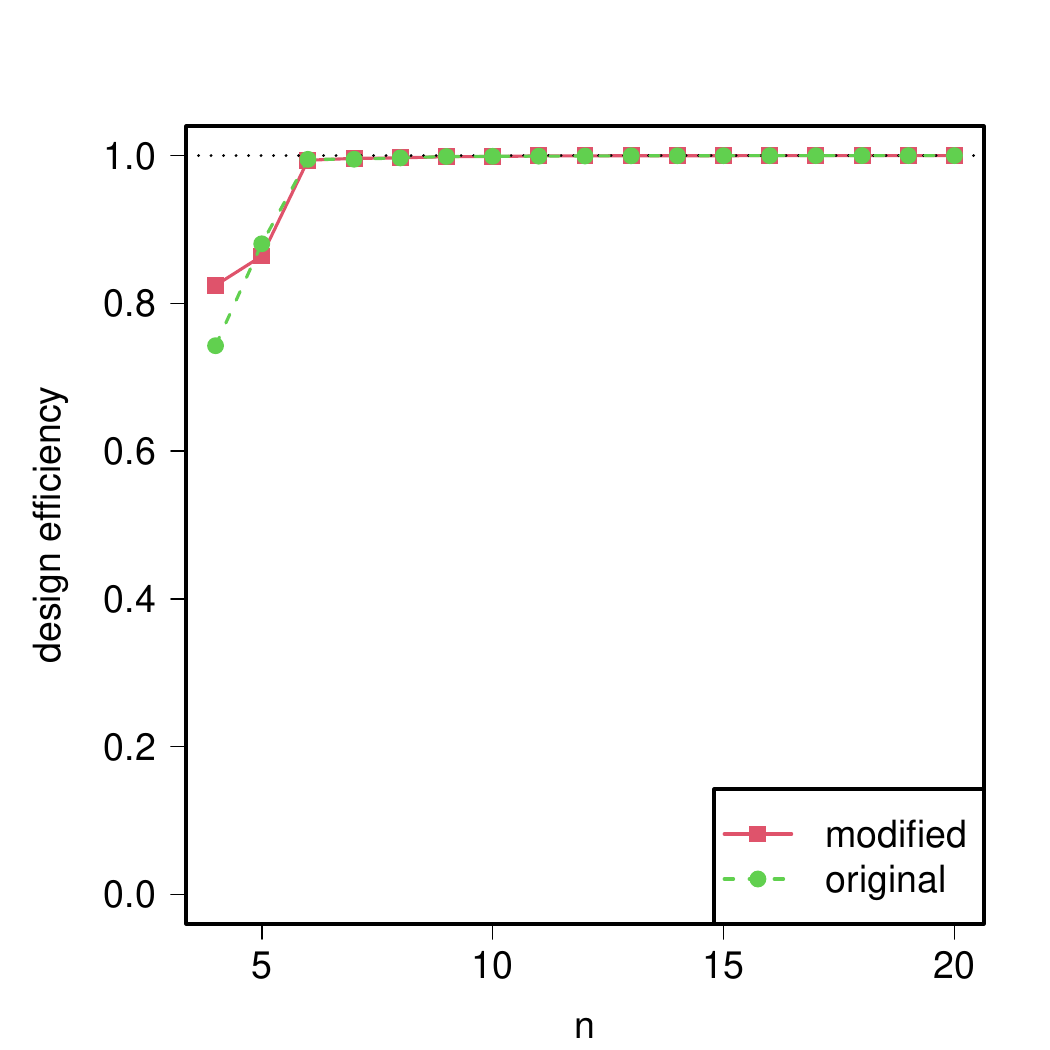}  
	\caption{Example~2: efficiencies of optimal exact $n$-point design relative to optimal virtual noise measure for original and modified formulation for D-optimality. \label{fig:efficiencies_example2}}
\end{figure}

As for the first example, the exact $n$-point design bounds resulting from the two virtual noise formulations are equally tight, see Figure~\ref{fig:efficiencies_example2}.

\subsection{Example 3: three-parameter model in two design dimensions with Gaussian kernel}\label{sec:example3}

For this example, we consider an example with two design variables. It has three parameters and uses a Gaussian kernel with fixed length-scale $\ell$.
\begin{eqnarray*}
	f^\T(x_1, x_2) & = & \left(1, x_1, x_2 \right), \\
	\Cov\{\varepsilon(x),\varepsilon( x')\} & = & \exp\left\{-\frac{1}{2 \ell^2}\left[(x_1 - x'_1)^2 + (x_2 - x'_2)^2 \right]\right\}, \\
	\mathcal{X} & = & \{-1, -0.8, -0.6, \ldots, 0.8, 1\} \times \{-1, -0.8, -0.6, \ldots, 0.8, 1\}.
\end{eqnarray*}

Since the covariance matrix equals the correlation matrix, the two formulations of the virtual noise variance are the same for this example. We apply the virtual noise method to three different settings for the length-scale parameter $\ell$: $\ell_1 = 1/(10 \sqrt{2})$ (weak correlation), $\ell_2 = 1/(2 \sqrt{5})$ (medium correlation), $\ell_3 = 1/(\sqrt{6})$ (strong correlation).

\begin{figure}[hbtp!]
	\centering
	\begin{tabular}{cc}
		\includegraphics[width=0.45\textwidth]{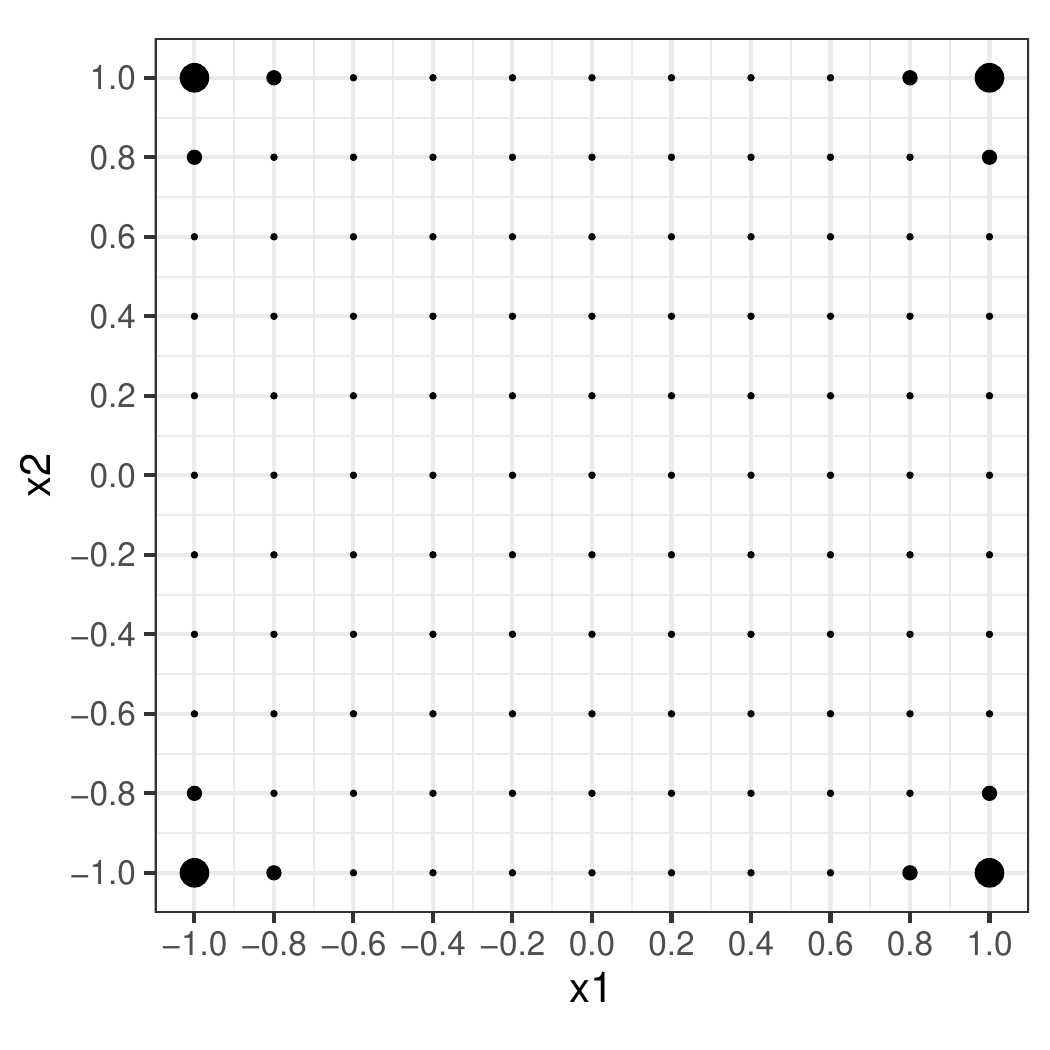} & 
		\includegraphics[width=0.45\textwidth]{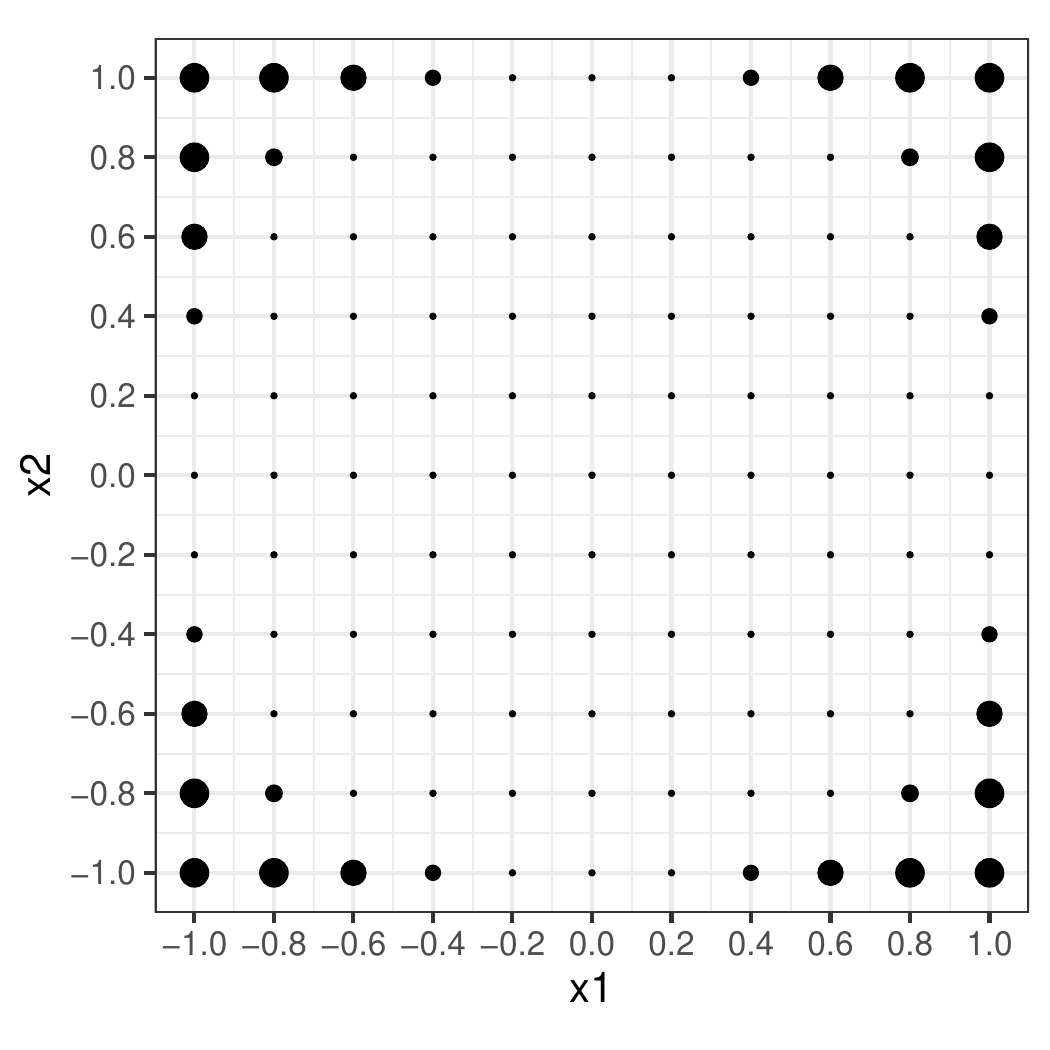} \\	
		\includegraphics[width=0.45\textwidth]{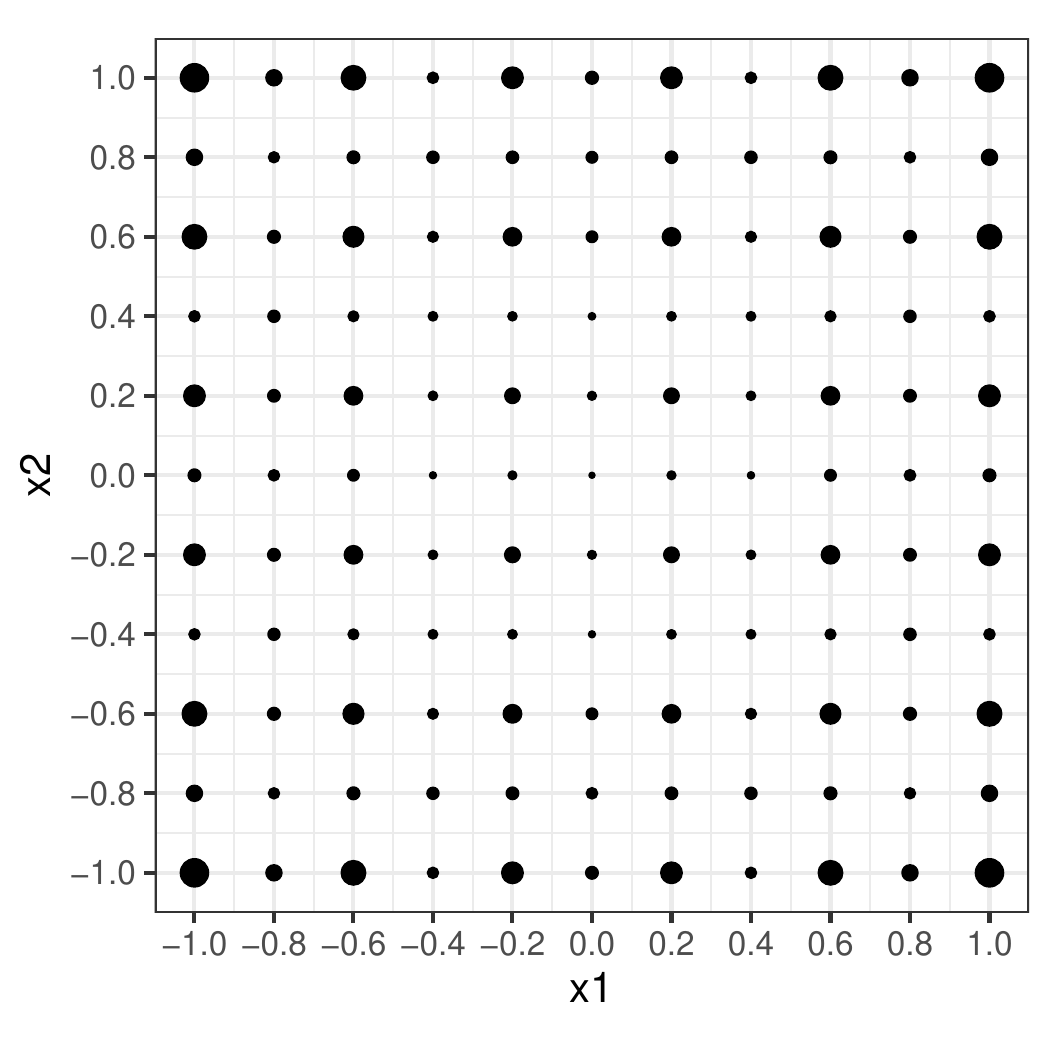} & 
		\includegraphics[width=0.45\textwidth]{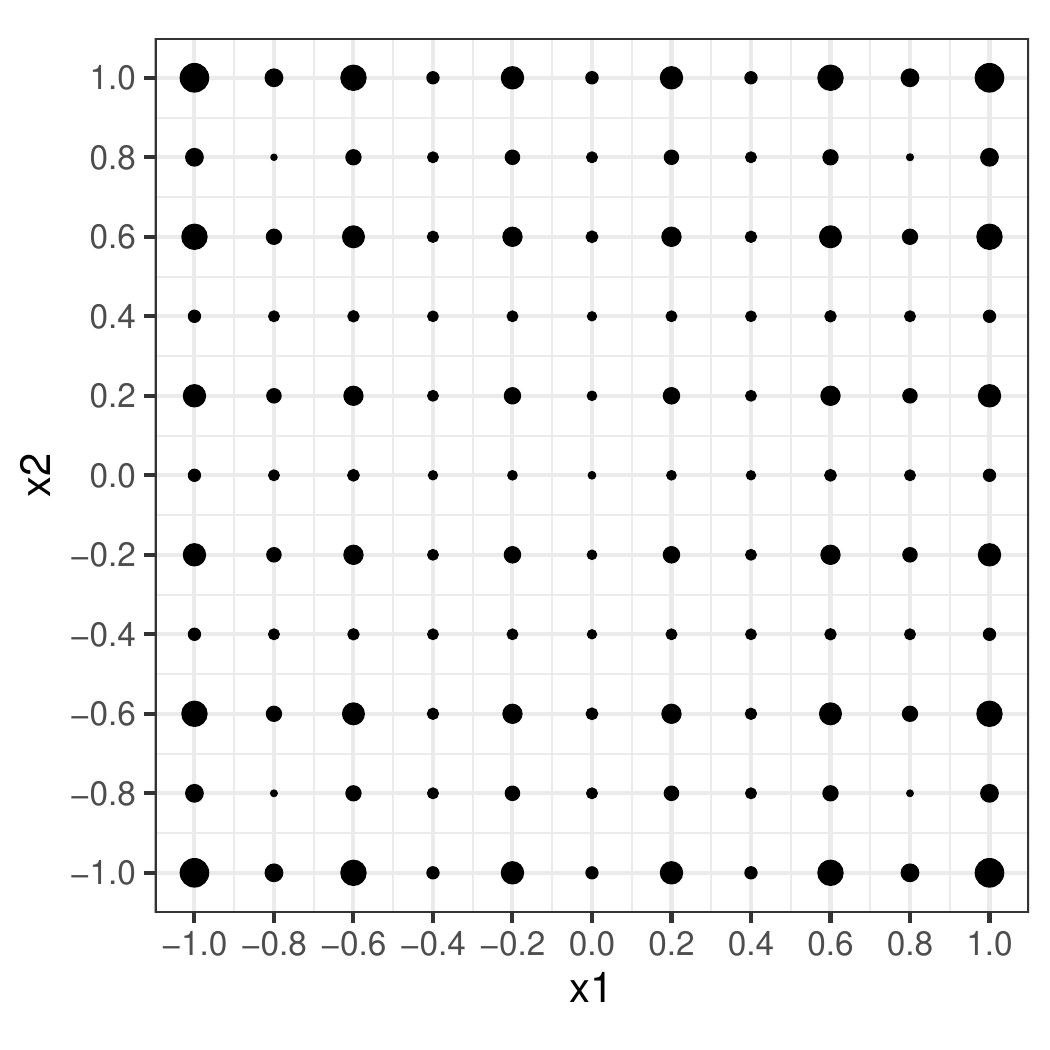} \\
		\includegraphics[width=0.45\textwidth]{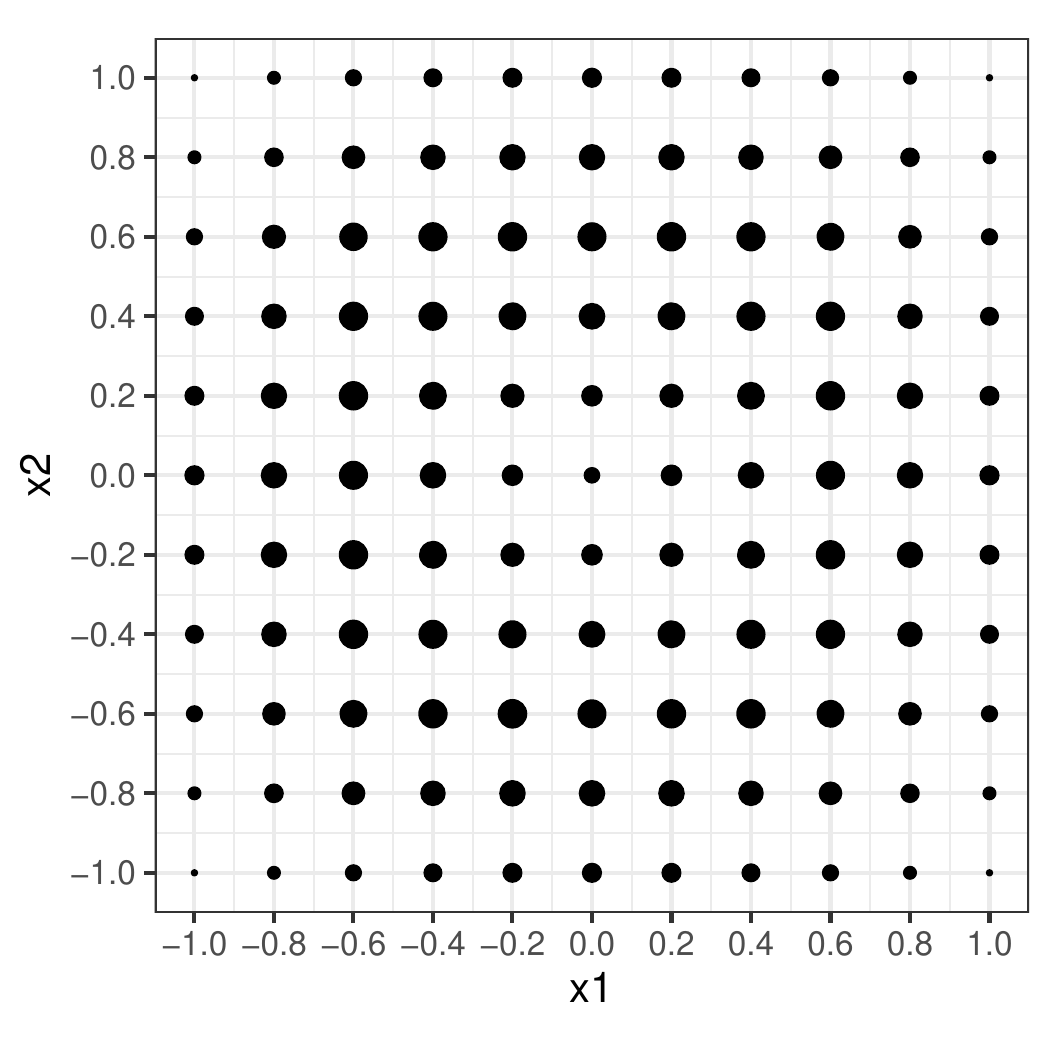} & 
		\includegraphics[width=0.45\textwidth]{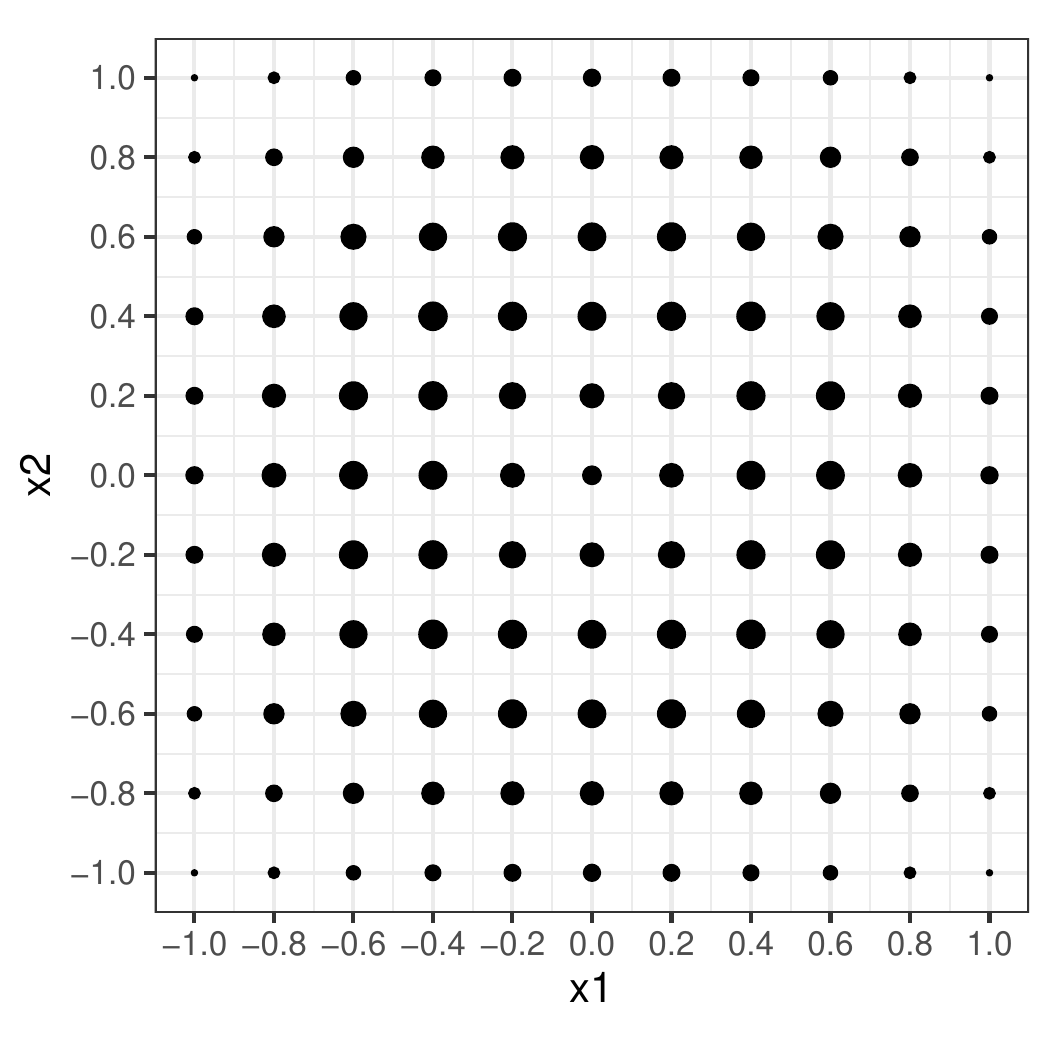} \\
	\end{tabular}
	\caption{Example~3: virtual noise design measures for $n = 5$ (left) and $n = 20$ (right) for D-optimality. Top: weak correlation, middle: medium correlation, bottom: strong correlation.  \label{fig:measures_example3}}
\end{figure}

The D-optimal virtual noise design measures for $n = 5$ and $n = 20$ are displayed in Figure~\ref{fig:measures_example3}. As expected, the higher the correlation, the more space-filing the design is. In the near-independence case, the measure is concentrated in the corners.

\begin{figure}[hbtp!]
	\centering
	\begin{tabular}{cc}
		\includegraphics[width=0.45\textwidth]{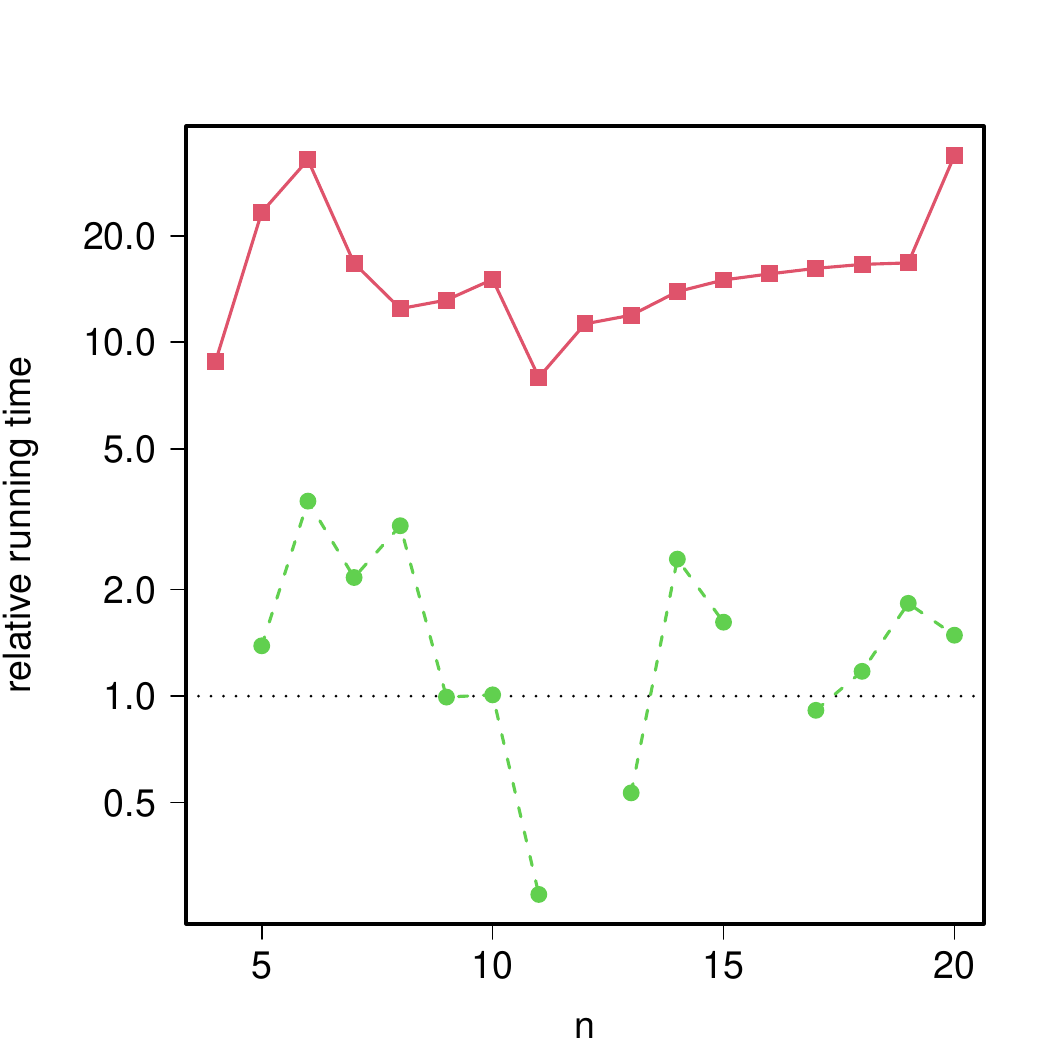} & 
		\includegraphics[width=0.45\textwidth]{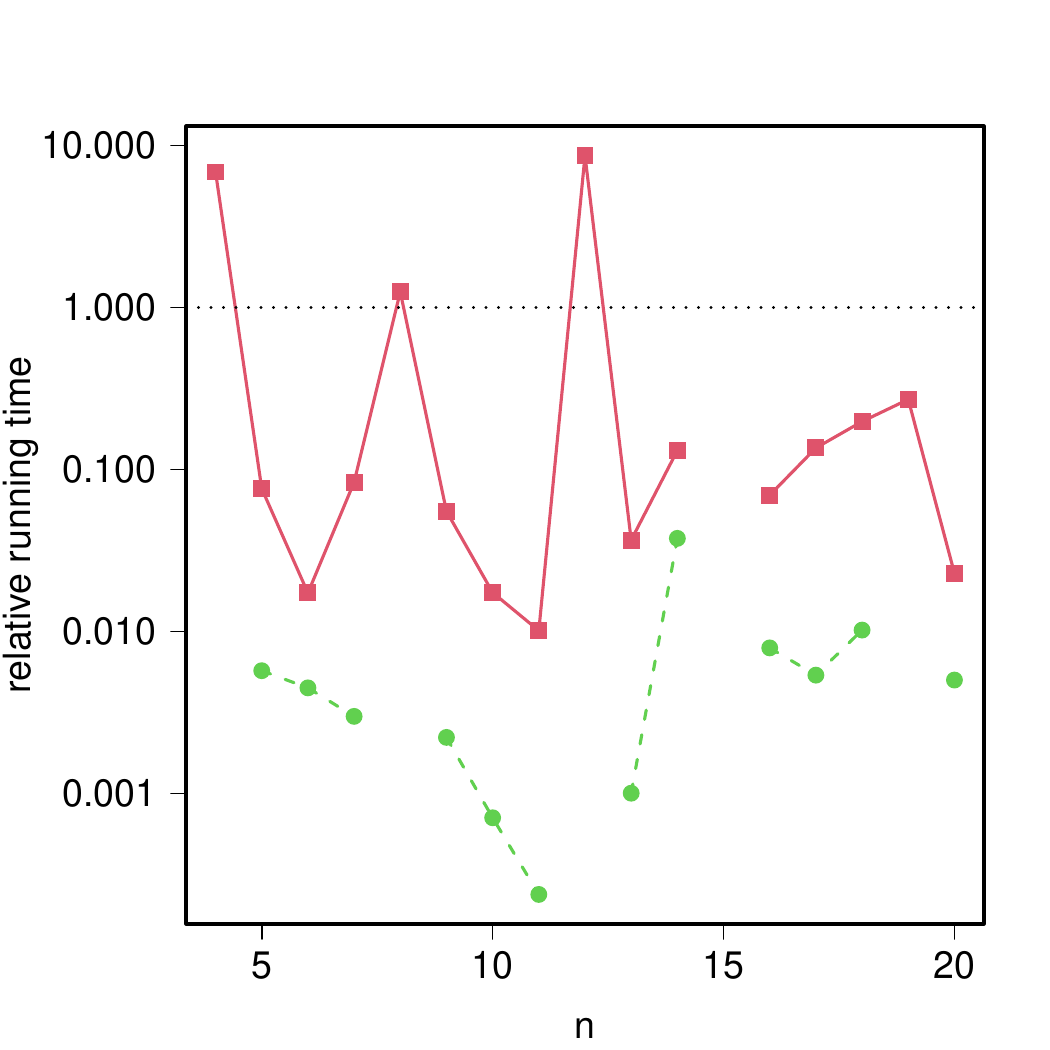} \\
		\includegraphics[width=0.45\textwidth]{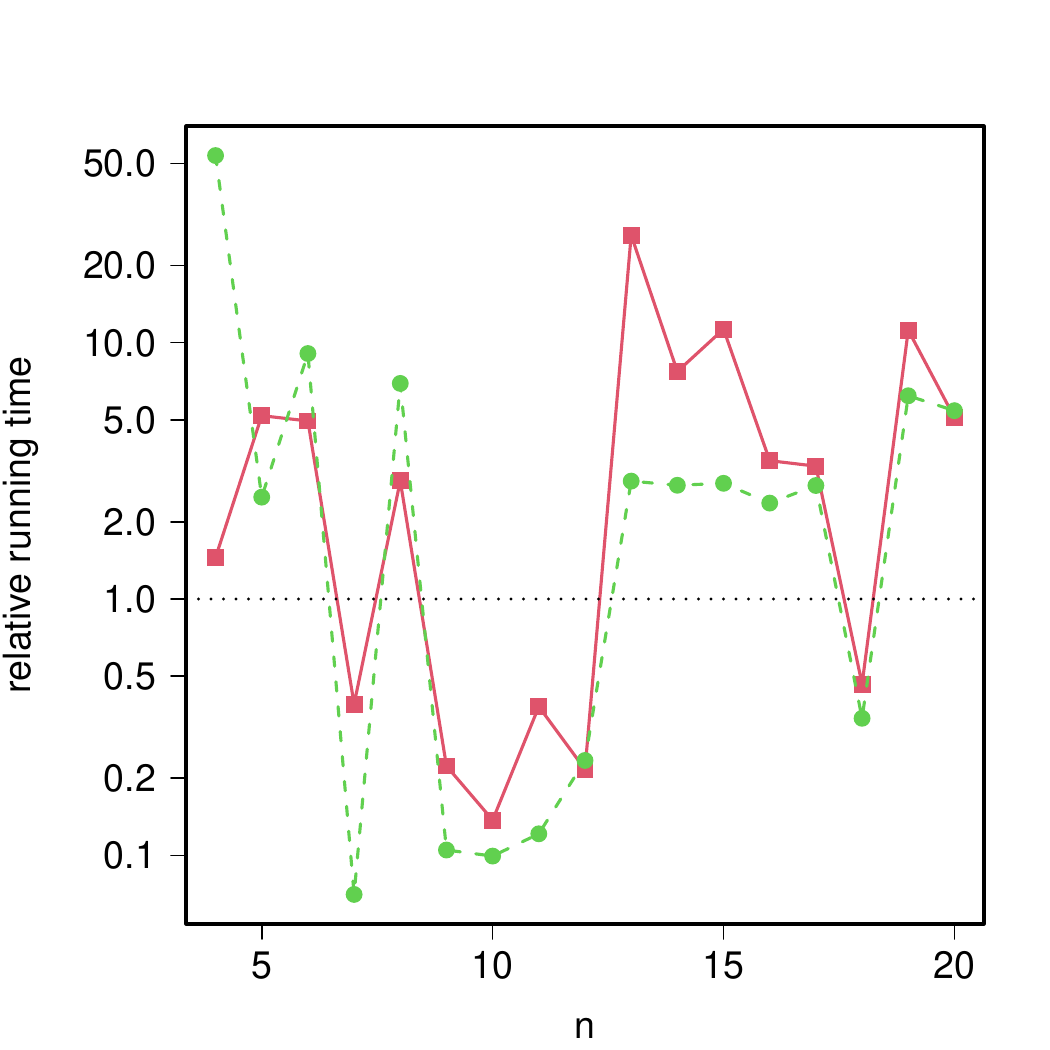} & 
		\includegraphics[width=0.45\textwidth]{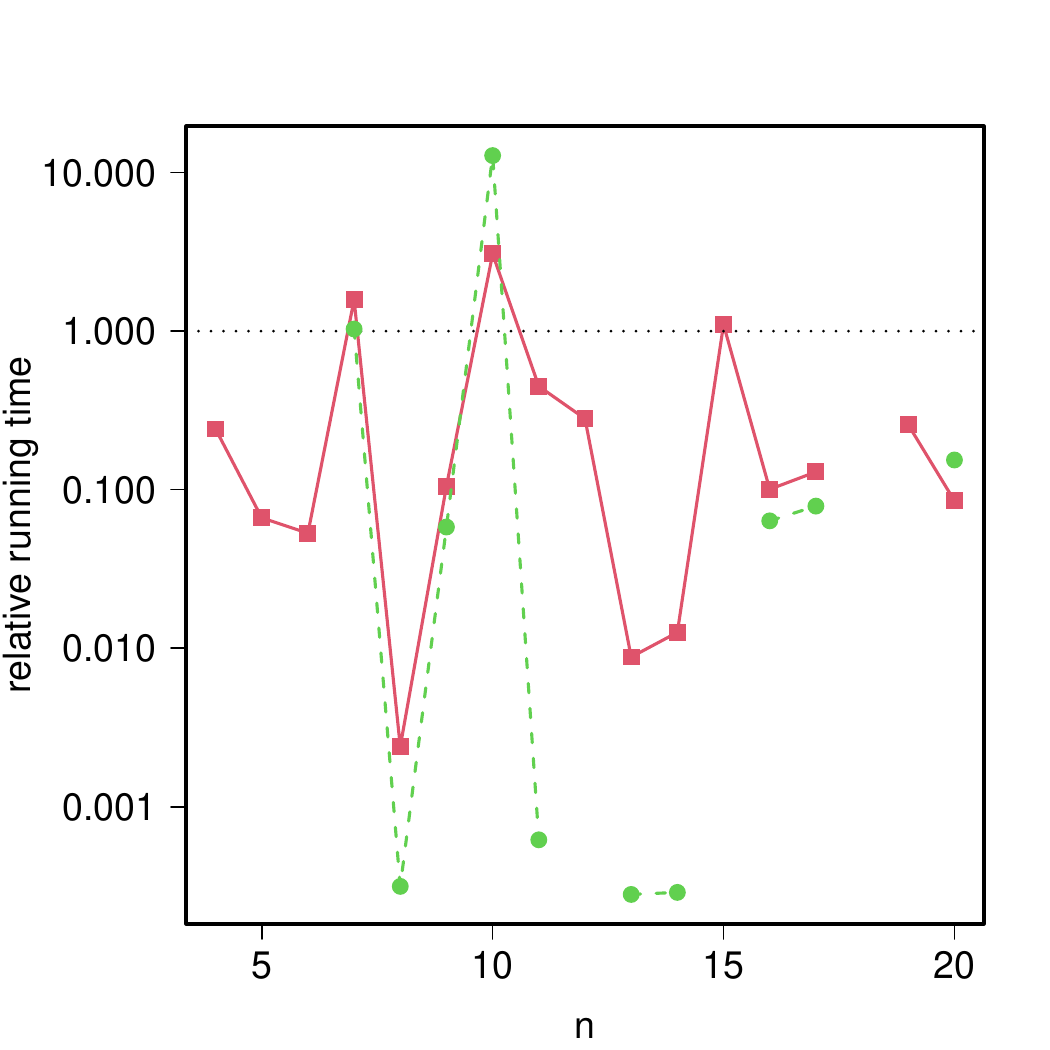} \\
		\includegraphics[width=0.45\textwidth]{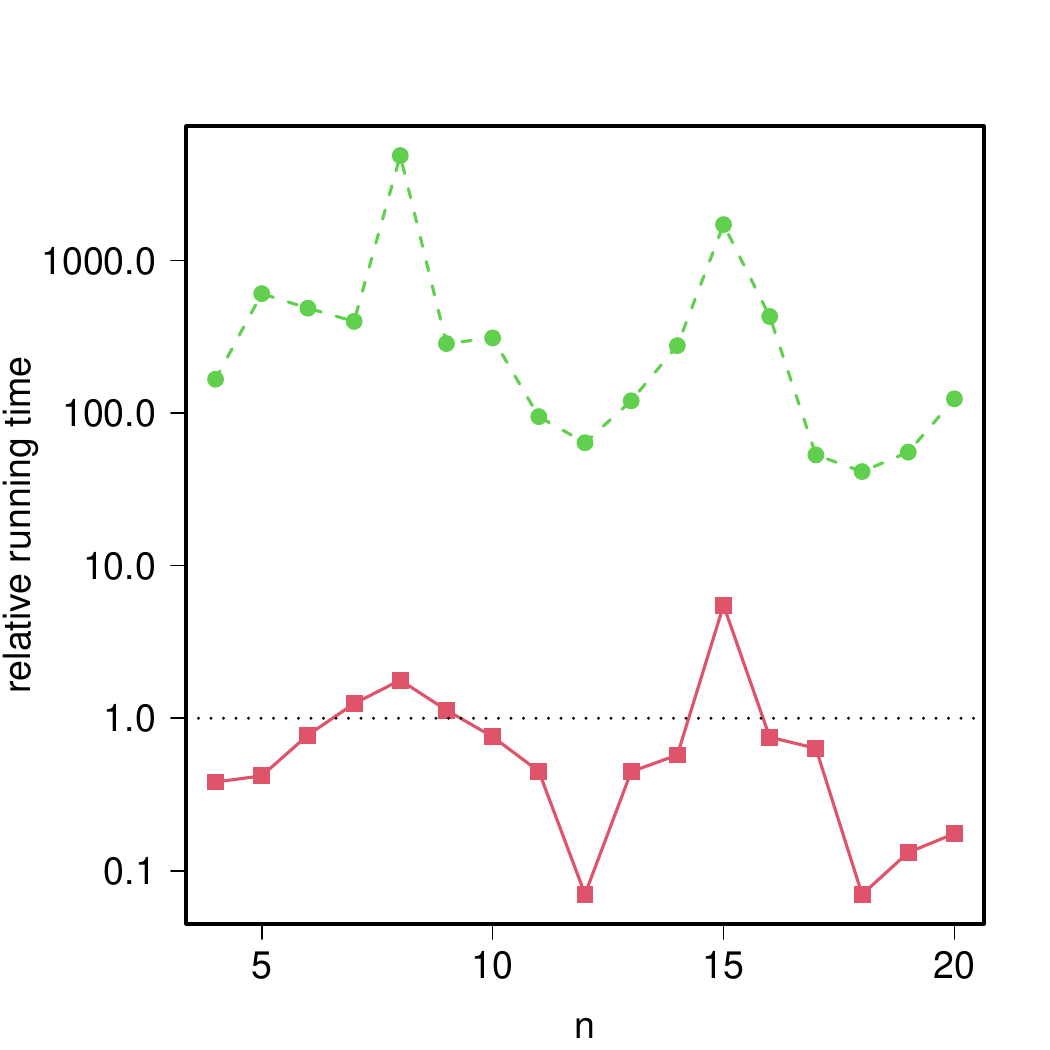} & 
		\includegraphics[width=0.45\textwidth]{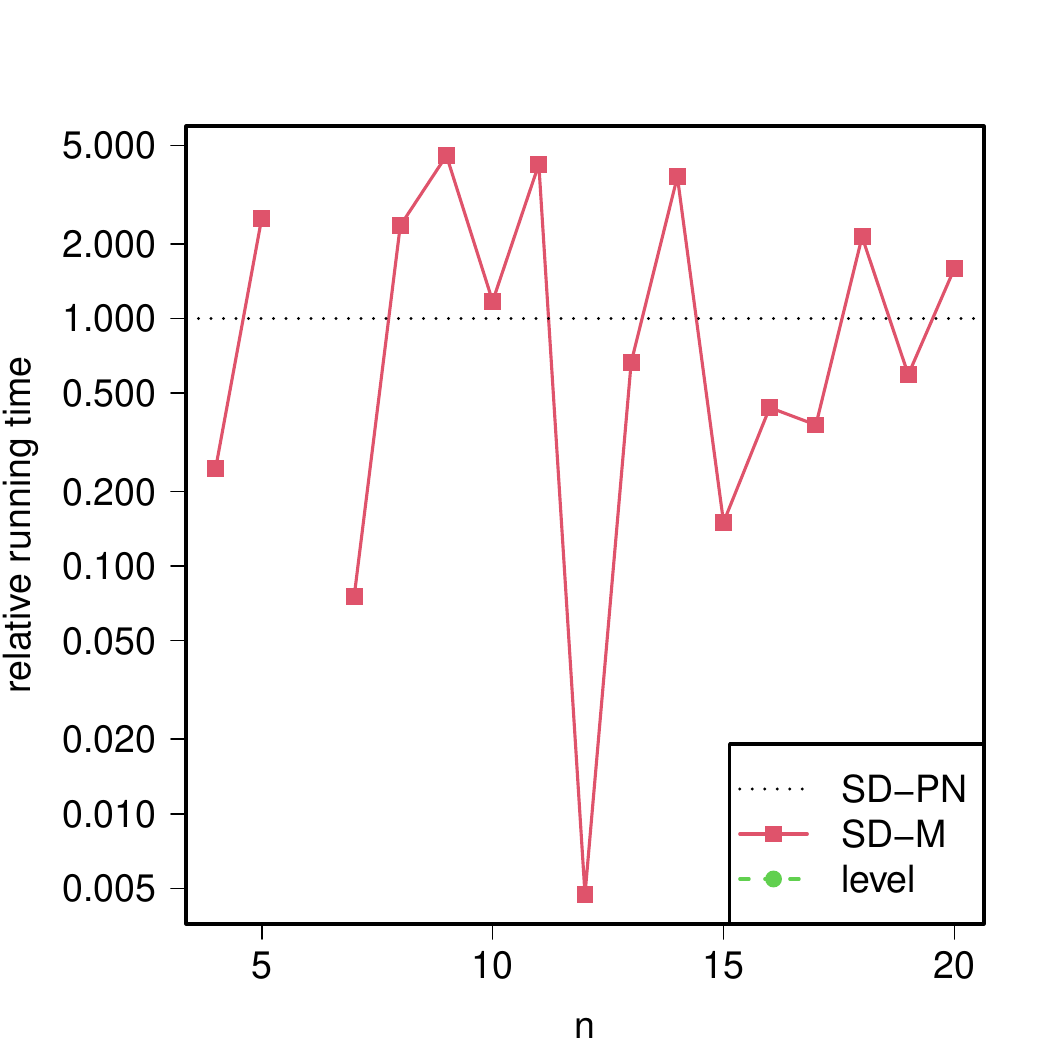} \\
	\end{tabular}
	\caption{Example~3: running times relative to SD-PN. Left: D-optimality, right: A-optimality, top: weak correlation, middle: medium correlation, bottom: strong correlation. Note that the y-axis uses a logarithmic scale. SD-PN is represented by the horizontal line at $1$ since it serves as the benchmark. \label{fig:runtimes_example3}}
\end{figure}

Comparing the running times of the algorithms in Figure~\ref{fig:runtimes_example3}, it is apparent that the level method struggles in the high-correlation setting. For A-optimality, all of the runs of the level method aborted due to numerical problems before achieving the required precision specified by SD-PN. SD-M is quite robust and works well compared to SD-PN in most cases, except for D-optimality in the low-correlation setting.

\begin{figure}[hbtp!]
	\centering
	\begin{tabular}{ccc}
		\includegraphics[width=0.3\textwidth]{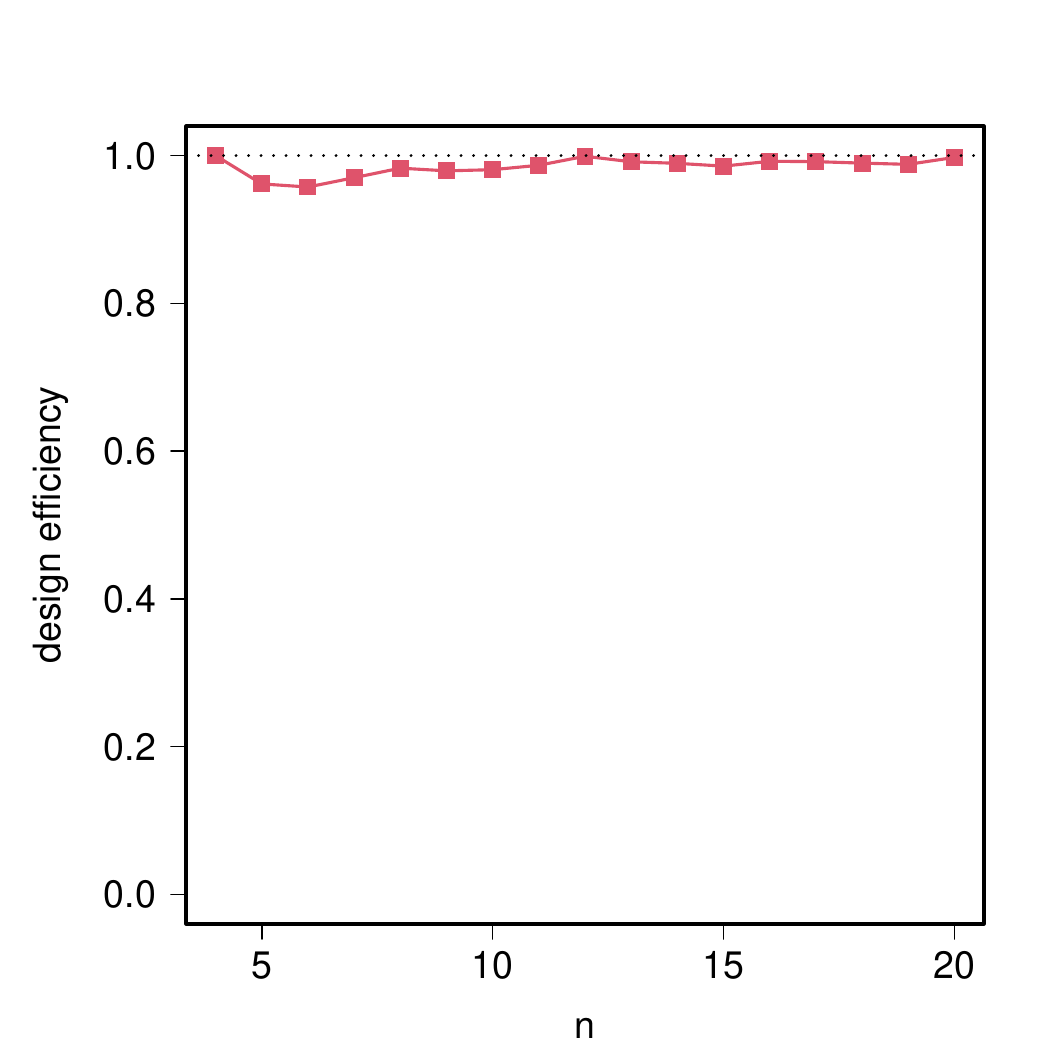} & 
		\includegraphics[width=0.3\textwidth]{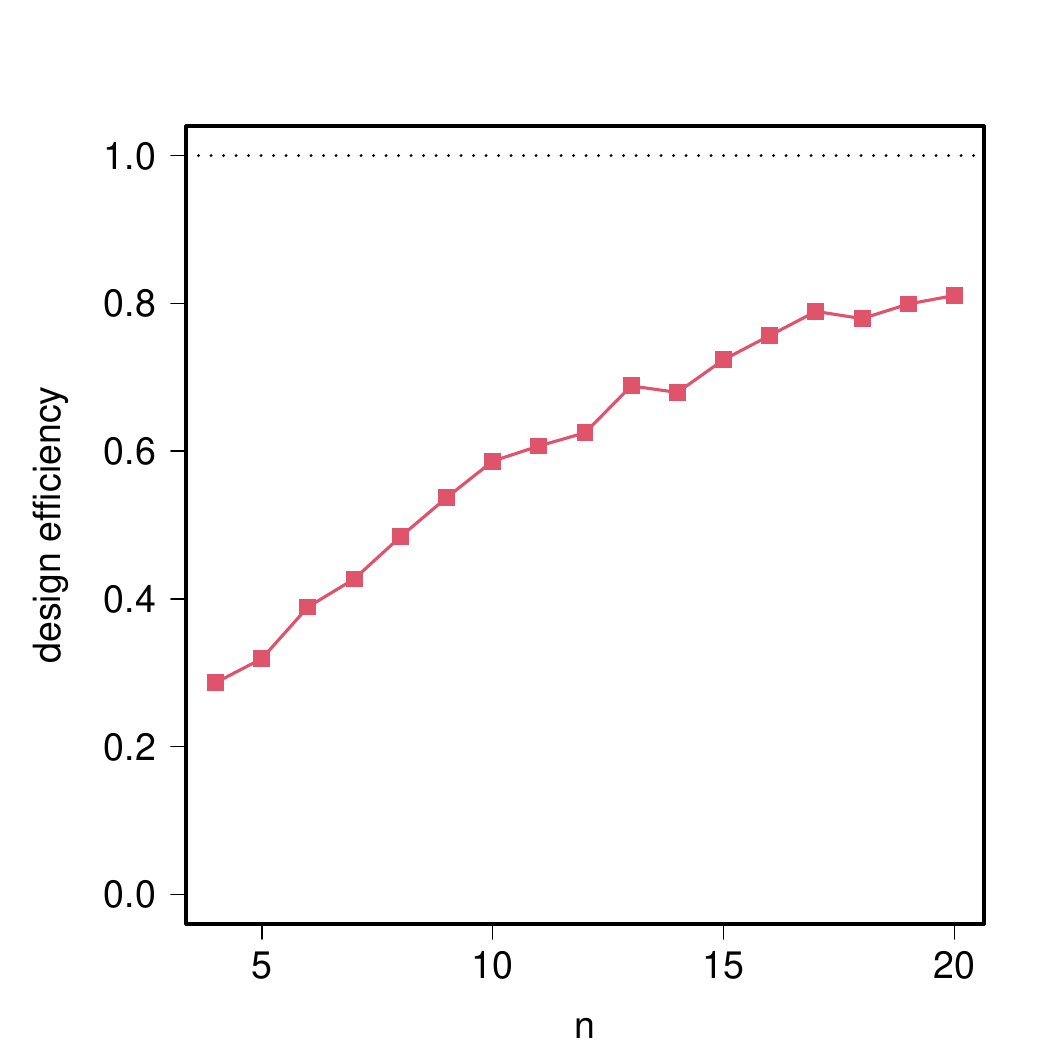} &
		\includegraphics[width=0.3\textwidth]{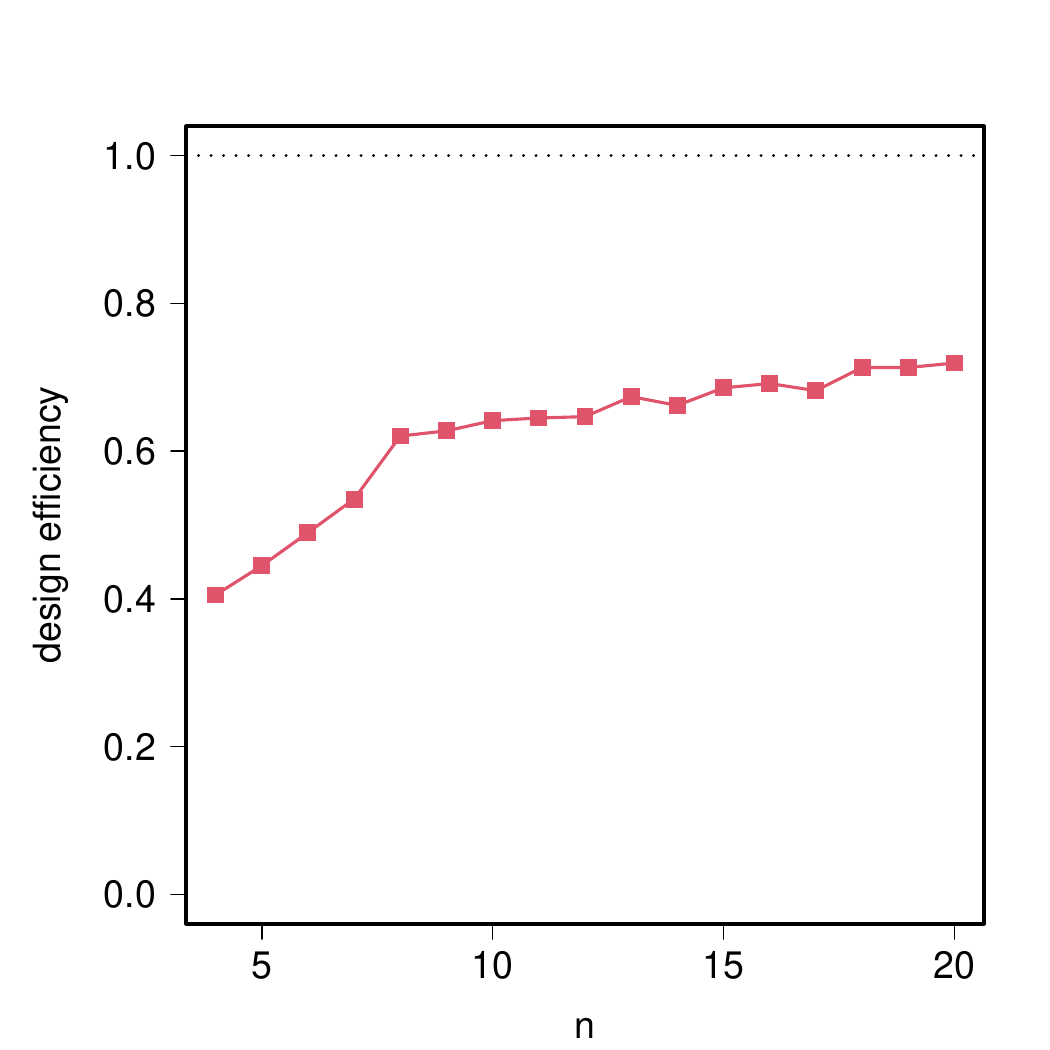}
	\end{tabular}
	\caption{Example~3: efficiencies of optimal exact $n$-point design relative to optimal virtual noise measure for D-optimality. Left: weak correlation, middle: medium correlation, right: strong correlation. \label{fig:efficiencies_example3}}
\end{figure}

Due to the rather uniform distribution of the design measure over the design space in the medium and high correlation settings, the criterion bounds resulting from the virtual noise model are quite loose. However, also in these examples the design efficiencies of the optimal exact $n$-point designs converge to $1$ as $n$ increases, albeit slowly, as can be seen in Figure~\ref{fig:efficiencies_example3}.

\section{Discussion and guidelines} 

\label{sec6}

The present paper complements \cite{pazman_correlated_2022} mainly by discussing and deepening its practical aspects and provides a firmer embedding into the existing literature, both for the classical uncorrelated case as well as modern convex relaxation techniques.

Thus we are now able to formulate guidelines on a more or less standardized procedure that could be used for general optimal design problems (including those with potentially correlated observations). We suggest to proceed as follows:
\begin{itemize}
	\item Specify a suitable regression model including correlations (e.g. a Gaussian process model), which includes the independent error assumption as a special case (to allow for seamless transition to the classical solutions).
	\item Utilize the virtual noise method to compute the criterion bound. Among the algorithms that we tested, it seems that the SD-M method is the most robust and reliable albeit not always the most efficient procedure for that purpose.
	\item Employ a suitable technique for finding the exact design from the literature. This could simply be by rounding from the design measure, although in most cases the algorithm by \cite{brimkulov_numerical_1980} proves to be superior to all other methods. As benchmarks one could also compute the exact design in the uncorrelated case and/or space-filling designs.
	\item Use the criterion bound to calculate efficiencies for various candidate designs and then select a final design based on that design's efficiency and additional practicability considerations.
\end{itemize}

\section*{Acknowledgements} 
M. Hainy and W. M\"uller were supported by the Linz Institute of Technology's grant LIFT\_C-2022-1-123 funded by the State of Upper Austria and Austria's Federal Ministry of Education, Science and Research. Computational resources and services used in this work were provided by the Scientific Computing Administration of the Johannes Kepler University, Linz, Austria.
We are most grateful for the helpful comments of D. Uci\'nski on an earlier version of the paper.

\newpage
\renewcommand{\baselinestretch}{1}
\normalsize
\bibliographystyle{asa}
\bibliography{20pazman}

\newpage

\titleformat{\section}{\large\bfseries}{Appendix~\thesection.}{2ex}{\large}

\appendix

\numberwithin{equation}{section}


\section{Proof of Theorem~\ref{th:special_case}}

\label{sec:proof_special_case}

W.l.o.g., we can assume that all $\xi(x_i) > 0$ for $i = 1,\ldots,N'$ and $\xi(x_i) = 0$ for $i = N'+1,\ldots,N$ (if $N' < N$). By the definition of the information matrix \eqref{eq:information_matrix_modified2}, the matrices $K'$, $W_{\tilde{\kappa}}'(\xi)$, and $\widetilde{F}'$ only contain the rows and columns corresponding to the first $N'$ observations.

Since the observations and errors are uncorrelated, the correlation matrix is the $N' \times N'$ identity matrix: $K' = I_{N'}$. The minimum eigenvalue of $I_{N'}$ is 1, so we can set $\tilde{\kappa} = 1$. From this, it follows that 
\begin{eqnarray*}
	K' + W_{\tilde{\kappa}}'(\xi) & = & K' + \diag\left\{\left(\tilde{\kappa} \left[ \frac{1}{n \, \xi(x_1)} - 1 \right], \ldots,  \tilde{\kappa} \left[ \frac{1}{n \, \xi(x_{N'})} - 1 \right]\right)\right\} \\
	& \underset{K'=I_{N'}, \, \tilde{\kappa} = 1}{=} & I_{N'} + \diag\left\{\left(\frac{1}{n \, \xi(x_1)}, \ldots, \frac{1}{n \, \xi(x_{N'})} \right)\right\} - I_{N'} \\
	& = & \diag\left\{\left(\frac{1}{n \, \xi(x_1)}, \ldots, \frac{1}{n \, \xi(x_{N'})} \right)\right\},
\end{eqnarray*}
where $\diag\{w\}$ denotes a diagonal matrix whose diagonal entries are composed of the elements of the vector $w$ such that $\left[\diag\{w\}\right]_{i,i} = w_i$.

Since $\Sigma' = \diag\left\{\left(\sigma^2(x_1), \ldots, \sigma^2(x_{N'})\right)\right\} = C'$, $\widetilde{F}' = (C')^{-1/2} F'$. We then have
\begin{eqnarray*}
	\widetilde{M}(\xi) & = & (\widetilde{F}')^{\T}  \left[ K' + W_{\tilde{\kappa}}'(\xi) \right]^{-1} \widetilde{F}' \\
	& = & (F')^{\T} (C')^{-1/2} \left[\diag\left\{\left(\frac{1}{n \, \xi(x_1)}, \ldots, \frac{1}{n \, \xi(x_{N'})} \right)\right\}\right]^{-1} (C')^{-1/2} F' \\
	& = & (f(x_1) \cdots f(x_{N'})) \: \diag\left\{\left(\frac{n \, \xi(x_1)}{\sigma^2(x_1)}, \ldots, \frac{n \, \xi(x_{N'})}{\sigma^2(x_{N'})} \right)\right\} \begin{pmatrix}
		f(x_1)^{\T} \\ 
		\vdots\\ 
		f(x_{N'})^{\T} 
	\end{pmatrix} \\
	& = & \sum_{i = 1}^{N'} f(x_i) \, \frac{n \, \xi(x_i)}{\sigma^2(x_i)} \, f(x_i)^{\T} =  n \sum_{x \in \X} \frac{1}{\sigma^2(x)} \, f(x) \, f(x)^{\T} \, \xi(x).
\end{eqnarray*}
The last equation follows because for all $x \in \X$ with $\xi(x) = 0$ also the contribution \linebreak $\left[n / \sigma^2(x)\right] f(x) \, f(x)^{\T} \, \xi(x)$ to the sum is $0$.
\qed


\section{Remainder of proof of Proposition~\ref{prop:liu_equality}}

\label{sec:proof_convex_relaxation}

In the case where $\xi \in \Xi\backslash\Xi_+$, i.e.\ $\xi(x_i) = 0$ for some $i \in \{1,\ldots,N\}$, we assume w.l.o.g.\ that $\xi(x_i) > 0$ for $i = 1,\ldots,N' < N$ and $\xi(x_i) = 0$ for $i = N'+1,\ldots,N$. Then we can partition the matrices $R = S^{-1}$ and $V = (n/\kappa) \, \diag\{\xi\}$ accordingly into
\begin{equation*}
S^{-1} = R = \begin{pmatrix}
R_{11} & R_{12} \\
R_{21} & R_{22}
\end{pmatrix}, \qquad
\frac{n}{\kappa} \, \diag\{\xi\} = V = \begin{pmatrix}
V_{11} & V_{12} \\
V_{21} & V_{22}
\end{pmatrix},
\end{equation*}
where $R_{11}$ is a positive definite $N' \times N'$ matrix, $R_{22}$ is a positive definite $(N-N') \times (N-N')$ matrix, $R_{12} = R_{21}^{\T}$ is an $N' \times (N-N')$ matrix, $V_{11}$ is an $N' \times N'$ diagonal matrix, and $V_{22}$ and $V_{12} = V_{21}^{\T}$ are matrices of dimension $(N-N') \times (N-N')$ and $N' \times (N-N')$, respectively, with all entries being $0$.

The matrix \begin{equation*}
T = S^{-1} +\frac{n}{\kappa} \, \diag\{\xi\}
\end{equation*}
can therefore be partitioned in the following way:
\begin{equation*}
T = R + V = \begin{pmatrix}
R_{11} + V_{11} & R_{12} + V_{12} \\
R_{21} + V_{21} & R_{22} + V_{22}
\end{pmatrix} = \begin{pmatrix}
T_{11} = R_{11} + V_{11} & T_{12} = R_{12} \\
T_{21} = R_{21} & T_{22} = R_{22}
\end{pmatrix}.
\end{equation*}

It is well known that the block inverse of a symmetric matrix is given by
\begin{eqnarray*}
	T^{-1} & = & \begin{pmatrix}
		G^{-1} & -G^{-1} T_{12} T_{22}^{-1} \\
		-T_{22}^{-1} T_{21} G^{-1} & T_{22}^{-1} + T_{22}^{-1} T_{21} G^{-1} T_{12} T_{22}^{-1}
	\end{pmatrix} \\
	& = & \begin{pmatrix}
		G^{-1} & -G^{-1} R_{12} R_{22}^{-1} \\
		-R_{22}^{-1} R_{21} G^{-1} & R_{22}^{-1} + R_{22}^{-1} R_{21} G^{-1} R_{12} R_{22}^{-1}
	\end{pmatrix},
\end{eqnarray*}
where $G = T_{11} - T_{12} T_{22}^{-1} T_{21} = R_{11} + V_{11} - R_{12} R_{22}^{-1} R_{21}$, we use the fact that $T_{11} = R_{11} + V_{11}$, $T_{12} = R_{12}$, $T_{21} = R_{21}$, $T_{22} = R_{22}$, and $R_{22}^{-1}$ exists because $R_{22}$ is positive definite. 

Let $S_{11}$ denote the upper left $N' \times N'$ block of $S$ corresponding to the points $x$ with $\xi(x) > 0$, which is positive definite. Note that $S_{11} = \left(R^{-1}\right)_{11} = \left(R_{11} - R_{12} R_{22}^{-1} R_{22}\right)^{-1}$. Since $S_{11}$ is positive definite, its inverse $S_{11}^{-1} = R_{11} - R_{12} R_{22}^{-1} R_{22}$ is also positive definite. Consequently, $G$ is positive definite and therefore invertible.

Next we compute $S^{-1} - S^{-1} T^{-1} S^{-1} = R - R T^{-1} R$. The result of these matrix operations is (after simplifications)
\begin{equation*}
R - R T^{-1} R = \begin{pmatrix}
X & 0 \\
0 & 0
\end{pmatrix},
\end{equation*}
where 
\begin{eqnarray*}
	X & = & R_{11} - R_{12} R_{22}^{-1} R_{21} - R_{11} G^{-1} R_{11} + R_{11} G^{-1} R_{12} R_{22}^{-1} R_{21} \\
	& & + R_{12} R_{22}^{-1} R_{21} G^{-1} R_{11}  - R_{12} R_{22}^{-1} R_{21} G^{-1} R_{12} R_{22}^{-1} R_{21} \\
	& = & R_{11} - R_{12} R_{22}^{-1} R_{21} - R_{11} G^{-1} \left[R_{11} - R_{12}R_{22}^{-1}R_{21} \right] + R_{12} R_{22}^{-1} R_{21} G^{-1} \left[R_{11} - R_{12}R_{22}^{-1}R_{21} \right] \\
	& = & R_{11} - R_{12} R_{22}^{-1} R_{21} - \left[R_{11} - R_{12}R_{22}^{-1}R_{21} \right] G^{-1} \left[R_{11} - R_{12}R_{22}^{-1}R_{21} \right] \\
	& = & S_{11}^{-1} - S_{11}^{-1} G^{-1} S_{11}^{-1} = S_{11}^{-1} - S_{11}^{-1} \left( S_{11}^{-1} + V_{11} \right)^{-1} S_{11}^{-1} \\
	& = & \left(S_{11} + V_{11}^{-1} \right)^{-1} = \left[C_{11} - \kappa \, I_{N'} + \frac{\kappa}{n} \, \diag\left\{\left(\xi(x_1)^{-1},\ldots,\xi(x_{N'})^{-1}\right)\right\} \right]^{-1} \\
	& = & \left[C_{11} + W_{\kappa}\left\{\left(\xi(x_1),\ldots,\xi(x_{N'})\right)\right\} \right]^{-1},
\end{eqnarray*}
in which $C_{11}$ denotes the upper left $N' \times N'$ block of the variance-covariance matrix $C$ corresponding to the points $x$ with $\xi(x) > 0$. The penultimate line is obtained by applying the matrix inversion formula $(A + BCD)^{-1} \equiv A^{-1} - A^{-1} B \left(C^{-1} + D A^{-1} B\right)^{-1} D A^{-1}$ with $A = S_{11}$, $C = V_{11}^{-1}$, and $B = D = I_{N'}$.

Finally, let us denote the first $N'$ rows of $F$ by $F_1$. The information matrix is then given by
\begin{eqnarray*}
	M(\xi) & = & F^{\T} \left[ S^{-1} - S^{-1} T^{-1} S^{-1} \right] F \\
	& = & F_1^{\T} X F_1 = F_1^{\T} \left[C_{11} + W_{\kappa}\left\{\left(\xi(x_1),\ldots,\xi(x_{N'})\right)\right\} \right]^{-1} F_1.
\end{eqnarray*}
This is exactly the form of the information matrix \eqref{eq:infmatrix_vn_orig} for the virtual noise model \eqref{eq:model_virtual_noise} with virtual noise variance \eqref{eq:variance_virtual_noise_original} when $\xi \in \Xi\backslash\Xi_+$.
\qed


\section{Derivations of gradients and Hessians}

\label{sec:deriv_gradients_Hessians}

To obtain the derivatives $\partial \, L(\xi) \bigl/ \partial \, \xi(\bar{x})$ for $\bar{x} \in \X$, we first compute
\begin{equation}
\frac{\partial \, Z(\xi)^{-1}}{\partial \, \xi(\bar{x})} = - Z(\xi)^{-1} J_{\bar{x}} (C - \kappa \, I_N) Z(\xi)^{-1}, \label{eq:Zinvderiv}
\end{equation}
where $J_{\bar{x}}$ denotes an $N \times N$ matrix where the entry in the row and column corresponding to point $\bar{x}$ is $1$ and all the other entries are $0$. We can use \eqref{eq:Zinvderiv} to obtain
\begin{eqnarray}
\frac{\partial \, L(\xi)}{\partial \, \xi(\bar{x})} & = & \frac{\partial}{\partial \, \xi(\bar{x})} \: \left[F^{\T} Z(\xi)^{-1} \diag\{\xi\} F \right] \notag \\
& = & F^{\T} \left[ \frac{\partial \, Z(\xi)^{-1}}{\partial \, \xi(\bar{x})} \,  \diag\{\xi\}  +  Z(\xi)^{-1} \, \frac{\partial \, \diag\{\xi\}}{\partial \, \xi(\bar{x})} \right] F \notag \\
& = & F^{\T} \left[ - Z(\xi)^{-1} J_{\bar{x}} (C - \kappa \, I_N) Z(\xi)^{-1}  \diag\{\xi\} + Z(\xi)^{-1}  J_{\bar{x}} \right] F \notag \\
& = & F^{\T} Z(\xi)^{-1} J_{\bar{x}} \left[I_N - (C - \kappa \, I_N) Z(\xi)^{-1}  \diag\{\xi\}\right] F \notag \\
& = & \frac{\kappa}{n} \, F^{\T} Z(\xi)^{-1} J_{\bar{x}} \left\{ Z(\xi)^{-1} \right\}^{\T} F. \label{eq:Lderiv}
\end{eqnarray}
The identity $I_N - (C - \kappa \, I_N) Z(\xi)^{-1}  \diag\{\xi\} \equiv (\kappa/n) \left\{ Z(\xi)^{-1} \right\}^{\T} \equiv \left\{(n/\kappa) Z(\xi)^{\T} \right\}^{-1}$, where $(n/\kappa) Z(\xi)^{\T} = (n/\kappa) (C - \kappa \, I_N) \, \diag\{\xi\} + I_N$, follows from the matrix inversion lemma $(A + BCD)^{-1} \equiv A^{-1} - A^{-1} B (C^{-1} + D A^{-1} B)^{-1} D A^{-1}$ with $A = I_N$, $B = (C - \kappa \, I_N)$, $C = (n/\kappa) \, I_N$, and $D = \diag\{\xi\}$.

We can now use Eq.~\eqref{eq:Lderiv} to derive the expressions for $\partial \, \Phi\{L(\xi)\} / \partial \, \xi(\bar{x})$. For D-optimality, we get
\begin{eqnarray*}
	\frac{\partial \, \Phi_D\{L(\xi)\}}{\partial \, \xi(\bar{x})} & = & \tr \left( \nabla_L \Phi_D\{L(\xi)\} \, \frac{\partial \, L(\xi)}{\partial \, \xi(\bar{x})} \right) \\
	& = & \psi_D \, \frac{\kappa}{n} \, \tr \left[L(\xi)^{-1} F^{\T} Z(\xi)^{-1} J_{\bar{x}} \left\{ Z(\xi)^{-1} \right\}^{\T} F \right] \\
	& = & \psi_D \, \frac{\kappa}{n} \, \tr \left[\left\{ Z(\xi)^{-1} \right\}^{\T} F L(\xi)^{-1} F^{\T} Z(\xi)^{-1} J_{\bar{x}}  \right] \\
	& = & \psi_D \, \frac{\kappa}{n} \, \left\{\left[Z(\xi)^{-1}\right]_{\cdot,\bar{x}} \right\}^{\T} F L(\xi)^{-1} F^{\T} \left[Z(\xi)^{-1}\right]_{\cdot,\bar{x}},
\end{eqnarray*}
where $\left[Z(\xi)^{-1}\right]_{\cdot,\bar{x}}$ denotes the column of $Z(\xi)^{-1}$ corresponding to the point $\bar{x}$.

It follows that $\nabla_{\xi} \Phi_D\{L(\xi)\}$ is given by the diagonal elements of the matrix
\begin{equation*}
\psi_D \, \frac{\kappa}{n} \, \left\{ Z(\xi)^{-1} \right\}^{\T} F L(\xi)^{-1} F^{\T} Z(\xi)^{-1}.
\end{equation*}

For A-optimality, the derivation is very similar. The expression for $\partial \, \Phi_A\{L(\xi)\} / \partial \, \xi(\bar{x})$ is obtained by replacing $\psi_D$ with $1$ and $L(\xi)^{-1}$ with $L(\xi)^{-2}$ in the derivation above. The result is
\begin{equation*}
\frac{\partial \, \Phi_A\{L(\xi)\}}{\partial \, \xi(\bar{x})} = \frac{\kappa}{n} \left\{\left[Z(\xi)^{-1}\right]_{\cdot,\bar{x}} \right\}^{\T} F L(\xi)^{-2} F^{\T} \left[Z(\xi)^{-1}\right]_{\cdot,\bar{x}},
\end{equation*}
and the gradient $\nabla_{\xi} \Phi_A\{L(\xi)\}$ consists of the diagonal elements of
\begin{equation*}
\frac{\kappa}{n} \left\{ Z(\xi)^{-1} \right\}^{\T} F L(\xi)^{-2} F^{\T} Z(\xi)^{-1}.
\end{equation*}

Some algorithms also require the Hessian $H(\xi) = \nabla_{\xi}^{\T} \left[\nabla_{\xi} \Phi\{L(\xi)\} \right]$. Let $\displaystyle H(\xi)_{\bar{x},\bar{y}} = \frac{\partial \, \Phi\{L(\xi)\}}{\partial \, \xi(\bar{x}) \: \partial \, \xi(\bar{y})}$ denote the element of the Hessian corresponding to the mixed derivative of the measures at the points $\bar{x}$ and $\bar{y}$.

We start with deriving the Hessian for $\Phi_{D1}\{L(\xi)\}$. Before we do that, we need the expression for
\begin{eqnarray}
\frac{\partial \, L(\xi)^{-1}}{\partial \, \xi(\bar{x})} & = & - L(\xi)^{-1} \frac{\partial \, L(\xi)}{\partial \, \xi(\bar{x})} L(\xi)^{-1} \notag \\
& = & - \frac{\kappa}{n} \, L(\xi)^{-1} F^{\T} Z(\xi)^{-1} J_{\bar{x}} \left\{ Z(\xi)^{-1} \right\}^{\T} F L(\xi)^{-1}. \label{eq:Linvderiv}
\end{eqnarray}

Now we can compute
\begin{eqnarray}
\frac{\partial \, \Phi_{D1}\{L(\xi)\}}{\partial \, \xi(\bar{x}) \: \partial \, \xi(\bar{y})} & = & \frac{\partial}{\partial \, \xi(\bar{y})} \left\{ \frac{\partial \, \Phi_{D1}\{L(\xi)\}}{\partial \, \xi(\bar{x})} \right\} \notag\\
& = & \frac{\partial}{\partial \, \xi(\bar{y})} \left\{ \frac{\kappa}{n} \, \tr \left[\left\{ Z(\xi)^{-1} \right\}^{\T} F L(\xi)^{-1} F^{\T} Z(\xi)^{-1} J_{\bar{x}}  \right] \right\} \notag \\
& = & \frac{\kappa}{n}\, \tr\biggl[ \frac{\partial  \left\{ Z(\xi)^{-1} \right\}^{\T}}{\partial \, \xi(\bar{y})} \, F L(\xi)^{-1} F^{\T} Z(\xi)^{-1} J_{\bar{x}} \notag \\
& & \qquad \; + \left\{ Z(\xi)^{-1} \right\}^{\T} F \, \frac{\partial \, L(\xi)^{-1}}{\partial \, \xi(\bar{y})} \, F^{\T} Z(\xi)^{-1} J_{\bar{x}} \notag \\
& & \qquad \; + \left\{ Z(\xi)^{-1} \right\}^{\T} F L(\xi)^{-1} F^{\T} \, \frac{\partial \, Z(\xi)^{-1}}{\partial \, \xi(\bar{y})} \, J_{\bar{x}} \biggr] \notag \\
& = & \frac{\kappa}{n} \, \tr\biggl[ \left\{ Z(\xi)^{-1} \right\}^{\T} F \, \frac{\partial \, L(\xi)^{-1}}{\partial \, \xi(\bar{y})} \, F^{\T} Z(\xi)^{-1} J_{\bar{x}} \notag \\
& & \qquad \; + 2 \left\{ Z(\xi)^{-1} \right\}^{\T} F L(\xi)^{-1} F^{\T} \, \frac{\partial \, Z(\xi)^{-1}}{\partial \, \xi(\bar{y})} \, J_{\bar{x}} \biggr]. \label{eq:PhiDmixedderiv}
\end{eqnarray}
The last equation follows from $$\tr \left[\frac{\partial  \left\{ Z(\xi)^{-1} \right\}^{\T}}{\partial \, \xi(\bar{y})} \, F L(\xi)^{-1} F^{\T} Z(\xi)^{-1} J_{\bar{x}}\right] = \tr \left[J_{\bar{x}} \frac{\partial  \left\{ Z(\xi)^{-1} \right\}^{\T}}{\partial \, \xi(\bar{y})} \, F L(\xi)^{-1} F^{\T} Z(\xi)^{-1} \right]$$ being equal to $$\tr \left[ \left\{ Z(\xi)^{-1} \right\}^{\T} F L(\xi)^{-1} F^{\T} \, \frac{\partial \, Z(\xi)^{-1}}{\partial \, \xi(\bar{y})} \, J_{\bar{x}} \right],$$ since $\tr(A) = \tr(A^{\T})$ for any square matrix $A$.

Plugging \eqref{eq:Zinvderiv} and \eqref{eq:Linvderiv} (with point $\bar{x}$ replaced by point $\bar{y}$) into \eqref{eq:PhiDmixedderiv} yields
\begin{eqnarray*}
	\frac{\partial \, \Phi_{D1}\{L(\xi)\}}{\partial \, \xi(\bar{x}) \: \partial \, \xi(\bar{y})} & = & 
	-\frac{\kappa}{n} \, \tr\biggl[ \frac{\kappa}{n} \underbrace{\left\{ Z(\xi)^{-1} \right\}^{\T} F L(\xi)^{-1} F^{\T} Z(\xi)^{-1}}_{P(\xi)} J_{\bar{y}} \underbrace{\left\{ Z(\xi)^{-1} \right\}^{\T} F L(\xi)^{-1} F^{\T} Z(\xi)^{-1}}_{P(\xi)} J_{\bar{x}} \notag \\
	& & \qquad + 2 \underbrace{\left\{ Z(\xi)^{-1} \right\}^{\T} F L(\xi)^{-1} F^{\T} Z(\xi)^{-1}}_{P(\xi)} J_{\bar{y}} \underbrace{(C - \kappa \, I_N) Z(\xi)^{-1}}_{R(\xi)} J_{\bar{x}} \biggr] \\
	& = & - \frac{\kappa}{n} \left[ \frac{\kappa}{n} \, P(\xi)_{\bar{x},\bar{y}} \, P(\xi)_{\bar{y},\bar{x}} + 2 \, P(\xi)_{\bar{x},\bar{y}} \, R(\xi)_{\bar{y},\bar{x}} \right] \\
	& = & - \frac{\kappa}{n} \left[ \frac{\kappa}{n} \, \left\{P(\xi)_{\bar{x},\bar{y}}\right\}^2 + 2 \, P(\xi)_{\bar{x},\bar{y}} \, R(\xi)_{\bar{x},\bar{y}} \right],
\end{eqnarray*}
where, for example, $P(\xi)_{\bar{x},\bar{y}}$ denotes the element of $P(\xi)$ in the row corresponding to point $\bar{x}$ and in the column corresponding to point $\bar{y}$. The last line follows because $P(\xi)$ and $R(\xi)$ are symmetric.

In summary, the Hessian for the first formulation of the D-optimality criterion is
\begin{equation*}
H_{D1}(\xi) = - \frac{\kappa}{n} \, P(\xi) \circ \left[ \frac{\kappa}{n} P(\xi) + 2 R(\xi)\right],
\end{equation*}
where $\circ$ denotes the element-wise (Hadamard) product.

For A-optimality, the derivation of the Hessian for $\Phi_{A}(\{L(\xi)\})$ requires
\begin{eqnarray}
\frac{\partial \, L(\xi)^{-2}}{\partial \, \xi(\bar{x})} & = & \frac{\partial \, L(\xi)^{-1} L(\xi)^{-1}}{\partial \, \xi(\bar{x})} \notag \\
& = & - L(\xi)^{-1} \frac{\partial \, L(\xi)}{\partial \, \xi(\bar{x})} L(\xi)^{-2} - L(\xi)^{-2} \frac{\partial \, L(\xi)}{\partial \, \xi(\bar{x})} L(\xi)^{-1} \notag \\
& = & - \frac{\kappa}{n} \, L(\xi)^{-1} F^{\T} Z(\xi)^{-1} J_{\bar{x}} \left\{ Z(\xi)^{-1} \right\}^{\T} F L(\xi)^{-2} \notag \\
& & - \frac{\kappa}{n} \, L(\xi)^{-2} F^{\T} Z(\xi)^{-1} J_{\bar{x}} \left\{ Z(\xi)^{-1} \right\}^{\T} F L(\xi)^{-1}. \label{eq:L2invderiv}
\end{eqnarray}

This result can be used to compute
\begin{eqnarray}
\frac{\partial \, \Phi_{A}\{L(\xi)\}}{\partial \, \xi(\bar{x}) \: \partial \, \xi(\bar{y})} & = & \frac{\partial}{\partial \, \xi(\bar{y})} \left\{ \frac{\partial \, \Phi_{A}\{L(\xi)\}}{\partial \, \xi(\bar{x})} \right\} \notag\\
& = & \frac{\partial}{\partial \, \xi(\bar{y})} \left\{ \frac{\kappa}{n} \, \tr \left[\left\{ Z(\xi)^{-1} \right\}^{\T} F L(\xi)^{-2} F^{\T} Z(\xi)^{-1} J_{\bar{x}}  \right] \right\} \notag \\
& = & \frac{\kappa}{n} \, \tr\biggl[ \frac{\partial  \left\{ Z(\xi)^{-1} \right\}^{\T}}{\partial \, \xi(\bar{y})} \, F L(\xi)^{-2} F^{\T} Z(\xi)^{-1} J_{\bar{x}} \notag \\
& & \qquad \; + \left\{ Z(\xi)^{-1} \right\}^{\T} F \, \frac{\partial \, L(\xi)^{-2}}{\partial \, \xi(\bar{y})} \, F^{\T} Z(\xi)^{-1} J_{\bar{x}} \notag \\
& & \qquad \; + \left\{ Z(\xi)^{-1} \right\}^{\T} F L(\xi)^{-2} F^{\T} \, \frac{\partial \, Z(\xi)^{-1}}{\partial \, \xi(\bar{y})} \, J_{\bar{x}} \biggr] \notag \\
& = & \frac{\kappa}{n}  \, \tr\biggl[ \left\{ Z(\xi)^{-1} \right\}^{\T} F \, \frac{\partial \, L(\xi)^{-2}}{\partial \, \xi(\bar{y})} \, F^{\T} Z(\xi)^{-1} J_{\bar{x}} \notag \\
& & \qquad \; + 2 \left\{ Z(\xi)^{-1} \right\}^{\T} F L(\xi)^{-2} F^{\T} \, \frac{\partial \, Z(\xi)^{-1}}{\partial \, \xi(\bar{y})} \, J_{\bar{x}} \biggr]. \label{eq:PhiAmixedderiv}
\end{eqnarray}
Similar to the derivation of the mixed derivative for D-optimality, the last equation follows from the equality of 
$$\tr \left[\frac{\partial  \left\{ Z(\xi)^{-1} \right\}^{\T}}{\partial \, \xi(\bar{y})} \, F L(\xi)^{-2} F^{\T} Z(\xi)^{-1} J_{\bar{x}}\right] = \tr \left[J_{\bar{x}} \frac{\partial  \left\{ Z(\xi)^{-1} \right\}^{\T}}{\partial \, \xi(\bar{y})} \, F L(\xi)^{-2} F^{\T} Z(\xi)^{-1} \right]$$ and $$\tr \left[ \left\{ Z(\xi)^{-1} \right\}^{\T} F L(\xi)^{-2} F^{\T} \, \frac{\partial \, Z(\xi)^{-1}}{\partial \, \xi(\bar{y})} \, J_{\bar{x}} \right].$$

Continuing from \eqref{eq:PhiAmixedderiv}, by plugging in \eqref{eq:Zinvderiv} and \eqref{eq:L2invderiv} (taking the derivatives with respect to $\xi(\bar{y})$ instead of $\xi(\bar{x})$), one obtains
\begin{eqnarray*}
	\frac{\partial \, \Phi_{A}\{L(\xi)\}}{\partial \, \xi(\bar{x}) \: \partial \, \xi(\bar{y})} & = & 
	- \frac{\kappa}{n} \, \tr\biggl[ \frac{\kappa}{n} \underbrace{\left\{ Z(\xi)^{-1} \right\}^{\T} F L(\xi)^{-1} F^{\T} Z(\xi)^{-1}}_{P(\xi)} J_{\bar{y}} \underbrace{\left\{ Z(\xi)^{-1} \right\}^{\T} F L(\xi)^{-2} F^{\T} Z(\xi)^{-1}}_{Q(\xi)} J_{\bar{x}} \notag \\
	&  & 
	\qquad \; + \frac{\kappa}{n} \underbrace{\left\{ Z(\xi)^{-1} \right\}^{\T} F L(\xi)^{-2} F^{\T} Z(\xi)^{-1}}_{Q(\xi)} J_{\bar{y}} \underbrace{\left\{ Z(\xi)^{-1} \right\}^{\T} F L(\xi)^{-1} F^{\T} Z(\xi)^{-1}}_{P(\xi)} J_{\bar{x}} \notag \\
	& & \qquad \;  + 2 \underbrace{\left\{ Z(\xi)^{-1} \right\}^{\T} F L(\xi)^{-2} F^{\T} Z(\xi)^{-1}}_{Q(\xi)} J_{\bar{y}} \underbrace{(C - \kappa \, I_N) Z(\xi)^{-1}}_{R(\xi)} J_{\bar{x}} \biggr] \\
	& = & -\frac{\kappa}{n} \, \tr \biggl[ \frac{\kappa}{n} \, J_{\bar{x}} P(\xi) J_{\bar{y}} Q(\xi) + \frac{\kappa}{n} \, Q(\xi) J_{\bar{y}} P(\xi) J_{\bar{x}}  \\
	&  & \qquad \; + 2 \, Q(\xi) J_{\bar{y}} R(\xi) J_{\bar{x}} \biggr] \\
	& = & -\frac{\kappa}{n} \, \tr \biggl[ \frac{2\kappa}{n} \, Q(\xi) J_{\bar{y}} P(\xi) J_{\bar{x}} + 2 \, Q(\xi) J_{\bar{y}} R(\xi) J_{\bar{x}} \biggr] \\
	& = & - \frac{2 \kappa}{n} \left[ \frac{\kappa}{n} \, Q(\xi)_{\bar{x},\bar{y}} \, P(\xi)_{\bar{y},\bar{x}} + Q(\xi)_{\bar{x},\bar{y}} \, R(\xi)_{\bar{y},\bar{x}} \right] \\
	& = & - \frac{2 \kappa}{n} \left[ \frac{\kappa}{n} \, Q(\xi)_{\bar{x},\bar{y}} \, P(\xi)_{\bar{x},\bar{y}} + Q(\xi)_{\bar{x},\bar{y}} \, R(\xi)_{\bar{x},\bar{y}} \right]
\end{eqnarray*}
due to the symmetry of $P(\xi)$, $Q(\xi)$, and $R(\xi)$. The third line from the bottom in the derivation above follows from the fact that $$\left[ J_{\bar{x}} P(\xi) J_{\bar{y}} Q(\xi) \right]^{\T} = Q(\xi) J_{\bar{y}} P(\xi) J_{\bar{x}}.$$

It follows that the Hessian of $\Phi_{A}\{L(\xi)\}$ is given by
\begin{equation*}
H_{A}(\xi) = - \frac{2 \kappa}{n} \, Q(\xi) \circ \left[ \frac{\kappa}{n} P(\xi) + R(\xi)\right].
\end{equation*}


\section{Kuhn-Karush-Tucker conditions for simplicial decomposition}

\label{sec:KKT_conditions}

Before running the CGP and the RMP, one needs to check whether the algorithm can stop. To that end, the Kuhn-Karush-Tucker conditions are examined. The following conditions must hold if the criterion is maximized at the current solution $\xi^{(k)}$:
\begin{equation}
\left[\nabla \, \Phi\left\{L\left(\xi^{(k)}\right)\right\}\right]_i \: \begin{cases}
\leq \lambda \quad \text{ if } \xi^{(k)}(x_i) = 0, \\
= \lambda \quad \text{ if } 0 < \xi^{(k)}(x_i) < 1/n, \\
\geq \lambda \quad \text{ if } \xi^{(k)}(x_i) = 1/n,
\end{cases} \label{eq:kuhn-karush-tucker}
\end{equation}
for $i = 1,\ldots,N$ and some scalar $\lambda$.

In practical computations, numerical tolerances are taken into account. The design measure at point $x_i$ is considered to be $0$ if $\xi^{(k)}(x_i) \leq \delta$ and it is considered to be $1/n$ if $\xi^{(k)}(x_i) \geq 1/n - \delta$ for some small $\delta > 0$. The following procedure is used to decide whether the Kuhn-Karush-Tucker conditions are approximately fulfilled. Denote the largest partial derivative in \eqref{eq:kuhn-karush-tucker} among all the points where $\xi^{(k)}(x_i) \approx 0$ by $\lambda_{\mathrm{low},\max}$, the smallest partial derivative among all the points where $\xi^{(k)}(x_i) \approx 1/n$ by $\lambda_{\mathrm{up},\min}$, the smallest and largest partial derivative among all the points where $\xi^{(k)}(x_i)$ is considered to be $> 0$ and $< 1/n$ by $\lambda_{\mathrm{int},\min}$ and $\lambda_{\mathrm{int},\max}$, respectively, and compute $\lambda_{\mathrm{int},\mathrm{avg}} = (\lambda_{\mathrm{int},\min} + \lambda_{\mathrm{int},\max})/2$. Given some small tolerance $\epsilon > 0$, the Kuhn-Karush-Tucker conditions are considered as being approximately satisfied if
\begin{eqnarray*}
	\lambda_{\mathrm{int},\max} - \lambda_{\mathrm{int},\min} & \leq & 2\epsilon, \\
	\lambda_{\mathrm{low},\max} - \lambda_{\mathrm{int},\mathrm{avg}} & \leq & \epsilon, \\
	\lambda_{\mathrm{int},\mathrm{avg}} - \lambda_{\mathrm{up},\min} & \leq & \epsilon.
\end{eqnarray*}


\section{Multiplicative algorithm}

\label{sec:multiplicative}

The restricted master problem \eqref{eq:SD_RMP2} with the constraint $w \in \setV_{k+1}$ can be solved with gradient-projection methods, see \cite{pronzato_design_2014} and \cite{bertsekas_nonlinear_2016}. These are iterative methods where the weights at step $l+1$ are updated according to 
\begin{equation}
w^{(l+1)} = w^{(l)} + \alpha_k \, d^{(l)}, \label{eq:weight_update}
\end{equation}
where $d^{(l)}$ is the direction and $\alpha_k$ controls the step size. Let $\phi(w)$ denote the function to be maximized. In the applications of this paper, $\phi(w) = \Phi\{L(X_{k+1} w)\}$. For so-called second-order methods, the direction $d^{(l)}$ is chosen to be the projection of $\Lambda \, \nabla_w \phi(w)$ onto the set $\mathcal{D}_{k+1} = \left\{d: \: \sum_{i=1}^{k+1} d_i = 0\right\}$ under the norm $\|x\|_{\Lambda^{-1}} = x^{\T} \Lambda^{-1} x$, where $\Lambda$ is a positive definite $(k+1) \times (k+1)$ matrix. Setting $\Lambda$ to the inverse Hessian gives the (projected) Newton method. The constraint $d^{(l)} \in \mathcal{D}_{k+1}$ guarantees that $w^{(l+1)} \in \setV_{k+1}$ if $w^{(l)} \in \setV_{k+1}$.

If all $w_i > 0$, the projected direction is given by
\begin{equation}
d^{(l)} = \Lambda \, \nabla_w \phi\left(w^{(l)}\right) - \frac{\iota^{\T} \Lambda \, \nabla_w \phi\left(w^{(l)}\right)}{\iota^{\T} \Lambda \iota} \, \Lambda \iota, \label{eq:projected_direction}
\end{equation}
where $\iota$ denotes a vector of ones. It can be shown that this is an ascent direction when $\alpha_k \rightarrow 0^+$.

To obtain the multiplicative algorithm, the matrix $\Lambda$ is set to $\Lambda = \diag\left\{w^{(l)}\right\}$ and the step size is chosen to be $\alpha_k = 1 / \left(\nabla^{\T} \phi\left(w^{(l)}\right)  w^{(l)}\right)$. Plugging these values into \eqref{eq:projected_direction} and \eqref{eq:weight_update}, the weight update becomes
\begin{eqnarray*}
	w^{(l+1)} & = & w^{(l)} + \frac{1}{\nabla^{\T}_w \phi\left(w^{(l)}\right) w^{(l)}} \, \left[\diag\left\{w^{(l)}\right\} \nabla_w \phi\left(w^{(l)}\right) - \frac{\nabla^{\T}_w \phi\left(w^{(l)}\right) w^{(l)}}{\underbrace{\iota^{\T} \, w^{(l)}}_{= 1}} \, w^{(l)}  \right] \\
	& = & w^{(l)} + \frac{1}{\nabla^{\T}_w \phi\left(w^{(l)}\right) w^{(l)}} \, \diag\left\{w^{(l)}\right\} \nabla_w \phi\left(w^{(l)}\right) - w^{(l)} \\
	& = & \frac{1}{\nabla^{\T}_w \phi\left(w^{(l)}\right) w^{(l)}} \, \diag\left\{w^{(l)}\right\} \nabla_w \phi\left(w^{(l)}\right).
\end{eqnarray*}

Therefore, component $i$ of the weights is updated according to
\begin{equation*}
w^{(l+1)}_i = w^{(l)}_i \cdot \frac{\left[\nabla_w \phi\left(w^{(l)}\right)\right]_i}{\nabla^{\T}_w \phi\left(w^{(l)}\right) w^{(l)}},
\end{equation*}
hence the name \emph{multiplicative algorithm}. One may generalize the weight updates to
\begin{equation*}
w^{(l+1)}_i = w^{(l)}_i \cdot \frac{\left\{\left[\nabla_w \phi\left(w^{(l)}\right)\right]_i\right\}^{\beta}}{\left\{\nabla^{\T}_w \phi\left(w^{(l)}\right)\right\}^{\beta} w^{(l)}} = w_i^{(l)} \cdot m_i\left(w^{(l)}, \beta\right)
\end{equation*}
for some $\beta > 0$, where the power is meant to be applied component-wise to the gradient vector $\nabla^{\T}_w \phi\left(w^{(l)}\right)$. \cite{yu_monotonic_2010} proves the monotonicity of these updates for the criterion $\Phi\{L(w)\}$ if $\beta \in (0,1]$ and if $-\Phi\left\{L^{-1}\right\}$ is monotonic with respect to Loewner's ordering and concave for all positive definite matrices $L$.

The stopping rule for the multiplicative algorithm is that all the update factors $m_i\left(w^{(l)}, \beta\right) \leq 1 + \tau$ for some small $\tau > 0$.

Finally, we state the formula for the gradient $\nabla_w \phi(w)$ when $\phi(w) = \Phi\{L(X_{k+1} w)\}$. It is
\begin{equation*}
\nabla_w \Phi\{L(X_{k+1} w)\} = X_{k+1}^{\T} \nabla_{\xi} \Phi\{L(\xi)\} \bigl|_{\xi = X_{k+1} w},
\end{equation*}
where the expressions for  $\nabla_{\xi} \Phi\{L(\xi)\}$ for the various criteria can be found in Section~\ref{sec:algorithms_preliminaries}.

\section{Projected Newton method}

\label{sec:projected_Newton}

We use the method by \cite{bertsekas_projected_1983} to solve the following optimization problem:
\begin{align}
\min_{w} &\;\: \phi(w) = -\Phi\{L(X_{k+1} w)\} \label{eq:PN_problem} \\
\text{subject to} &\;\: \sum_{i=1}^{k+1} w_i = 1,  \notag \\
&\;\: w_i \geq 0, \quad i = 1,\ldots,k+1. \notag
\end{align}

This optimization problem can be simplified by eliminating the restriction $\sum_{i=1}^{k+1} w_i = 1$. Assume w.l.o.g.\ that the elements of $w$ and the corresponding columns of $X_{k+1}$ are arranged in step $l$ such that the highest index belongs to the largest element in the current weight vector $w^{(l)}$, i.e., $w^{(l)}_{k+1} = \max_{i \in \{1,\ldots,k+1\}} w^{(l)}_i$. The optimization is now performed over the $k$-dimensional vector $y = \left(y_1,\ldots,y_k\right)^{\T} = \left(w_1,\ldots,w_k\right)^{\T}$ and the objective function in step $l$ is therefore
\begin{equation}
h_l(y) = \phi\left(y_1,\ldots,y_k, 1 - \sum_{i=1}^k y_i\right). \label{eq:reduced_objective}
\end{equation}
Since the constraint $\sum_{i=1}^k y_i \leq 1$ is not active in a neighborhood of $y^{(l)}$ by construction of $y$, the optimization problem at step $l$ becomes (at least locally near $y^{(l)}$)
\begin{align*}
\min_{y} &\;\: h_l(y) \label{eq:PN_problem_reduced} \\
&\;\: y_i \geq 0, \quad i = 1,\ldots,k. \notag
\end{align*}
To solve this optimization problem, one step of the iterative update procedure
\begin{equation*}
y^{(l+1)} = y^{(l)}(\alpha_l) = \max\left\{0, \, y^{(l)} - \alpha_l \, \Lambda_l \, \nabla_y \, h_l\left(y^{(l)}\right) \right\} \label{eq:PN_update}
\end{equation*}
is conducted. After obtaining $y^{(l+1)}$, $w^{(l+1)}$ is constructed by setting $w^{(l+1)}_i = y^{(l+1)}_i$ for $i=1,\ldots,k$, and $w^{(l+1)}_{k+1} = 1-\sum_{i=1}^k y^{(l+1)}_i$. Then the Kuhn-Karush-Tucker conditions are checked for $\nabla_w \, \phi\left(w^{(l+1)}\right)$, similar to checking whether the simplicial decomposition method has converged, see Appendix~\ref{sec:KKT_conditions}. If the solution has not converged yet, the algorithm enters its next iteration.

The block of the matrix $\Lambda_l$ corresponding to the elements of $y^{(l)}$ which are greater than $0$ is the inverse of the Hessian of $h_l\left(y^{(l)}\right)$ corresponding to those elements of $y^{(l)}$. For the block of the matrix $\Lambda_l$ corresponding to the elements of $y^{(l)}$ which are equal to $0$, only the diagonal elements of the Hessian are used and inverted. All other elements of $\Lambda_l$ are $0$.

In fact, to avoid zigzagging behavior, all the elements of $y^{(l)}$ for which $y^{(l)}_i \leq \mu \cdot \left[\nabla_y \, h_l\left(y^{(l)}\right)\right]_i$ for some constant $\mu > 0$ when $\left[\nabla_y \, h_l\left(y^{(l)}\right)\right]_i > 0$ are considered as being $0$, so the corresponding off-diagonal elements of the Hessian are set to $0$.

In order to ensure that $\sum_{i=1}^{k} y^{(l+1)}_i \leq 1$, the step size $\alpha_l$ must not exceed $\bar{\alpha}_l$, where
\begin{equation*}
\bar{\alpha}_l = \sup\left\{ \alpha \: \biggl| \: \sum_{i=1}^k y^{(l)}_i(\alpha) \leq 1 \right\}.
\end{equation*}
A simple search algorithm of complexity $\mathcal{O}(k)$ can be applied to determine $\bar{\alpha}_l$.

\cite{bertsekas_projected_1983} prove that this projected Newton scheme always achieves descent and terminates at a global minimum for a convex function $\phi(w)$ if $\Lambda_l$ is chosen as described above and $\alpha_l \leq \bar{\alpha}_l$.

We find the actual value $\alpha_l \leq \bar{\alpha}_l$ by following an Armijo rule. The Armijo rule is an iterative step size reduction scheme where the step size is decreased until the improvement of the objective function exceeds some threshold, where the threshold decreases over time. This scheme ensures that the improvements are significant enough such that the gradient descent scheme is guaranteed to converge. In our example the threshold is composed of two different parts, depending on whether the elements of $y^{(l)}$ are considered to be $0$ or greater than $0$. Let the vector consisting of the elements of $y^{(l)}$ being $> 0$ be denoted by $\bar{y}^{(l)}$, the block of $\Lambda_l$ pertaining to those elements by $\bar{\Lambda}_l$, and the gradient vector with respect to $\bar{y}$ by $\nabla_{\bar{y}} \, h_l\left(y^{(l)}\right)$. The search direction with respect to $\bar{y}^{(l)}$ is given by $\bar{d}^{(l)} = -\bar{\Lambda}_l \nabla_{\bar{y}} \, h_l\left(y^{(l)}\right)$. On the other hand, denote by $\tilde{y}^{(l)}$ the elements of $y^{(l)}$ that are $\approx 0$ and denote by $\nabla_{\tilde{y}} \, h_l\left(y^{(l)}\right)$ the corresponding elements of the gradient.

The Armijo rule is defined by the parameters $\sigma > 0$, $\beta \in (0,1)$, and an initial step size $\bar{s}_l$. We take $\bar{s}_l = \min\{1, \bar{\alpha_l}\}$, which is a common initial step size. The step size $\alpha_l$ is set to $\alpha_l = \beta^m \bar{s}_l$, where $m$ is the first nonnegative integer for which the improvement
\begin{equation*}
h_l\left(y^{(l)}\right) - h_l\left(y^{(l)}\left(\beta^m \bar{s}_l\right)\right) \geq -\sigma \beta^m \bar{s}_l \, \nabla_{\bar{y}}^{\T} \, h_l\left(y^{(l)}\right) \bar{d}^{(l)} + \sigma \, \nabla_{\tilde{y}}^{\T} \, h_l\left(y^{(l)}\right) \left(\tilde{y}^{(l)} - \tilde{y}^{(l)} \left(\beta^m \bar{s}_l\right)\right).
\end{equation*}
The first term on the right-hand side is the standard Armijo threshold for unconstrained optimization, whereas the second term is the threshold for the Armijo rule along the projection arc.

To compute the gradient and Hessian of $h_l(y)$ defined by Eq.~\eqref{eq:reduced_objective}, the first and second partial derivatives with respect to $y_i$ and $y_j$ for $i,j = 1,\ldots,k$ need to be computed. They are given by
\begin{equation*}
\frac{\partial \, h_l(y)}{\partial \, y_i} = \frac{\partial \, \phi(w)}{\partial \, w_i} - \frac{\partial \, \phi(w)}{\partial \, w_{k+1}}
\end{equation*}
and
\begin{equation*}
\frac{\partial^2 \, h_l(y)}{\partial \, y_i \, \partial \, y_j} = \frac{\partial^2 \, \phi(w)}{\partial \, w_i \, \partial \, w_j} - \frac{\partial^2 \, \phi(w)}{\partial \, w_i \, \partial \, w_{k+1}} - \frac{\partial^2 \, \phi(w)}{\partial \, w_{k+1} \, \partial \, w_{j}} + \frac{\partial^2 \, \phi(w)}{\partial \, w_{k+1} \, \partial \, w_{k+1}},
\end{equation*}
where the derivatives of $\phi(\cdot)$ are evaluated at $w = \left(y_1,\ldots,y_k, 1-\sum_{i=1}^k y_i\right)^{\T}$.

To conclude, we provide the formula for the Hessian of $\phi(w) = -\Phi\{L(X_{k+1} w)\}$. It is 
\begin{equation*}
\nabla_w^{\T}\left[\nabla_w \phi(w)\right] = -X_{k+1}^{\T} H(\xi) \bigr|_{\xi = X_{k+1} w} X_{k+1},
\end{equation*}
where $H(\xi)$ denotes the Hessian with respect to the design measures for the respective criterion $\Phi\{\cdot\}$ derived in Appendix~\ref{sec:deriv_gradients_Hessians}.


\section{Criterion update functions for finding exact designs}

\label{sec:criterion_updates}

\cite{brimkulov_numerical_1980} derive the formula for updating the determinant of the information matrix in the correlated setting.

For an exact design $\tau = \{x_1,\ldots,x_n\}$, the covariance matrix $C$ has entries 
\begin{equation*}
C_{ij} = \Cov\{\varepsilon(x_i), \varepsilon(x_j)\}; \qquad i,j = 1,\ldots,n,
\end{equation*}
and the $n \times p$ model matrix is $F(\tau) = (f(x_1) \cdots f(x_n))^{\T}$. The information matrix is then given by
\begin{equation*}
M(\tau) = F(\tau)^{\T} C(\tau)^{-1} F(\tau),
\end{equation*}
see Equation~\eqref{eq:information_matrix_exact}.

We want to add the point $\bar{x}$ to the design $\tau$ to obtain the augmented design $\bar{\tau} = \{\tau, \bar{x}\}$. The determinant of the information matrix for the augmented design can be computed by the following updating formula:
\begin{equation*}
\det \{M(\bar{\tau})\} = \det \{M(\tau)\} \cdot \left[1 + \frac{\tilde{f}(\bar{x})^\T M(\tau)^{-1} \tilde{f}(\bar{x})}{\tilde{\sigma}^2(\bar{x})} \right],
\end{equation*}
where 
\begin{eqnarray}
\tilde{\sigma}^2(\bar{x}) & = & \Var\{\varepsilon(\bar{x})\} - c(\tau,\bar{x})^\T C^{-1}(\tau) c(\tau,\bar{x}), \label{eq:brimkulov_variance}\\
\tilde{f}(\bar{x}) & = & f(\bar{x}) - F(\tau)^\T C^{-1}(\tau) c(\tau,\bar{x}), \label{eq:brimkulov_covariates}
\end{eqnarray}
with 
\begin{equation*}
c(\tau,\bar{x}) = \left[\Cov\{\varepsilon(x_1),\varepsilon(\bar{x})\}, \ldots, \Cov\{\varepsilon(x_n),\varepsilon(\bar{x})\} \right]^{\T}.
\end{equation*}
Therefore, adding the design point $\bar{x}$ for which
\begin{equation*}
\phi_D(\bar{x},\tau) = \frac{\tilde{f}(\bar{x})^\T M(\tau)^{-1} \tilde{f}(\bar{x})}{\tilde{\sigma}^2(\bar{x})}
\end{equation*}
is maximized with respect to $\bar{x}$ gives the design with the maximum determinant of the information matrix across all designs where the design $\tau$ is augmented by one design point.

For A-optimality, Proposition~1 of \cite{liu_sensorselection_2016} establishes that
\begin{equation*}
\tr\{M(\tau)^{-1}\} - \tr\{M(\bar{\tau})^{-1}\} = \frac{[\tilde{\sigma}^2(\bar{x})]^{-1} \cdot \tilde{f}(\bar{x})^\T M(\tau)^{-2} \tilde{f}(\bar{x})}{1 + [\tilde{\sigma}^2(\bar{x})]^{-1} \cdot \tilde{f}(\bar{x})^\T M(\tau)^{-1} \tilde{f}(\bar{x})} \geq 0,
\end{equation*}
where $\tilde{\sigma}^2(\bar{x})$ and $\tilde{f}(\bar{x})$ are given by \eqref{eq:brimkulov_variance} and \eqref{eq:brimkulov_covariates}, respectively. Therefore, maximizing the function
\begin{equation*}
\phi_A(\bar{x},\tau) = \frac{[\tilde{\sigma}^2(\bar{x})]^{-1} \cdot \tilde{f}(\bar{x})^\T M(\tau)^{-2} \tilde{f}(\bar{x})}{1 + [\tilde{\sigma}^2(\bar{x})]^{-1} \cdot \tilde{f}(\bar{x})^\T M(\tau)^{-1} \tilde{f}(\bar{x})}
\end{equation*}
with respect to $\bar{x}$ yields the design which augments the design $\tau$ by one point that minimizes the trace of the inverse information matrix.

\end{document}